\documentclass[twocolumn,showpacs,prx,amsfonts,amsmath,amssymb,floatfix]{revtex4} 
\usepackage{color}
\usepackage{hhline}
\usepackage{mathrsfs}
\usepackage{graphicx}
\usepackage{dcolumn}
\usepackage{bm}
\usepackage{multirow}
\usepackage{booktabs}
\usepackage{afterpage}
\usepackage{amsmath}
\usepackage{ulem}

\begin{document}

\title{Interference of the Bloch phase in layered materials with stacking shift}

\author{Ryosuke Akashi$^{1}$}
\thanks{akashi@cms.phys.s.u-tokyo.ac.jp}
\author{Yo Iida$^{1}$}
\author{Kohei Yamamoto$^{1}$}
\author{Kanako Yoshizawa$^{2}$}
\affiliation{$^1$Department of Physics, the University of Tokyo, Hongo, Bunkyo-ku 113-8656, Japan}
\affiliation{$^2$Research Organization for Information Science and Technology, Kobe 650-0047, Japan}

\date{\today}
\begin{abstract}
In periodic systems, electronic wave functions of the eigenstates exhibit the periodically modulated Bloch phases and are characterized by their wave numbers ${\bf k}$. We theoretically address the effects of the Bloch phase in general layered materials with stacking shift. When the interlayer shift and the Bloch wave vector ${\bf k}$ satisfy certain conditions, interlayer transitions of electrons are prohibited by the interference of the Bloch phase. We specify the manifolds in the ${\bf k}$ space where the hybridization of the Bloch states between the layers is suppressed in accord with the stacking shift. These manifolds, named stacking-adapted interference manifolds (SAIM), are obviously applicable to general layered materials regardless of detailed atomic configuration within the unit cell. We demonstrate the robustness and usefulness of the SAIM with first-principles calculations for layered boron nitride, transition-metal dichalcogenide, graphite, and black phosphorus. We also apply the SAIM to general three-dimensional crystals to derive special {\bf k}-point paths for the respective Bravais lattices, along which the Bloch-phase interference strongly suppresses the band dispersion. Our theory provides a general and novel view on the anisotropic electronic kinetics intrinsic to the periodic-lattice structure. 
\end{abstract}
\pacs{}

\maketitle
\section{Introduction}
\begin{figure}[t]
 \begin{center}
  \includegraphics[scale=.42]{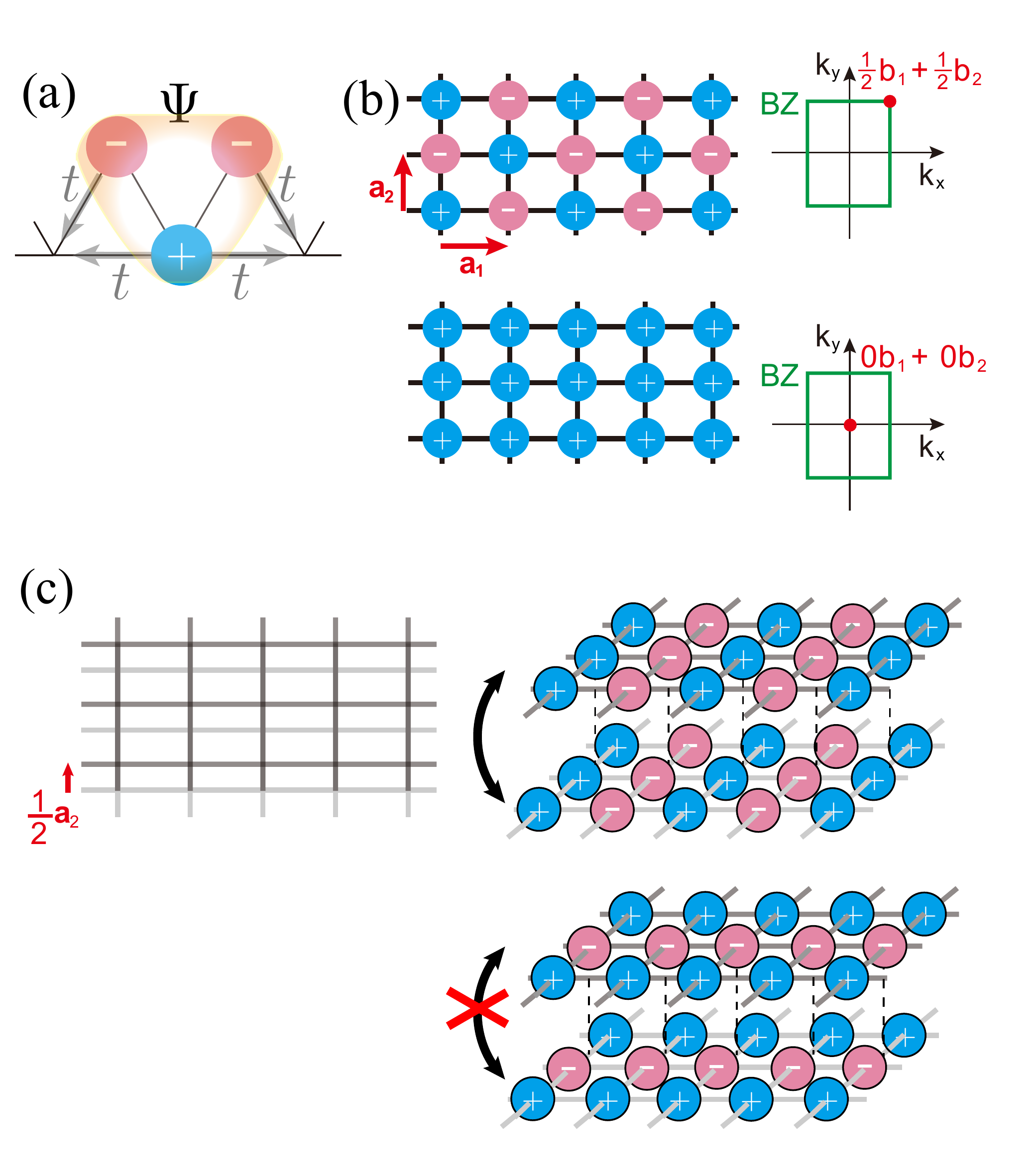}
  \caption{Interference of electronic wave function, demonstrated by simple model wave functions whose values at sites are represented by $\pm 1$. (a) spatially confined electronic eigen state on the triangle-chain lattice~\cite{Tasaki-review-PTP1998}. $t$ denotes the intersite hopping amplitude. The combination of the hopping geometry and the spatial phase structure of the wave function yields zero transition amplitude to the neighboring sites. (b) Spatial phase configuration of the Bloch state on the rectangular lattice. The corresponding crystal wave numbers are indicated in the BZ. (c) Interlayer hybridization between the Bloch states at the neighboring layers. When the second layer is shifted from the original one, the hybridization is canceled for some Bloch states because of the phase interference.}
  \label{fig:schematic}
 \end{center}
\end{figure}

When an electron moves in the condensed matter, interplay of its wave property and atomic configuration often yields nontrivial interference effects, which impose constraints on the electronic dynamics. A celebrated example of such interference effects is the flat band~\cite{Tasaki-review-PTP1998}. Despite nonzero hopping integrals, spatially confined electronic eigenstates emerge in exotic structures such as delta-chain, decorated square, and kagome lattices~[Fig.~\ref{fig:schematic}(a)]. There, possible intriguing phenomena due to electronic correlation have been discussed extensively: The flat-band ferromagnetism~\cite{Lieb-PRL1989, Mielke-JPhysA-1, Mielke-JPhysA-2, Tasaki-PRL1992,Mielke-Tasaki-MathPhys1993}, fractional quantum hall effect~\cite{Bergholtz-Liu-review2013}, etc. Theoretical formulation of the interference effects thus provide us with guiding principles to control the kinetics of electrons, which should facilitate exploration of novel electronic phenomena emerging from the competition of kinetic and local-correlation effects. In this work, we discuss a general effect of interference of the Bloch wave function governing the one-body characteristics of electrons.

The electronic wave function of the eigenstate in the periodic potential has the form~\cite{Seitz-grouptheory1936}
\begin{eqnarray}
\psi_{n{\bf k}}({\bf r})
=
\exp(i{\bf k}\cdot {\bf r})
u_{n{\bf k}} ({\bf r})
,
\label{eq:Bloch-func}
\end{eqnarray}
with ${\bf k}$ being the crystal wavenumber defined in the Brillouin zone (BZ). $u_{{n{\bf k}}}({\bf r})$ denotes the lattice-periodic part which satisfies $u_{n{\bf k}}({\bf r}+{\bf R})=u_{n{\bf k}}({\bf r})$ for any lattice vectors ${\bf R}$. Since the plane-wave part, or the Bloch phase [Fig.~\ref{fig:schematic}(b)], does only concern the periodic structure of the system, this should be utilizable to devise interference effects applicable to a wide range of materials.

In this study, we theoretically address such possible interference effects in general layered structures. Theoretical outcomes of our study are directly applicable to recently synthesized atomically thin materials such as graphene~\cite{Geim-Novoselov-graphene2004}, boron nitride (BN)~\cite{Geim-Novoselov-thinlayer}, transition-metal dichalcogenides (TMD)~\cite{Geim-Novoselov-thinlayer, Park-Lee-1Tprime-MoTe2,Zhou-Dresselhaus-1Tprime-MoTe2,Naylor-Johnson-1Tprime-MoTe2-monolayer} and phosphorene~\cite{Li-Zhang-phosphorene-transistor-nnano, Liu-Tomanek-phosphorene-mobility-acsnano}. Toward development of novel nanoelectronics devices, numerous experimental and theoretical investigations have been carried out on their electronic properties depending on the number and stacking geometry of the layers~\cite{CastroNeto-graphene-review2009,Xu-Heinz-review-NPhys2014, Qian-review-2DMater}. Generally the neighboring layers interact with each other through various mechanisms; electrostatic force, the van der Waals force, electronic interlayer hybridization and charge redistribution~\cite{Tomanek-DMC-phosphorene}, etc. Here we focus on the hybridization from the viewpoint of the interference of the wave functions. When two layers are placed along with each other, their Bloch states hybridize between the layers. Because of the phase of the wave function, this hybridization is prohibited for some combination of ${\bf k}$ and the relative geometry of the layers [Fig.~\ref{fig:schematic}(c)]. This effect should impose {\bf k}-point and stacking-geometry dependent constraints for the interlayer interaction of the electronic states. As a matter of fact, in graphite, it has been known that the interlayer hybridization at special points of the BZ is much affected by the stacking pattern~\cite{McClure-model-graphite1957,Slonczewski-Weiss-model-graphite1958,McClure-rhombo-graphite1969,Charlier-Gonze-multilayer-graphite1994}. By carefully examining electronic band structures in the literature, one can find the quasi-two-dimensional states due to this effect in a variety of works not only on graphite~\cite{Kresse-graphite-BN-band1996, Coehoorn-Wold-TMDC-bandstruct-PES,Dawson-Bullett-TMTe2, McClure-model-graphite1957,Slonczewski-Weiss-model-graphite1958,McClure-rhombo-graphite1969,Charlier-Gonze-multilayer-graphite1994,Partoens-multilayer2006, MacDonald-graphene-chiraldecomp,Chiu-Shyu-r-graphite, Liu-Shen-BN-stack, Aoki-Amawashi-SSComm}. In addition, it has recently been experimentally demonstrated that this interference can be utilized for controlling the interlayer motion of electrons and holes by the stacking geometry~\cite{Suzuki2014, Akashi2015}. However, these features have been discussed in distinct material-specific contexts. Our aim in this study is to establish a general theory which not only relates the above-mentioned phenomena but also applies to any layered materials. Such theory should give us a unified insight into how the electronic states in the layered structures are controllable, which will also facilitate efficient nanoelectronics materials design.

\begin{figure}[t]
 \begin{center}
  \includegraphics[scale=.5]{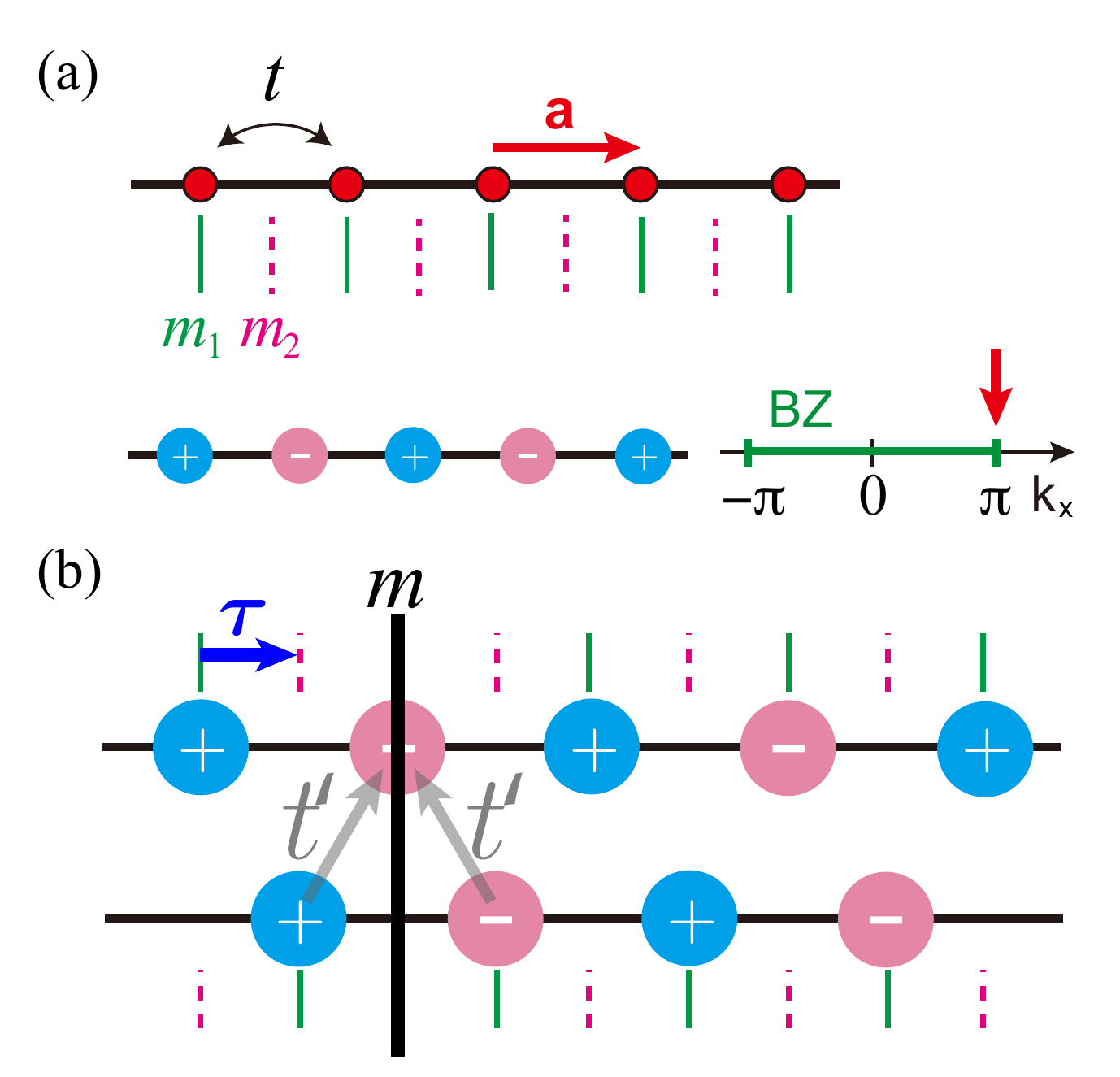}
  \caption{Hybridization between the Bloch states at one-dimensional chains. (a) the single-band tight-binding model on one-dimensional chain, which is invariant under mirror reflections with respect to $m_{1}$ and $m_{2}$. $t$ denotes the intersite hopping amplitude and ${\bm a}$ denotes the primitive translation vector. The phase configuration of the Bloch state with $k$=$\pi$ is displayed. (b) hybridization between the $k$=$\pi$ Bloch states at neighboring chain. The two chains are parallelly stacked with shift $\tau$ so that the inequivalent mirror planes of the neighboring chains coincide with each other. $t'$ denotes the intersite hopping amplitude for the neighboring sites belonging to different chains.}
  \label{fig:1d-chain}
 \end{center}
\end{figure}

In this paper, we present a theory how the Bloch phase governs the possible/impossible interlayer hybridization, which respects only the space group and stacking shift of the layers. For theoretical simplicity, we concentrate on the stacking {\it shift} and leave the stacking {\it twist} beyond our scope. Still, we will see below that the applicability of our theory is broad. 

We specifically formulate special manifolds in the two-dimensional reciprocal space as functions of the stacking shift [Fig.~\ref{fig:schematic} (c)]---the Bloch states in those manifolds cannot hybridize between the layers. These manifolds are determined regardless of the cell periodic part of the wavefunction $u_{n{\bf k}}({\bf r})$ and summarized as the stacking-adapted interference manifolds (SAIM). The characters of the Bloch states in any of the SAIM can be simply and sensitively controlled by the stacking shift between the layers via the change of the interlayer hybridization. To demonstrate this, we calculate the electronic states in layered BN, TMD and graphite from the first principles and relate their properties to the interlayer stacking shift. We show that, once the electronic properties within the SAIM is known for the monolayer form of these materials, we can predict their stacking dependences with only the information of the interlayer shift vectors. Our results show the usefulness of the SAIM for the control of the electronic states in the layered materials. 

In addition to the direct application, we believe that the SAIM is a useful concept to derive various non-trivial electronic properties in general periodic structures. We exemplify this point by decomposing general three-dimensional lattices into stacked layers. It is consequently revealed that special {\bf k}-point paths in the BZ are present where the band dispersion is anomalously suppressed regardless of the atomic configuration within the unit cell. We show the robustness of these paths by examining the first-principles calculations for semiconducting lithium hydride (LiH) and sodium chloride (NaCl) with rocksalt-type crystal structure. These paths, which we name the Bloch-phase induced flat-band paths (BIFP), indicate hidden anisotropic characteristics of the Bloch states intrinsic to the three-dimensional lattice structure.

This paper is organized as follows. In Sec.~\ref{sec:one-dim}, we discuss the Bloch-phase interference in stacked one-dimensional chain as a simple example for later generalization. The general theory for arbitrary layers and derivation of the SAIM are presented in Sec.~\ref{sec:general-theory}. The first-principles electronic structure calculations for BN, TMD, graphite and black phosphorus and their interpretation based on the SAIM are presented in Sec.~\ref{sec:layered-semicon}. The consideration on the general three-dimensional lattice is exhibited in Sec.~\ref{sec:three-dim}, where the BIFP is introduced through an analysis on the tight-binding model on the face-centered cubic (FCC) lattice and the first-principles band dispersions along the BIFP are examined for LiH and NaCl. Miscellaneous topics are discussed in Sec.~\ref{sec:discussion}. Section~\ref{sec:summary} is devoted to summary and conclusions of the present work. The full lists of the SAIM and BIFP are provided in Appendices~\ref{sec:list-net-map} and \ref{sec:list-BZ}.


\section{Introductory example: one-dimensional chain}
\label{sec:one-dim}

We exemplify the present concept with an $s$-wave single orbital tight-binding model on a one-dimensional chain [Fig.~\ref{fig:1d-chain} (a)] with periodic boundary condition. The primitive lattice vector is denoted by $a$. The electronic eigenstates of this system is characterized by the one-dimensional Bloch wavenumber $k$ $(-\pi < k \leq \pi)$. 

Next, we put an identical chain in parallel with the original one with finite shift $\tau$ [Fig.~\ref{fig:1d-chain}(b)]. Since the primitive lattice vector of the total system remains $a$, inter-chain hybridization occurs only between the Bloch states having the same $k$. Although there seems to be no general reason for prohibiting such hybridization, it is not the case for $k=\pi$: When $\tau=a/2$, the hybridization between the $k=\pi$ states of the two systems is suppressed. This is illustrated in Fig.~\ref{fig:1d-chain}(b): the hopping amplitudes from the two neighboring sites of the other chain cancel with each other due to their opposite phases of the wave function.

Let us discuss this fact more formally. For the two-chain system with shift vector $\tau$, we define the Bloch eigenstates for the $i$th chain as $| i, k \rangle$ ($i$=1, 2). The hybridization amplitude between the Bloch states of the respective chains $t_{k}(\tau)$ is then defined by $t_{k}(\tau)=\langle 1, k | H(\tau) | 2,  k \rangle$, where $H(\tau)$ denotes the total Hamiltonian including the hopping between the sites belonging to different chains $t'$ [Fig.~\ref{fig:1d-chain}(b)]. Here and hereafter, we use the terms ``hopping" and ``hybridization" to describe the transition between the atomic (or molecular) orbitals and Bloch states, respectively. The individual chain is invariant under mirror reflection with respect to planes perpendicular to the chain $m$. Note that there are two inequivalent types of planes which cannot be connected with the lattice vector: ones crossing the sites (plane 1) and ones crossing between the sites (plane 2) [See Fig.~\ref{fig:1d-chain}(b)]. When $\tau=a/2$, plane 1 for the one chain and plane 2 for the other chain coincide with each other and therefore the total Hamiltonian retains the global mirror symmetry $m$ [Fig.~\ref{fig:1d-chain}(b)].  We then get
\begin{eqnarray}
t_{\pi}(a/2)
&=&\langle 1, \pi | H(a/2) | 2, \pi \rangle
\nonumber \\
&=&\langle 1, \pi |mm H(a/2)mm | 2, \pi \rangle
\nonumber \\
&=&[\langle 1, \pi |m][m H(a/2)m][m | 2, \pi \rangle]
\nonumber \\
&=&-t_{\pi}(a/2)
,
\end{eqnarray}
where we have used $[m H(a/2)m]=H(a/2)$ and $[m | 1, \pi \rangle]=| 1, \pi \rangle$ and $[m | 2, \pi \rangle]=-| 2, \pi \rangle$. The last line forces $t_{\pi}(a/2)$ to be zero. 

Thus, the inter-chain hybridization between the $k=\pi$ Bloch states are prohibited when $\tau=a/2$. The key factors in the proof are the following: (i) There are two types of mirror planes which cannot be connected with the lattice vector, (ii) the $k=\pi$ state is the eigenstate of the mirror reflection, whose eigenvalue depends on the planes, and (iii) when $\tau=a/2$, the common mirror planes correspond to plane 1 and plane 2 for the first and second chains, respectively. We next extend the present discussion for general two-dimensional crystals. 

\section{General theory on interlayer hybridization}
\label{sec:general-theory}
\subsection{Stacking-adapted interference manifolds}
\begin{figure}[t]
 \begin{center}
  \includegraphics[scale=.5]{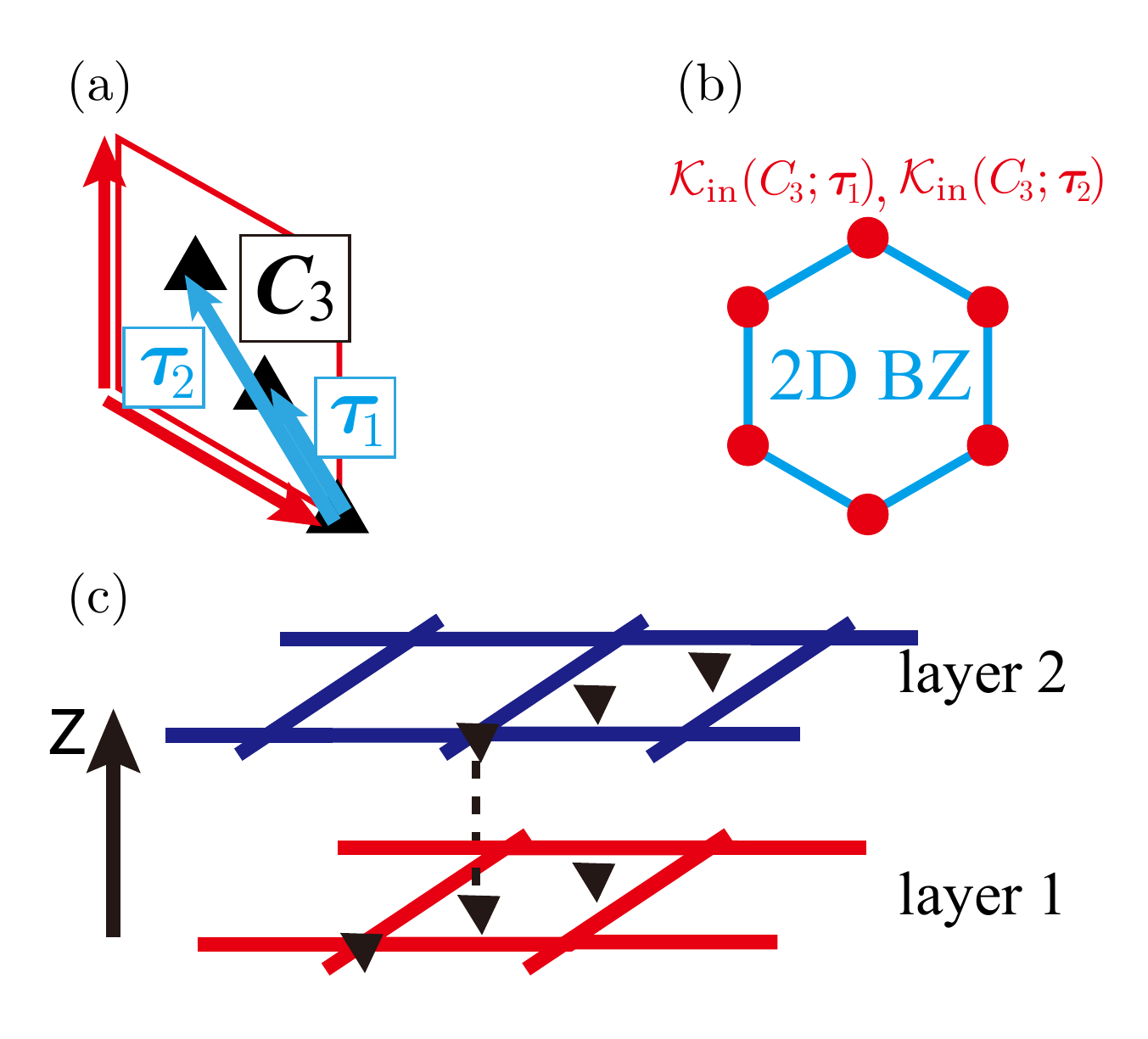}
  \caption{Example of the theorem on the interlayer hybridization. (a) The hexagonal net has three inequivalent $C_{3}$ axes, which are connected by fractional vectors ${\bm \tau}_{1}$ and ${\bm \tau}_{2}$. (b) $\mathcal{K}_{\rm in}(C_{3}; {\bm \tau})$ with ${\bm \tau}$$=$${\bm \tau}_{1}$ or ${\bm \tau}_{2}$, indicated by red or dark points in the hexagonal BZ. (c) When an identical layer is stacked along the original one with in-plane shift ${\bm \tau}$$=$${\bm \tau}_{1}$ or ${\bm \tau}_{2}$, the global $C_{3}$ axis corresponds to inequivalent ones for the respective layers. The interlayer hybridization is consequently suppressed for ${\bf k}\in\mathcal{K}_{\rm in}(C_{3};{\bm \tau})$.}
  \label{fig:2D-scheme}
 \end{center}
\end{figure}

We discuss the two-dimensional periodic system based on its classification by the net, which is a line graph connecting the points generated by the two-dimensional primitive translation vectors (see Fig.~\ref{fig:2D-scheme}(a) for example) and therefore the two-dimensional variant of the Bravais lattice~\cite{footnote-net-origin}. Every periodic two-dimensional crystal structure corresponds to either of the five nets: square, hexagonal, diamond, rectangular, or oblique nets. The definitions of the nets are provided in Appendix~\ref{sec:def-nets} for convenience. Note that the concept of net is also applicable to when the atoms are arranged within a nonzero range in the direction perpendicular to the net. We also employ the ``in-plane" symmetry operation $S$; namely, the operation on the two-dimensional space spanned by the net [$S:(x,y,z) \rightarrow (x'(x,y), y'(x,y),z)$]. In the discussions below, the set of the in-plane symmetry $S$ is limited to the  orthogonal operation, which keep the vector inner product invariant. There is a one-to-one correspondence between the possible set of $S$ compatible with the crystal structure and the corresponding net; for example, only the rotation by $\pi$ ($C_{2}$) is compatible with the oblique net. 


Let us next consider a two-dimensional periodic system which has a certain in-plane symmetry $S$. We do not address the one-dimensional case explicitly since it is trivial~\cite{footnote-1d}. For some combinations of $S$ and net, there are two or more possible reference axis/planes in the unit cell, which cannot be connected to each other by the primitive translation vectors. Define the fractional translation connecting these references by $T_{\bm \tau}$, where ${\bm \tau}$ represents the shift vector. Labeling the symmetry operations with respect to different references by $S_{1}, S_{2}, \cdots, S_{n}$, these elements are related by
\begin{eqnarray}
S_{2}
&=&
T_{{\bm \tau}_{1}}
S_{1}
T^{-1}_{{\bm \tau}_{1}}
,
\nonumber \\
S_{3}
&=&
T_{{\bm \tau}_{2}}
S_{1}
T^{-1}_{{\bm \tau}_{2}}
,\nonumber \\
&\cdots&
,\nonumber \\
S_{n}
&=&
T_{{\bm \tau}_{n-1}}
S_{1}
T^{-1}_{{\bm \tau}_{n-1}}
.
\end{eqnarray}
An example of the hexagonal net with $S$=$C_{3}$ is illustrated in Fig~\ref{fig:2D-scheme}(a). This net is formed by the lattice vectors ${\bf a}_{1}$ and ${\bf a}_{2}$ (red or dark arrows) and there are three possible rotation axes (filled triangles) in the unit cell. These axes are connected by the vectors ${\bm \tau}_{1}$ and ${\bm \tau}_{2}$ (cyan or light arrows). 

\begin{figure*}[htbp]
 \begin{center}
  \includegraphics[scale=.6]{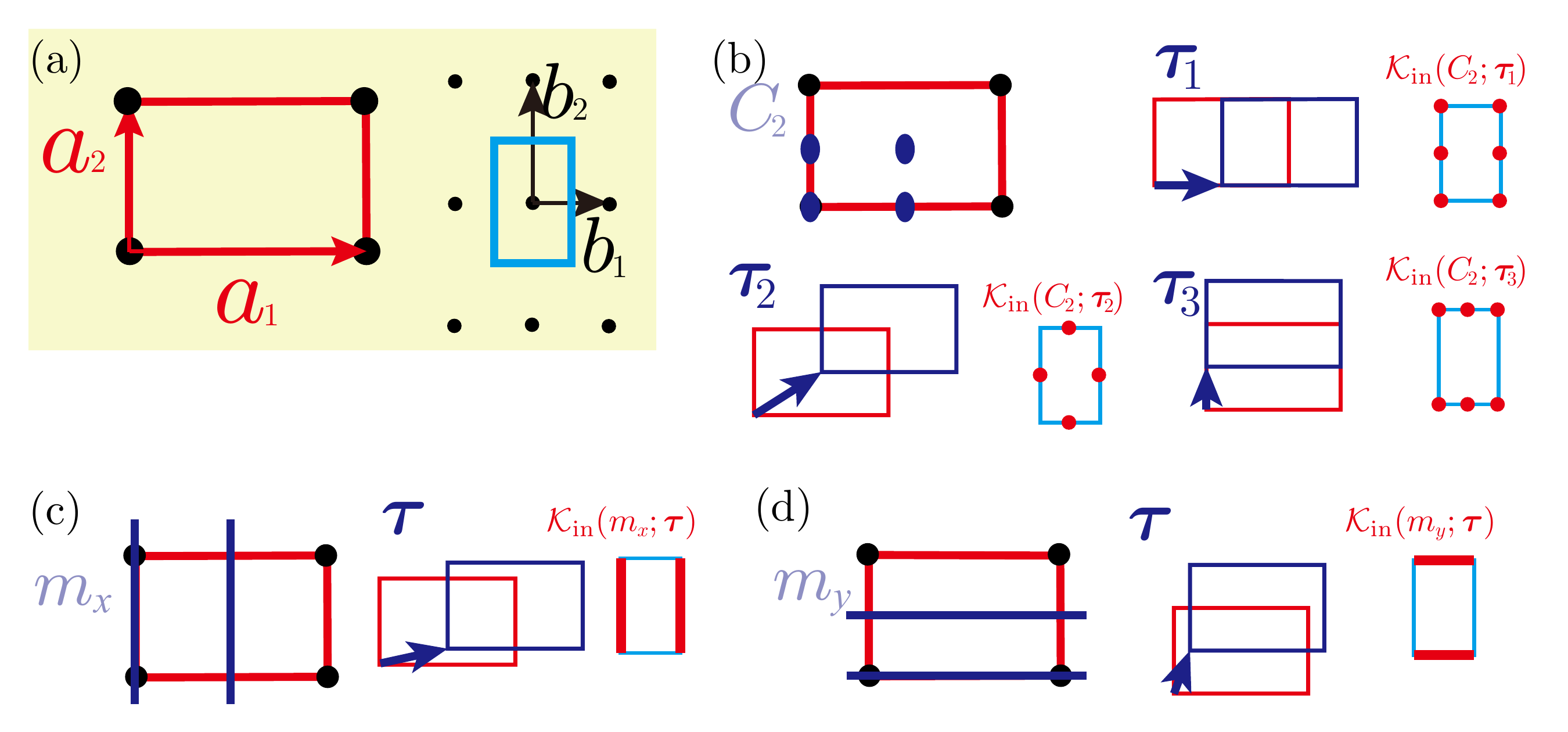}
  \caption{$\mathcal{K}_{\rm in}(S; {\bm \tau})$ for the rectangular net. (a) the primitive translation vectors $a_{1}$ and $a_{2}$, the corresponding reciprocal vectors $b_{1}$ and $b_{2}$, and the BZ. (b)--(d) possible lists of $\mathcal{K}_{\rm in}(S; {\bm \tau})$ when (b) $S=C_{2}$, (c) $S=m_{x}$ and (d) $S=m_{y}$, respectively. The red or dark points and bold lines in the BZ indicate $\mathcal{K}_{\rm in}(S; {\bm \tau})$. For (c) and (d), the same $\mathcal{K}_{\rm in}(S; {\bm \tau})$ applies for arbitrary setting of ${\bm \tau}$ with respect to the component parallel to the reflection plane.}
  \label{fig:rectangle-sum}
 \end{center}
\end{figure*}

Here we introduce an electronic one-body Hamiltonian $h$ for the two-dimensional periodic layer, which is characterized by real-space ionic potential. Regardless of the form of the potential, the corresponding eigenstates, the Bloch states, take the form $|n{\bf k} \rangle\equiv e^{i{\bf k}\cdot \hat{{\bf r}}} |u_{n{\bf k}}\rangle$, where $|u_{n{\bf k}}\rangle$ denotes the cell-periodic part composed of the atomic orbitals. Note that the wave functions introduced in Eq.~(\ref{eq:Bloch-func}) are given by $\psi_{n{\bf k}}({\bf r})=\langle {\bf r} |n{\bf k} \rangle$ and $u_{n{\bf k}}({\bf r})=\langle {\bf r} |u_{n{\bf k}} \rangle$, respectively. Later in this subsection, we focus only one non-degenerate band and omit the band index $n$ ($|n{\bf k}\rangle$$\rightarrow$$|{\bf k}\rangle$ and $|u_{n{\bf k}}\rangle$$\rightarrow$$|u_{{\bf k}}\rangle$ ). When $\psi_{{\bf k}}({\bf r})$ is an eigenfunction of $S$, 
\begin{eqnarray}
S_{j}\psi_{{\bf k}}({\bf r})
&=&\exp[i(\varphi^{\rm pw}_{j}({\bf r})+\varphi^{\rm cell}_{j}({\bf r}))]\psi_{{\bf k}}({\bf r})
\nonumber \\
&\equiv&
\exp(i\varphi_{j})\psi_{{\bf k}}({\bf r}),
\\
&& (j=1,2,\dots,n) \nonumber
\label{eq:Bloch-def}
\end{eqnarray}
 where $\varphi^{\rm pw}_{j}({\bf r})$ denotes the contribution of the plane-wave part ($S_{j}e^{i{\bf k}\cdot {\bf r}}\equiv e^{i\varphi^{\rm pw}_{j}({\bf r})} e^{i{\bf k}\cdot {\bf r}}$) and $\varphi^{\rm cell}_{j}({\bf r})$ that of the cell-periodic part ($S_{j}u_{{\bf k}}({\bf r})=e^{i\varphi^{\rm cell}_{j}({\bf r})}u_{{\bf k}}({\bf r})$). The following formula is then derived (see Appendix \ref{sec:derivation} for detail) 
\begin{eqnarray}
\varphi^{\rm pw}_{j}({\bf r})-\varphi^{\rm pw}_{1}({\bf r})
&=&
({\bf k}-S{\bf k})\cdot{\bm \tau_{j-1}}
\label{eq:trans-pw}
,
\\
\varphi^{\rm cell}_{j}({\bf r})-\varphi^{\rm cell}_{1}({\bf r})
\label{eq:trans-cell}
&=&0,
\end{eqnarray}  
and consequently we get
\begin{eqnarray}
S_{j} \psi_{{\bf k}}({\bf r})
&=&
T_{{\bm \tau}_{j-1}}
S_{1}
T^{-1}_{{\bm \tau}_{j-1}}
\psi_{{\bf k}}({\bf r})
\nonumber \\
&=&
\exp[i({\bf k}-S{\bf k})\cdot{\bm \tau_{j-1}}+ i \varphi_{1}]\psi_{{\bf k}}({\bf r})
\label{eq:Bloch-trans}
.
\end{eqnarray}

In the BZ, we then define the stacking-adapted interference manifolds (SAIM) $\mathcal{K}_{\rm in}(S; {\bm \tau})$ by the following:
\begin{eqnarray}
&&{\bf k}\in \mathcal{K}_{\rm in}(S; {\bm \tau})
\Leftrightarrow
\nonumber \\
&&({\bf k}-S{\bf k})\cdot {\bm \tau}\neq 0 \ {\rm mod} \ 2\pi
\label{eq:K-in-def}
\\
 &&\ \  {\rm and} \ \  {\bf k}-S{\bf k} = n_{1}{\bm b}_{1}+n_{2}{\bm b}_{2} (\exists n_{1}, n_{2} \in \mathbb{Z})
,
\label{eq:eigen}
\end{eqnarray}
where ${\bm b}_{1}$ and ${\bm b}_{2}$ are the reciprocal vectors. The condition Eq.~(\ref{eq:eigen}) is required by the assumption that $|{\bf k} \rangle$ is an eigenstate of $S$. With ${\bm \tau}$ being a vector connecting different references of $S$, $\mathcal{K}_{\rm in}(S; {\bm \tau})$ is, if nonempty, located at the boundary of the BZ~\cite{inui}. Figure~\ref{fig:2D-scheme}(b) illustrates $\mathcal{K}_{\rm in}(C_{3}; {\bm \tau}_{1})$ [=$\mathcal{K}_{\rm in}(C_{3}; {\bm \tau}_{2})$] for the hexagonal net. 

Finally, we stack an identical layer with shift ${\bm \tau}$ and label the Bloch states of the $i$th layer as $|i, {\bf k}\rangle$ [See Fig.~\ref{fig:2D-scheme}(c), e.g.]. A short summary of the consequence is the following: The hybridization between the Bloch states of different layers are canceled for ${\bf k}\in \mathcal{K}_{\rm in}(S;{\bm \tau})$ due to the interference of the Bloch phase.

The total one-body Hamiltonian $H({\bm \tau})$ has the form $H({\bm \tau})=h_{1}+h_{2}+h_{12}({\bm \tau})+v_{1}({\bm \tau})+v_{2}({\bm \tau})$. Here, $h_{1}$ and $h_{2}$ denote the Hamiltonian for the first and second layers in isolated forms and $v_{1}({\bm \tau})$ and $v_{2}({\bm \tau})$ are the layer-diagonal potential terms due to the presence of the neighboring layers. $h_{12}({\bm \tau})$ represents all the hopping terms between the layers. Since the lattice translation vectors of the total and individual systems are common, regardless of ${\bm \tau}$, the total Hamiltonian $H({\bm \tau})$ and the interlayer hopping term $h_{12}({\bm \tau})$ are block-diagonalized for each ${\bf k}$. Using $\{|i, {\bf k} \rangle\}$ $(i=1,2)$ as a basis set, the $2\times 2$ matrix representation of $H({\bm \tau})$ is
\begin{eqnarray}
H({\bm \tau})
=
\left(
\begin{array}{cc}
 \langle 1,{\bf k}| H({\bm \tau}) |1,{\bf k} \rangle & \langle 1,{\bf k}| H({\bm \tau}) |2,{\bf k} \rangle \\
 \langle 2,{\bf k}| H({\bm \tau}) |1,{\bf k} \rangle & \langle 2,{\bf k}| H({\bm \tau}) |2,{\bf k} \rangle 
 \end{array}
 \right )
 .
\end{eqnarray} 
When ${\bm \tau}={\bm \tau}_{1}, {\bm \tau}_{2}, \cdots$, the total Hamiltonian is invariant under {\it global} symmetry operation $\mathcal{S}$, which corresponds to $S_{1}$ and $T_{-{\bm \tau}}S_{1}T^{-1}_{-{\bm \tau}}$ for the first and second layer, respectively. When ${\bf k}\in\mathcal{K}_{\rm in}(S;{\bm \tau})$, for the non-diagonal element, we get
\begin{eqnarray}
\langle 1,{\bf k}| H({\bm \tau}) |2,{\bf k} \rangle
&=&
\langle 1,{\bf k}| h_{12}({\bm \tau}) |2,{\bf k} \rangle
\nonumber \\
&=&
\langle 1,{\bf k}|\mathcal{S}^{-1}\mathcal{S} h_{12}({\bm \tau}) \mathcal{S}^{-1} \mathcal{S} |2,{\bf k}\rangle
\nonumber \\
&=&
[\langle 1,{\bf k}|S^{-1}_{1}] h_{12}({\bm \tau}) [T_{-{\bm \tau}}S_{1}T^{-1}_{-{\bm \tau}} |2,{\bf k} \rangle]
\nonumber \\
&=&
e^{-{\rm i} ({\bf k}\!-\!S{\bf k})\cdot {\bm \tau}}
 \langle 1,{\bf k}| h_{12}({\bm \tau}) |2,{\bf k} \rangle
.
\label{eq:theorem-proof}
\end{eqnarray}
Here we have used Eqs.~(\ref{eq:Bloch-def}) and (\ref{eq:Bloch-trans}). For ${\bf k}\in\mathcal{K}_{\rm in}(S; {\bm \tau})$, $\exp[-{\rm i} ({\bf k}\!-\!S{\bf k})\cdot {\bm \tau}] \neq 1$ and the only allowed value for $\langle 1,{\bf k} | h_{12}({\bm \tau}) |2,{\bf k} \rangle$ is zero. The prefactor $e^{-{\rm i} ({\bf k}\!-\!S{\bf k})\cdot {\bm \tau}}$ comes solely from the plane-wave part of the wave function (see Eqs.(\ref{eq:trans-pw}) and (\ref{eq:trans-cell})) and therefore represents the interference of the Bloch phase~\cite{comment-v12}.  

The manifold $\mathcal{K}_{\rm in}(S; {\bm \tau})$ is uniquely determined according to the kind of the net, symmetry element $S$ and the shift vector ${\bm \tau}$. For example, we show $\mathcal{K}_{\rm in}(S; {\bm \tau})$ for the rectangular net in Fig.~\ref{fig:rectangle-sum}. The lattice and reciprocal vectors and the corresponding BZ are shown in panel (a). The possible symmetry elements compatible with this net are the two-fold rotation ($C_{2}$) and mirror reflections with respect to the $y-z$ and $z-x$ planes ($m_{x}$ and $m_{y}$, respectively). When the whole crystal is $C_{2}$ symmetric, there are four inequivalent $C_{2}$ axes in the unit cell and there are three vectors ${\bm \tau}_{1}$---${\bm \tau}_{3}$ connecting them [Fig.~\ref{fig:rectangle-sum}(b)]. Elementary calculations give us the following: 
\begin{eqnarray}
\mathcal{K}_{\rm in}(C_{2}; {\bm \tau}_{1})
&=&
\left\{
{\bf k} |\ \ {\bf k}= 
{\bm b}_{1}/2, \ \ {\bm b}_{1}/2 +{\bm b}_{2}/2 
\right\}
,
\\
\mathcal{K}_{\rm in}(C_{2}; {\bm \tau}_{2})
&=&
\left\{
{\bf k} |\ \ {\bf k}= 
{\bm b}_{1}/2, \ \ {\bm b}_{2}/2
\right\}
,
\\ 
\mathcal{K}_{\rm in}(C_{2}; {\bm \tau}_{3})
&=&
\left\{
{\bf k} |\ \ {\bf k}= 
 {\bm b}_{2}/2, \ \  {\bm b}_{1}/2 +{\bm b}_{2}/2
\right\}
.
\end{eqnarray}
When the crystal is $m_{x}$ symmetric, on the other hand, there are two inequivalent mirror planes in the unit cell~[Fig.~\ref{fig:rectangle-sum}(c)]. The vector connecting them can be taken arbitrarily with respect to the component parallel to the planes: ${\bm \tau}={\bm a}_{1}/2+s{\bm a}_{2}$ with $-1/2\leq s \leq 1/2$. 
Regardless of $s$, we get 
\begin{eqnarray}
\mathcal{K}_{\rm in}(m_{x}; {\bm \tau})
&=&
\left\{
{\bf k} |{\bf k}= 
{\bm b}_{1}/2\!+\!t{\bm b}_{2} 
\right\}
\ (-1/2\! \leq \! t \! \leq \!1/2)
.
\end{eqnarray}
A similar discussion applies to $S=m_{y}$. With $\tau={\bm a}_{2}/2+s{\bm a}_{1}$,
\begin{eqnarray}
\mathcal{K}_{\rm in}(m_{y}; {\bm \tau})
&=&
\left\{
{\bf k} |{\bf k}= 
t{\bm b}_{1}\!+\!{\bm b}_{2}/2 
\right\}
\ (-1/2 \! \leq \! t \! \leq \! 1/2)
.
\end{eqnarray}
The complete list of non-empty $\mathcal{K}_{\rm in}(S; {\bm \tau})$ for all the nets is available in Appendix~\ref{sec:list-net-map}.

We thus introduced a concept of SAIM $\mathcal{K}_{\rm in}(S; {\bm \tau})$ for the respective nets and established its relation to interlayer hybridization of the Bloch states, which can be summarized as follows. Suppose a single-band monolayer system which is invariant under in-plane symmetry operation $S$. When an identical layer is stacked side by side, a theorem on interlayer hybridization is derived. Namely, 
\begin{eqnarray}
\langle 1,{\bf k}| H({\bm \tau}) |2,{\bf k} \rangle=0
\ ({\bf k}\in \mathcal{K}_{\rm in}(S; {\bm \tau}))
\label{eq:summary-singleband}
\end{eqnarray}
where ${\bm \tau}$ and $H({\bm \tau})$ denote the in-plane shift between the adjacent layers and the total Hamiltonian, respectively. If there are two or more symmetry references for $S$ within the unit cell, $ \mathcal{K}_{\rm in}(S; {\bm \tau})$ is non-empty with ${\bm \tau}$ set to connect the different symmetry references. Note that this theorem does not concern the character of the cell-periodic parts of the Bloch states.

\subsubsection*{Layers related by symmetry operation}
\label{sec:layer-inv}
Frequently, we find multilayer structures where the neighboring layers are not exactly identical but related with each other by some symmetry operation $\Sigma$; namely, inversion, reflection, etc.. Extending the present theory to such cases is in principle possible if the corresponding symmetry operation does not change the shape and orientation of the two-dimensional BZ. For example, three-dimensional hexagonal BN consists of two kinds of BN layers which are the inverse of the others. Although the atomic configurations are different for the two adjacent layers, their BZs are common. We here address an extension to such systems to enhance the applicability of the present theory.

First, we need to redefine the ``shift" between the layers. Namely, we set the positions of the nets for the two layers so that their nodes correspond to the centers of the cell-periodic part of the Bloch states $|u_{k} \rangle$. The in-plane shift ${\bm \tau}$ is defined with these nets. Next, we see if ${\bf k}\in\mathcal{K}_{\rm in}(S; {\bm \tau})$ is transformed by $\Sigma$ by either of the following:
\begin{eqnarray}
{\bf k} \xrightarrow{\Sigma} &&{\bf k} 
\label{eq:condition1}
\\
{\bf k} \xrightarrow{\Sigma} && -{\bf k} 
.
\label{eq:condition2}
\end{eqnarray}
In the former case, the character of $|u_{\bf k} \rangle$ for the first and second layers are the same, whereas in the latter $|u_{\bf k} \rangle$ for the second layer is complex conjugate of the first since  they are related by $\mathcal{T}\Sigma$, with $\mathcal{T}$ being the time reversal. Equation (\ref{eq:summary-singleband}) is obviously applicable if either of the following is satisfied: (I) ${\bf k}\in \mathcal{K}_{\rm in}(S; {\bm \tau})$ is transformed by Eq.~(\ref{eq:condition1}), or (II) ${\bf k}\in \mathcal{K}_{\rm in}(S; {\bm \tau})$ is transformed by Eq.~(\ref{eq:condition2}) {\it and} $|u_{\bf k} \rangle$ is a real representation of $S$. Otherwise, whether the interlayer hybridization is zero or nonzero depends on the specific character of $|u_{\bf k} \rangle$. In summary, the extension is achieved with only two additional steps for the cell-periodic part: (a) Redefine the in-plane shift between the layers based on the positions of the orbital and (b) examine if it is real representation of $S$. 

When we apply the present results to realistic models based on the crystal structure of the material, step (b) above requires a burdensome calculation to specify the character of $|u_{{\bf k}} \rangle$. To facilitate efficient applications, we here exemplify situations where Eq.~(\ref{eq:summary-singleband}) is obviously applicable: When $\Sigma$ is the mirror plane parallel to the layers; when neither $C_{3}$ nor $C_{4}$ is compatible with the net; $|u_{\bf k} \rangle$ is formed by the atomic orbital with zero orbital angular momentum with respect to the axis perpendicular to the layers ($s$, $p_{z}$, $d_{z^2}$, $\dots$); etc. 

\subsection{Multilayer and bulk systems}
When a number of identical or symmetry-related layers are stacked so that the shape of their corresponding two-dimensional BZs are the same, the total Hamiltonian remains block-diagonalized with respect to the crystal wavenumber defined for the original layers. By arranging the interlayer shift vectors, desired combinations of hybridizations among the layers can be suppressed for ${\bf k}\in \mathcal{K}_{\rm in}(S; {\bm \tau})$. 

Next, suppose an infinite number of the layers are stacked so that a certain ${\bf k}\equiv {\bf k}_{\rm in}$ satisfies ${\bf k}_{\rm in}\in \mathcal{K}_{\rm in}(S; {\bm \tau})$ for all the shift vectors between the adjacent layers. The eigenstates formed by $\{ |i, {\bf k}_{\rm in}\rangle\}$ $(i=1,2,\dots)$ are then labeled by the crystal wavenumber perpendicular to the layers $k_{z}$. However, their energies are almost independent of $k_{z}$ because of the absence of the interlayer hybridization; namely, the flat band dispersion is seen along $k_{z}$ from ${\bf k}_{\rm in}$. Note that there is still a possibility of interlayer hybridization between distant layers. For example, when the hexagonal nets are stacked with uniform shift ${\bm \tau}={\bm \tau}_{1}$ ([Fig.~\ref{fig:2D-scheme}(a)], the $3n$-nearest neighbor layers ($n=1,2,\cdots$) are regarded as unshifted from the original layer. The hybridization amplitudes among $\{ |3n, {\bf k}_{\rm in} \rangle \}$, $\{ |3n+1, {\bf k}_{\rm in} \rangle \}$ and $\{ |3n+2, {\bf k}_{\rm in} \rangle \}$ can then be nonzero, respectively. In summary, the band dispersion in $k_{z}$ direction tends to be much flatter through ${\bf k}_{\rm in}$ than through other ${\bf k}$ points because of the canceled interlayer hybridizations.

\subsection{Multiband and degeneracy}
\label{sec:multiband-degene}
In the above discussions, we have concentrated on the intra-band hybridization by limiting ourselves to the single band case. For general mutiband systems, interlayer hybridizations between different orbitals are possible even for ${\bf k}\in\mathcal{K}_{\rm in}(S; {\bm \tau})$, if their orbital characters are different. Nevertheless, we can ignore such inter-orbital hybridizations when the respective bands in focus are non-degenerate and well separated from other bands in energy at ${\bf k}\in\mathcal{K}_{\rm in}(S; {\bm \tau})$ in the monolayer system. This point is elaborated in Appendix~\ref{sec:multi-etc}. Finally, the following rule of thumb is established: In general multilayer systems, the Bloch states at ${\bf k}\in \mathcal{K}_{\rm in}(S; {\bm \tau})$ for each layer are well protected from interlayer hybridizations if only their energy eigenvalues are nondegenerate and well separated from other states. In Sec.~\ref{sec:layered-semicon}, we demonstrate that this seemingly naive rule of thumb indeed applies to a wide range of multilayer systems.

\subsection{Implication of stacking-adapted interference manifolds}
\label{sec:stack-dependence-comment}
We have seen that the interlayer hybridization is generally canceled by the Bloch phase for ${\bf k}$ within $\mathcal{K}_{\rm in}(S; {\bm \tau})$. This is valid even in the multiband case for the intraband hybridization. These results imply a notable fact: When we examine the electronic structure of layered systems with changing the interlayer shift, the Bloch states around $\mathcal{K}_{\rm in}(S; {\bm \tau})$ with any $S$ and ${\bm \tau}$ {\it always} exhibit noticeable variation. This special feature provides us with a guiding principle to explore the shift-dependent electronic property of layered systems, that we should examine the electronic states around $\mathcal{K}_{\rm in}(S; {\bm \tau})$. For the other ${\bf k}$ points, the shift-vector dependence crucially differs by systems and its general features are probably difficult to derive. For example, in the single-band system formed by $s$-wave atomic orbitals, the interlayer hybridization obviously varies little by changing the interlayer shift for ${\bf k}$ far from $\mathcal{K}_{\rm in}(S; {\bm \tau})$. This is reasonably inferred since the overlap integral of the cell-periodic parts of the Bloch functions in the neighboring layers have positive definite values regardless of the shift. In this case, one finds the appreciable shift-vector dependence only in the vicinity of $\mathcal{K}_{\rm in}(S; {\bm \tau})$ with any $S$.

\section{Applications to layered semiconductors}
\label{sec:layered-semicon}
In the previous section, we have presented a theory relating the interlayer hybridization of the Bloch states and the in-plane shift between adjacent layers based on a general modeling of layered materials. The significant consequence is that whether the hybridization is canceled or not can be discussed only by examining the relative shift of adjacent layers. We next show that our theory indeed applies to real materials. First-principles electronic structure calculation is employed for this purpose, where the inter-site hopping is treated without modeling. We take for example the layered BN, graphite, group-VI TMD and black phosphorus. The structures of the former three in the monolayer form commonly correspond to the hexagonal net. They form various multilayer structures with shifts ${\bm \tau}$ so that the $C_{3}$ axes for the neighboring layers are shared, and then $\mathcal{K}_{\rm in}(C_{3};{\bm \tau})$ corresponds to the corners of the hexagonal BZ, $K$ and $K'$ [Fig.~\ref{fig:2D-scheme} (b)]. On the other hand, black phosphorus monolayer (phosphorene) forms in the rectangular net structure. According to the in-plane symmetry and interlayer shift, $\mathcal{K}_{\rm in}$ possibly corresponds to the either edges of the BZ(Fig.~\ref{fig:rectangle-sum}).

We first examine the electronic states in bilayer BN with various stacking geometries, showing that whether the interlayer hybridization is suppressed or not is correctly predicted by simply seeing how the two layers are shifted from each other. Next, anomalously flat band dispersion predicted from the present theory is examined in the bulk BN. We also address molybdenum disulfide MoS$_{2}$. Interlayer hybridization in multilayer MoS$_{2}$ has been recently discussed in the literature~\cite{Liu-review2015, Akashi2015}, which is actually an example of the present general theorem. Fi, we work on layered graphite. The two-fold degeneracy at the $K$ and $K'$ points in the monolayer form defies straightforward application of the present theorem as discussed in Sec.~\ref{sec:multiband-degene}. Nevertheless, we see that a minimal extension enables us simple geometric analysis on the interlayer hybridization. The possible interlayer hybridizations and the band structure along the $K$--$H$ line (along the $k_{z}$ direction passing through the $K$ point) for these systems have been discussed before in various preceding works, though in separate contexts. Our theory rederives results consistent with them from a unified viewpoint. Finally, through the comparison of the electronic structures in the black phosphorus with the experimentally observed shifted stacking and pathological unshifted stacking, we see that our theory is well applicable to the structure other than the hexagonal one. The detailed conditions for the first-principles calculations are given in Appendix ~\ref{sec:abinitio-detail}.

\subsection{Boron nitrides}
\subsubsection{Bilayer}
\label{sec:BN-bi}
\begin{figure*}[t]
 \begin{center}
  \includegraphics[scale=.45]{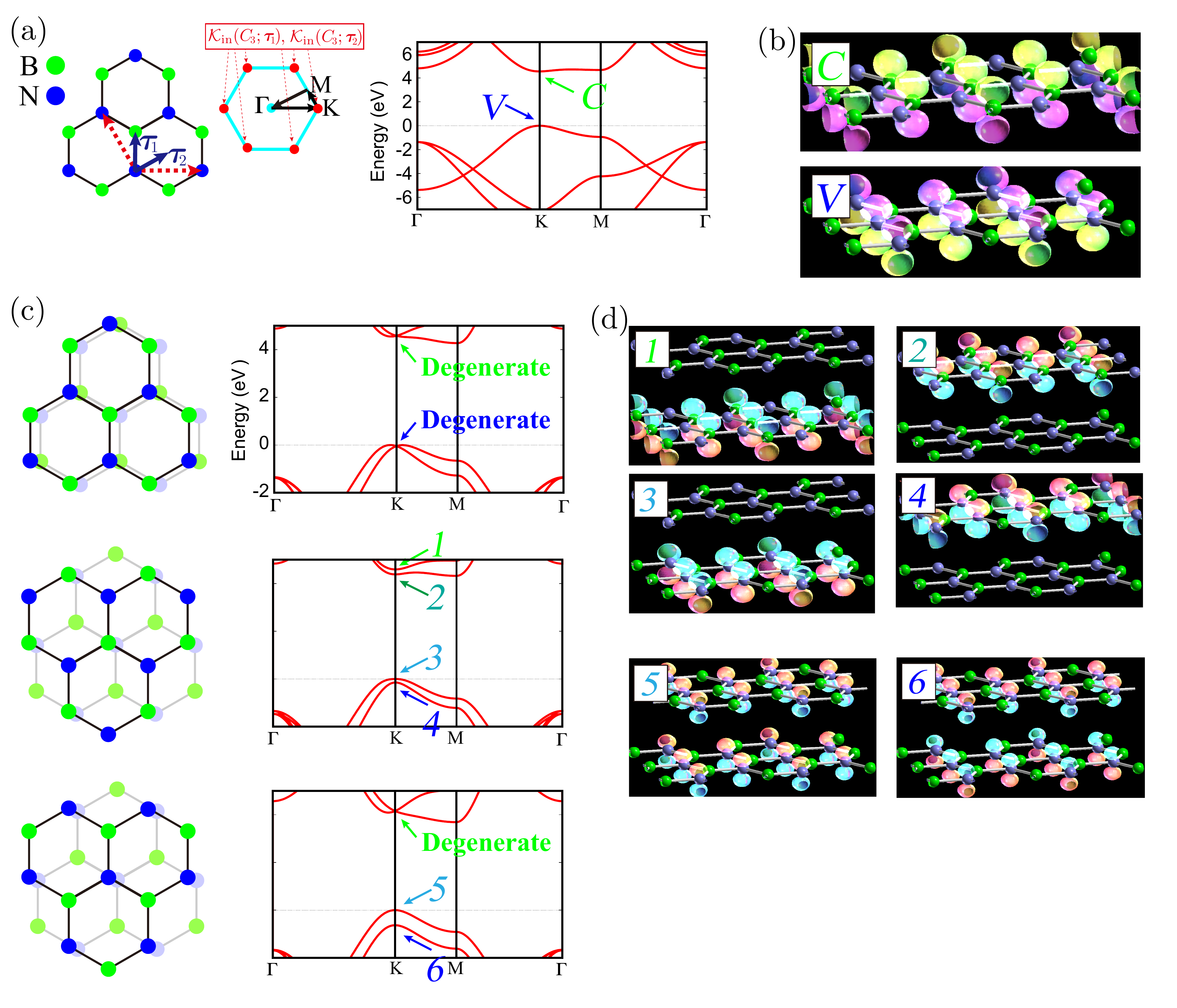}
  \caption{Electronic structure of bilayer BN. (a) The monolayer crystal structure, the corresponding BZ and calculated band structure. Primitive lattice vectors ${\bf a}_{1}$ and ${\bf a}_{2}$ and possible shift vectors ${\bm \tau}_{1}$ and ${\bm \tau}_{2}$ are also shown. In the BZ, $k$-point path and $\mathcal{K}_{\rm in}(C_{3}; {\bm \tau}_{1})$ and $\mathcal{K}_{\rm in}(C_{3}; {\bm \tau}_{2})$ are indicated. (b) Calculated Bloch wave functions for the conduction (C) and valence (V) states at the $K$ point. Spatial dependence of the plane-wave part ${\rm exp}(i{\bf k}\cdot {\bf r})$ has been ignored. (c) the bilayer structures and their calculated band structures. (d) Calculated Bloch wave functions for the states indicated in panel (c). Those for the degenerate states are not shown.}
  \label{fig:BN-bi-sum}
 \end{center}
\end{figure*}

Here we investigate the bilayers of BN with three types of stacking labeled $H$, $R$ and $H_{\rm s}$, which are depicted in Fig.~\ref{fig:BN-bi-sum}(c). $H$ corresponds to the $h$-BN bilayer; $R$ type corresponds to the neighboring two layers in the bulk rhombohedral BN; $H_{\rm s}$ corresponds to the $h$-BN bilayer with shift, where nitrogen atoms in the upper layer are placed over those in the lower. The former two have been experimentally realized~\cite{Warner-Buchner-BNbilayer-topography2010}, whereas the latter stacking has been studied only theoretically and found to be unstable~\cite{Ooi-Adams-variousBN2006, Constantinescu-Kuc-BNstacking-PRL2013}. 

The crystal and electronic band structures of the monolayer BN are displayed in Fig.~\ref{fig:BN-bi-sum}(a). The crystal structure belongs to the hexagonal net, where bold dashed arrows indicate the primitive lattice vectors. The monolayer is $C_{3}$-symmetric and therefore there are two additional $C_{3}$ axes in the unit cell indicated by ${\bm \tau}_{1}$ and ${\bm \tau}_{2}$. The valence-top and conduction-bottom states at the $K$ and $K'$ points are well separated in energy from other states, which justifies the neglect of inter-orbital hybridization between the layers. 

The above pieces of information are enough for analyzing the interlayer hybridization in the $R$ stacking because the two layers are related only by translation. The in-plane shift between the two layers ${\bm \tau}$ yields $\mathcal{K}_{\rm in}(C_{3}; {\bm \tau})=\{K, K'\}$ and therefore interlayer hybridization amplitudes must be respectively zero for the conduction-bottom and valence-top states at the $K$ and $K'$ points. This is confirmed by observing the wave functions of the $K$-point valence-top and conduction-bottom states: Due to the inequivalence of the electrostatic potentials for the states in the upper and lower layers, small band splitting is seen in the valence-top and conduction-bottom states. As clearly seen in Fig.~\ref{fig:BN-bi-sum}(d), the calculated wave functions for these states (1--4) are localized around each layer, indicating the zero hybridization between them.

The analysis for the $H$ and $H_{\rm s}$ stackings requires additional information for the monolayer band structure because the second layer is an inverse of the first. Namely, the conduction-bottom and valence-top states at the $K$ and $K'$ points are formed by the B- and N-$2p_{z}$ orbitals, respectively. The present cases then correspond to case (II) in Sec.~\ref{sec:layer-inv}, which allows us to rely again on the simple geometry-based argument regardless of the specific orbital characters. 

Introducing the hexagonal nets for the B and N sites, respectively, we see that in the $H$ stacking the B-net (N-net) of the upper layer is shifted from that of the lower layer by ${\bm \tau}_{1}$(${\bm \tau}_{2}$). This shift yields zero hybridizations at the $K$ and $K'$ points for the both states. This is immediately confirmed by the calculated band structure: The $K$ and $K'$-point valence-top and conduction-bottom states are doubly degenerate [Fig.~\ref{fig:BN-bi-sum}(c)]. Since the degeneracy is lifted when either the potential difference between the layers or the interlayer hybridization is nonzero, the degeneracy indicates the zero interlayer hybridization amplitudes. In the $H_{\rm s}$ case, on the other hand, the B-nets are shifted from each other whereas the N-nets are not. Then, only the conduction-bottom states should be protected from the interlayer hybridization. This is also confirmed with the calculated band structure displayed in panel (c), where only the valence-top states split. The wave functions of the valence-top states are equally distributed to the two layers, which indicates the formation of the bonding/antibonding states due to the interlayer hybridization. The conduction-bottom states are doubly degenerate, indicating the zero interlayer hybridization.

\begin{figure}[t]
 \begin{center}
  \includegraphics[scale=.35]{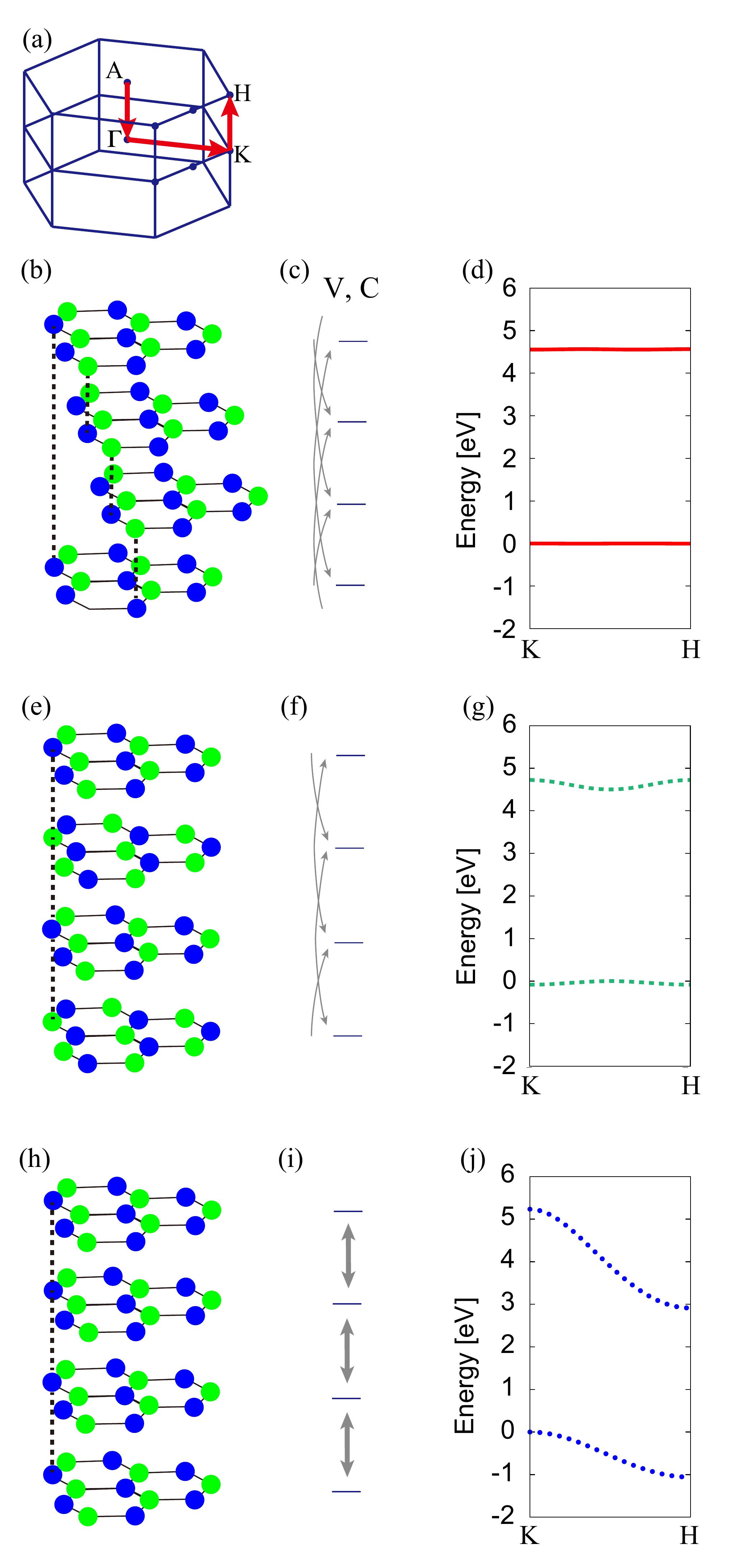}
  \caption{(a) the hexagonal BZ. (b), (e) and (h) crystal structures of the bulk BN with the rhombohedral, hexagonal, and AA stacking. (c), (f) and (i) the selection rules for interlayer hybridizations, where the allowed hybridizations are indicated by arrows. (d), (g) and (j) the first-principles band structures along the $K$--$H$ line. The shade on the band indicates the two-fold degeneracy.}
  \label{fig:BN-bulk-sum}
 \end{center}
\end{figure}

\begin{figure}[t]
 \begin{center}
  \includegraphics[scale=.52]{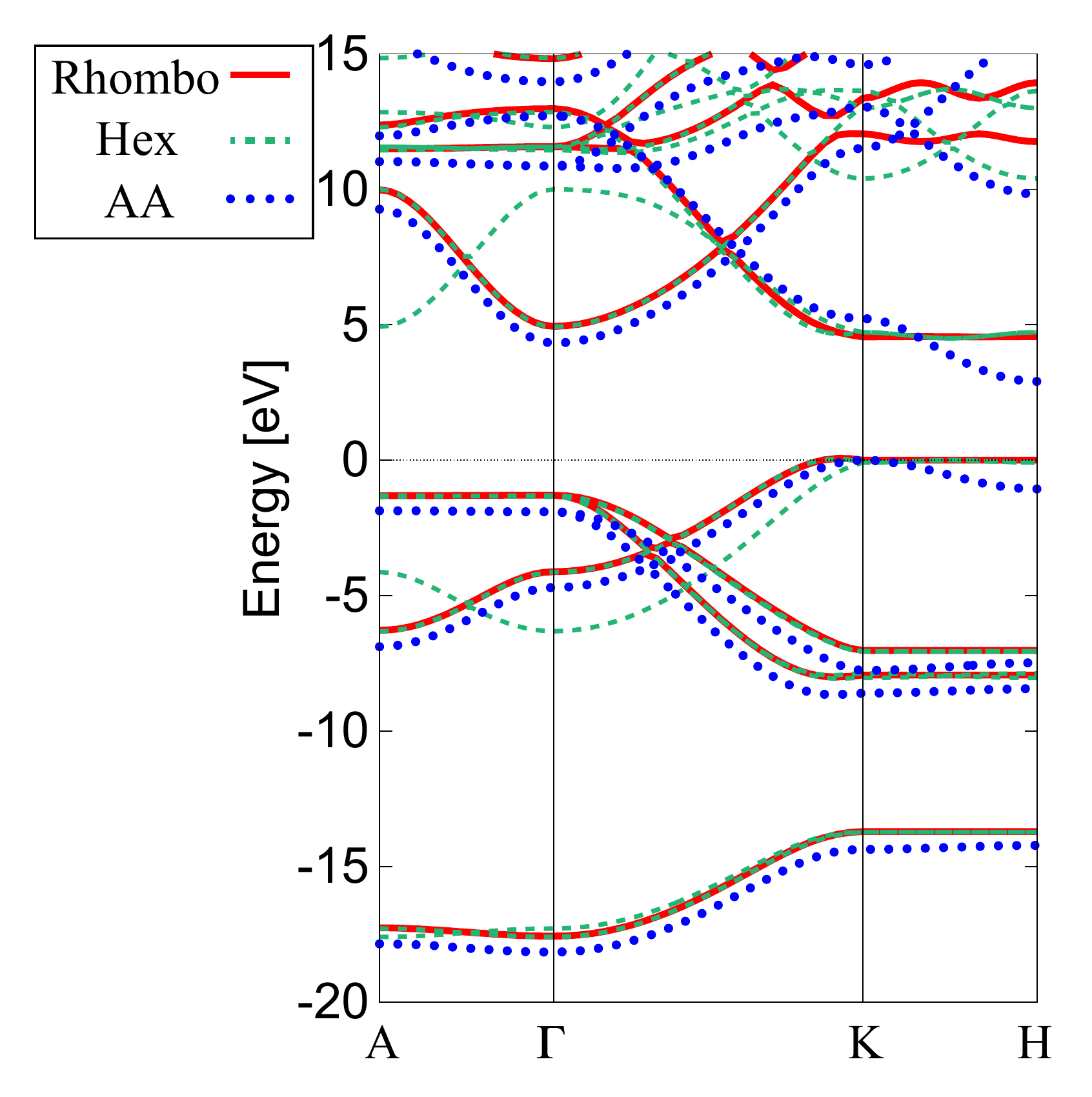}
  \caption{First-principles band structures in the bulk BN for the different stackings.}
  \label{fig:BN-bulk-band}
 \end{center}
\end{figure}

\subsubsection{Bulk}
\label{sec:BN-bulk}
We have thus demonstrated the relevance of our theory relating the stacking geometry and interlayer hybridization with the bilayer systems. Next, we proceed to the infinite-layer case. Although the band structure in this system has been studied~\cite{Kresse-graphite-BN-band1996}, we revisit it in view of the present theory, deriving which hybridization processes dominate the bandwidth for various stacking geometries. 

We address the three types of structure: (i) the rhombohedral BN, where the monolayers are infinitely stacked with constant shift ${\bm \tau}_{1}$ [Fig.~\ref{fig:BN-bulk-sum}(b)]; (ii) the hexagonal BN, where the layers are $H$-stacked (See Figs~\ref{fig:BN-bi-sum}(c) and \ref{fig:BN-bulk-sum}(e)); (iii) the AA BN, where the layers are stacked without shift or inversion [Fig.~\ref{fig:BN-bulk-sum}(h)]. For the $K$- and $K'$-point states, the rules of interlayer hybridization established in Sec.~\ref{sec:BN-bi} straightforwardly yield the complete list of possible/impossible hybridizations for all the three stackings. In the rhombohedral BN, hybridizations to nearest-neighbor and next-nearest-neighbor layers are prohibited because of the relative shifts by ${\bm \tau}_{1}$ and ${\bm \tau}_{2}$, respectively. The most relevant hybridization allowed is therefore those between the third-nearest neighbors [Fig.~\ref{fig:BN-bulk-sum}(b)]. In the hexagonal BN, on the other hand, the hybridizations between the next-nearest neighbor layers are allowed because they are unshifted [Fig.~\ref{fig:BN-bulk-sum}(e)]. Finally, in the AA BN, hybridization between the neighboring layers is obviously allowed [Fig.~\ref{fig:BN-bulk-sum}(h)]. We depict the geometrically allowed hybridizations for the three stackings by arrows in Fig.~\ref{fig:BN-bulk-sum}(c), (f), and (h). Therefore, it is anticipated that the bandwidth along the $K$--$H$ line has the smallest value in the rhombohedral structure, followed in order by the hexagonal and AA. The first-principles band structures indeed show this trend [Fig.~\ref{fig:BN-bulk-sum}(d), (g), (j)]. 

We also display the calculated band structures in the $A$--$\Gamma$--$K$--$H$ path in a broader energy range in Fig.~\ref{fig:BN-bulk-band}. Apart from the Brillouin-Zone folding with the hexagonal stacking, the band structures for the three stackings are very similar in the $A$--$\Gamma$--$K$ path. Significant differences are seen only in the $K$--$H$ path, which indicates that the phase interference effect due to the stacking geometry is relevant only around the $K$-point states. Interestingly, in the $K$--$H$ path, the other valence bands are also flat in the rhombohedral and hexagonal cases, demonstrating the generality of the present hybridization rule.

\begin{figure}[t]
 \begin{center}
  \includegraphics[scale=.32]{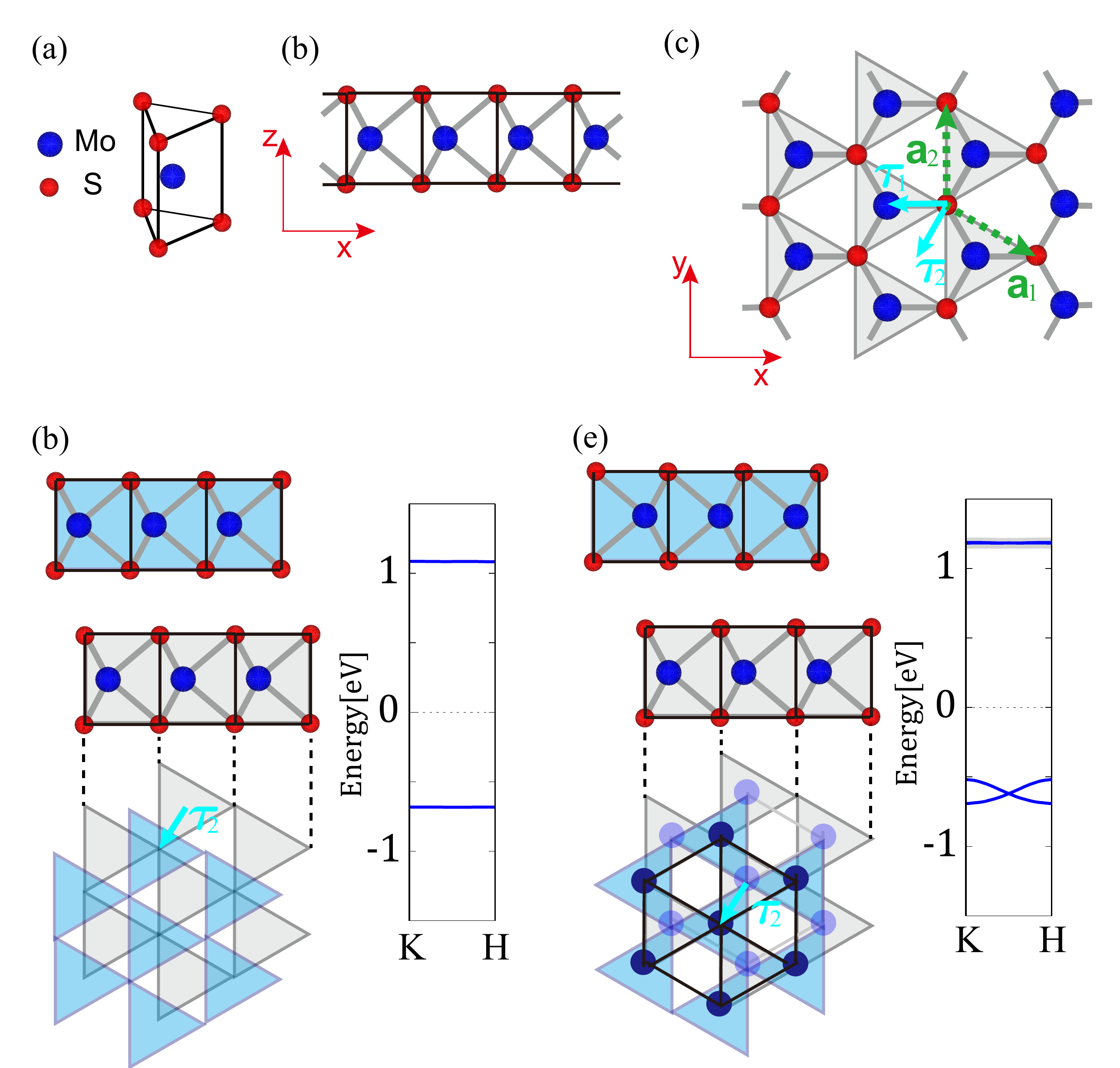}
  \caption{(a) unit cell of MoS$_{2}$ monolayer, (b) side view, (c) top view, where ${\bf a}_{1}$ and ${\bf a}_{2}$ denote the primitive lattice vectors. Inequivalent $C_{3}$ axes are indicated by vectors ${\bm \tau}_{1}$ and ${\bm \tau}_{2}$. (d) [(e)] the 3R (2H) stacking and corresponding bulk band structure along the $K$--$H$ line. The shade on the band indicates the two-fold degeneracy. The band structures are taken from Ref.~\onlinecite{Suzuki2014}.}
  \label{fig:TMDC-sum}
 \end{center}
\end{figure}

\subsection{Transition-metal dichalcogenides}
\label{sec:TMD}
The TMD $MX_{2}$ ($M$=Mo, W; $X$=S, Se, Te) mono and multilayer have been recently much in focus for its valley-contrasted phenomena~\cite{Xu-Heinz-review-NPhys2014}. Many features of TMD are common to the BN in their monolayer form: Seen from above, the former structure is identical to the latter. The only difference is that the atoms are placed in a non-zero range with respect to the $z$ direction in TMD (Figs.~\ref{fig:TMDC-sum} (a), (b), and (c)). The valence-band top and conduction-band bottom are located at the $K$ and $K'$ points of the hexagonal BZ. Most studies on the TMD multilayer concern either the 2H or 3R stacking depicted in Figs.~\ref{fig:TMDC-sum} (d) and (e). In prior to this work, for MoS$_{2}$, some of the present authors have found that the interlayer hybridization of the conduction-band bottom states is zero with both the 2H and 3R stackings, whereas that for the valence-band top states is zero with only the 3R stacking~\cite{Suzuki2014} (Figs.~\ref{fig:TMDC-sum} (d) and (e)). This has been proved to be the consequence of the phase interference~\cite{Liu-review2015,Akashi2015}. We revisit this topic to see that the previous results are re-derived in a simplified fashion with the present theory.

First, the monolayer structure belongs to the hexagonal net. The crystal is $C_{3}$ symmetric and therefore there are two possible shift vectors ${\bm \tau}_{1}$ and ${\bm \tau}_{2}$ [Fig.~\ref{fig:TMDC-sum} (c)] which yield $\mathcal{K}_{\rm in}(C_{3}; {\bm \tau})=\{K, K'\}$. The valence-top and conduction-bottom states are non-degenerate and well separated from other bands. These pieces of information are enough for the 3R stacking. We straightforwardly get to the consequence that the interlayer hybridization is prohibited for both the valence-top and conduction-bottom states. For the 2H stacking, where the upper layer is inverted from the lower, we need additional information: The cell-periodic parts of the valence-band top and conduction-band bottom Bloch states are formed by Mo $4d_{x^{2}-y^{2}}+d_{xy}$ orbital with angular momentum $l_{z}$$=$$-2$ and $4d_{z^{2}}$ orbital with $l_{z}$$=$$0$, respectively~\cite{Xiao-MoS2-PRL}. The case (II) in Sec.~\ref{sec:layer-inv} applies only for the conduction-bottom. We next introduce a hexagonal net for the Mo sites. In the 2H stacking, the Mo-nets are shifted from each other by ${\bm \tau}={\bm \tau}_{2}$ [Fig.~\ref{fig:TMDC-sum}(e)]. The argument based on the geometry immediately yields zero hybridization for the conduction-bottom states~\cite{comment-valence}. The present theory thus gives a simple way of understanding the possible/impossible interlayer hybridizations of the valley states of the TMD.

\subsection{Graphite}
\label{sec:graphite}

\begin{figure}[t]
 \begin{center}
  \includegraphics[scale=.37]{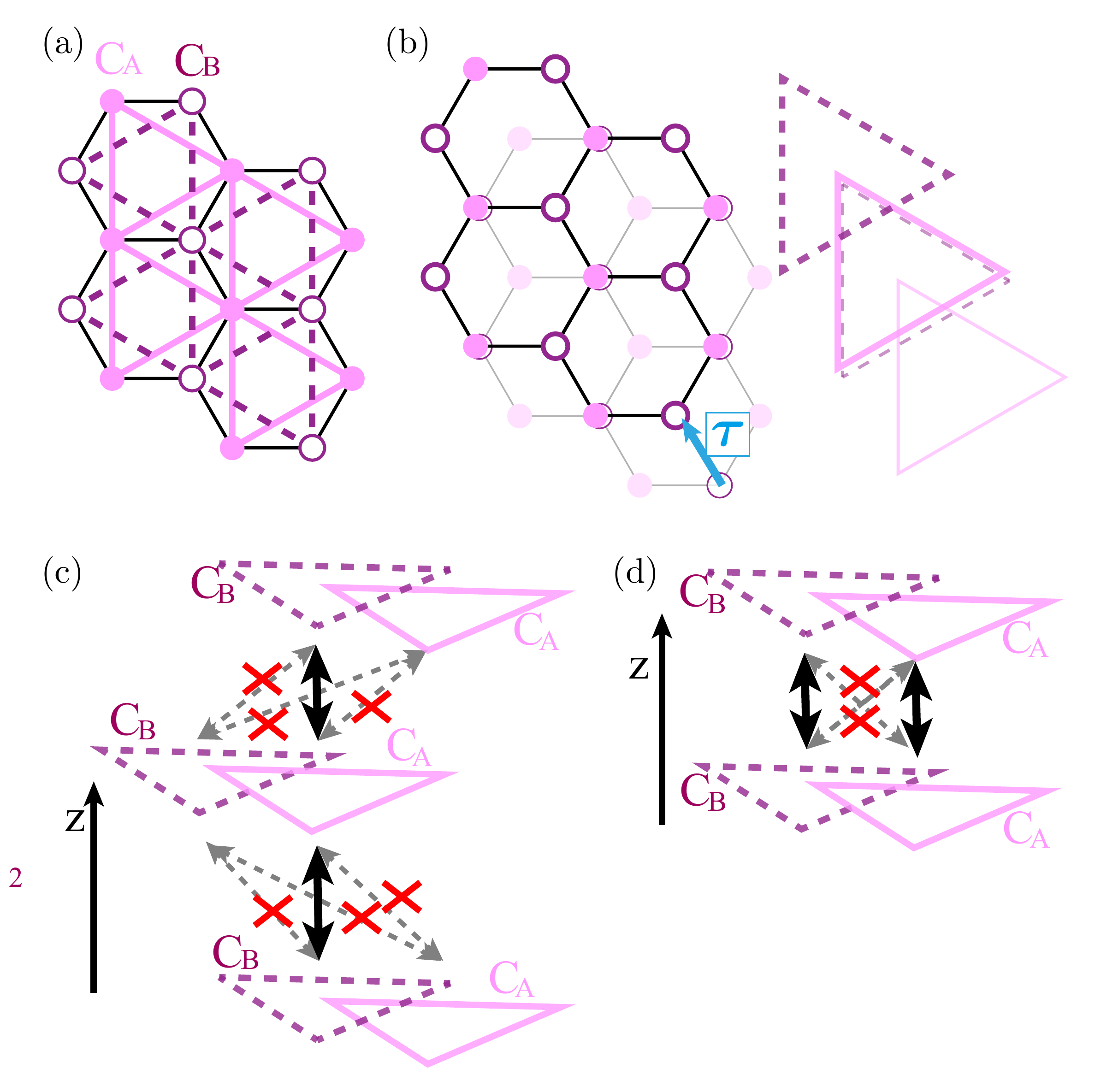}
  \caption{(a) site-selective nets for inequivalent C atoms in graphite monolayer. (b) bilayer with shift ${\bm \tau}$ (left) and corresponding relations of the nets (right). The bottom layer is depicted in lighter colors. (c) (d) the interlayer hybridization rules for the $K$-($K'$-)point states derived from the site-selective nets. The allowed (forbidden) hybridizations are depicted in solid (dashed) arrows.}
  \label{fig:graphite-rule}
 \end{center}
\end{figure}

In the band structure of graphene monolayer, the celebrated Dirac cone emerges at the $K$ and $K'$ points. Since the states at the Dirac cone are two-fold degenerate, interlayer hybridizations among all of the degenerate states inevitably become relevant. Still, we can apply the analysis based on the layer geometries. The key property is that the Dirac-cone states are formed by the C-$2p_{z}$ orbitals at the inequivalent C sites [C$_{A}$ and C$_{B}$ in Fig.~\ref{fig:graphite-rule}(a)]. We then introduce the auxiliary hexagonal nets for the C$_{A}$ and C$_{B}$ sites, respectively [Fig.~\ref{fig:graphite-rule}(a)]. Once we know that the two orbitals both belong to the same representation of $C_{3}$, we can later forget about the specific character of the orbitals and focus only on the geometric relations between the C$_{A}$- and C$_{B}$-nets. The analysis below corresponds to a simplified counterpart of Charlier {\it et al.} (Ref.~\onlinecite{Charlier-Gonze-multilayer-graphite1994})

Let us consider the stacked bilayer of graphene with shift ${\bm \tau}$ as depicted in Fig.~\ref{fig:graphite-rule}(b). Extracting the nets of the two layers, we see that the C$_{A}$-net of the upper layer is stacked onto the C$_{B}$-net of the lower layer without shift, whereas the shift vectors for the other combinations are either ${\bm \tau}$ or $-{\bm \tau}$. Therefore, the only allowed interlayer hybridization for the $K$- and $K'$-point states is that between the C$_{A}$ state of the upper and C$_{B}$ state of the lower [Fig.~\ref{fig:graphite-rule}(c)]. Note that for the bilayer stacked without shift, we straightforwardly get to the rule that the interlayer hybridizations across the C$_{A}$ and C$_{B}$ orbitals are prohibited [Fig.~\ref{fig:graphite-rule}(d)].
\begin{figure}[t]
 \begin{center}
  \includegraphics[scale=.45]{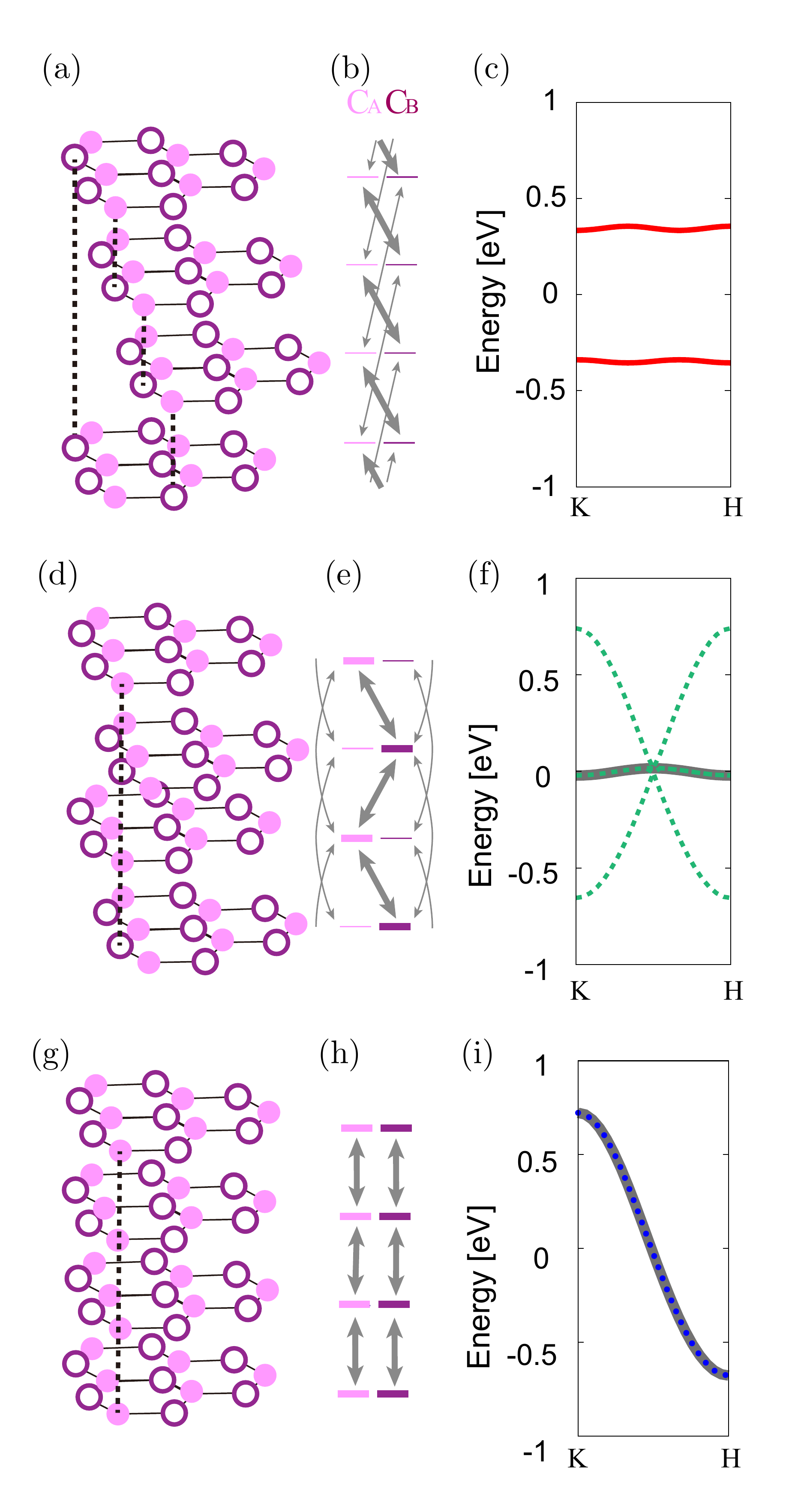}
  \caption{(a), (d) and (g) crystal structures of the bulk graphite with the rhombohedral, hexagonal, and AA stacking. (b), (e) and (h) the selection rules for interlayer hybridizations, where we show with gray arrows the allowed minimal hybridizations necessary for connecting all the states through the layers. The states forming channels dominated by the nearest-neighbor hybridization are depicted with thick horizontal lines. (c) (f) and (i) the first-principle band structures along the $K$--$H$ line, where thick lines indicate two-fold degeneracy.}
  \label{fig:graphite-sum}
 \end{center}
\end{figure}

On the basis of the rules derived above, we predict the band structures of bulk graphite in the $K$--$H$ path with three types of stacking: Rhombohedral (with shift ${\bm \tau}$, ${\bm \tau}$, ${\bm \tau}\cdots$ ), Bernal (${\bm \tau}$, $-{\bm \tau}$, ${\bm \tau}\cdots$), and AA (without shift) stacking [Fig.~\ref{fig:graphite-sum}(a), (d), (g)]. 

Let us begin with the rhombohedral stacking. We first depict the allowed hybridizations between the neighboring layers by thick arrows in Fig.~\ref{fig:graphite-sum}(b) based on Fig.~\ref{fig:graphite-rule}(c). These hybridizations yield localized bonding and antibonding states. A dispersive band should not be formed, however, since there is no channel passing through the whole crystal. Next, we add the allowed hybridizations between the next-nearest neighbor layers as thin arrows based on Fig.~\ref{fig:graphite-rule}(c). With these hybridizations, channels through the crystal are formed. Therefore, there should emerge bands whose bandwidths are governed by the second-nearest neighbor hybridization. The prediction is confirmed by the first-principle calculation as depicted in Fig.~\ref{fig:graphite-sum}(c). The inter-band splitting reflects the bonding/antibonding nature dominated by the nearest-neighbor hybridization.

\begin{figure*}[t]
 \begin{center}
  \includegraphics[scale=.35]{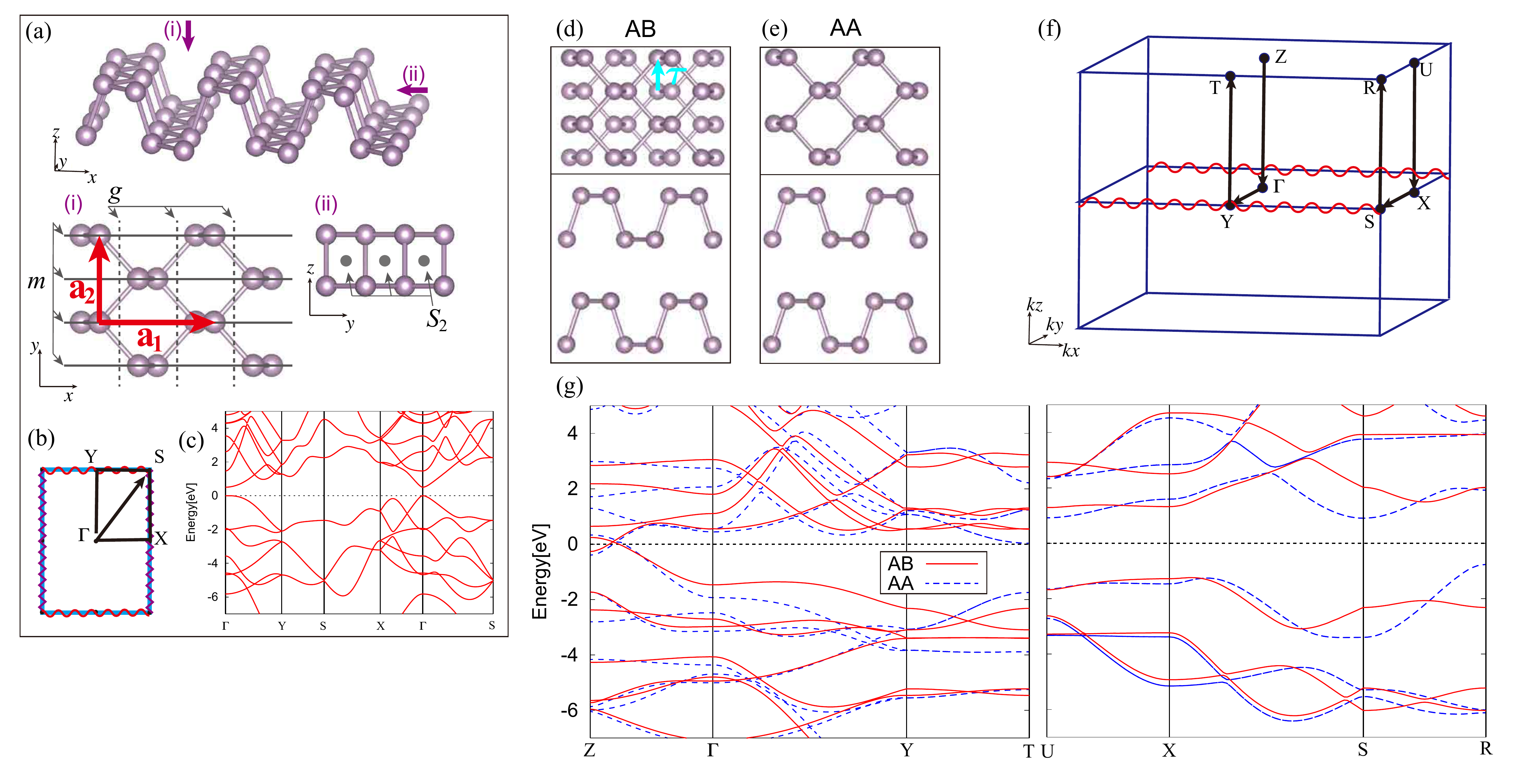}
  \caption{(a) Structure of the black phosphorene. In the (i) top and (ii) side views, symmetry elements are shown: Mirror $m$, glide-mirror $g$ and two-fold screw axis $S_{2}$. Vectors ${\bf a}_{1}$ and ${\bf a}_{2}$ denote the two-dimensional primitive translation vectors. (b) The corresponding BZ, where $\mathcal{K}_{\rm in}(m;{\bf a}_{2}/2)$ and $\mathcal{K}_{\rm in}(g;{\bf a}_{1}/2)$ are indicated by wavy and zigzag lines, respectively. (c) Band structure of the black phosphorene. (d), (e) Top (upper) and side views (lower) of bulk black phosphorus with AB (interlayer shift ${\bm \tau}={\bf a}_{2}/2$) and AA (${\bm \tau}=0$) stacking, respectively. (f) BZ for the orthorhombic structure, where $\mathcal{K}_{\rm in}(m;{\bf a}_{2}/2)$ in the monolayer form is indicated by the wavy lines. (g) Band structures of the bulk black phosphorus for the AB and AA stackings. Note that our calculated band structure shows gap closing at the Z point, at odds with the experiments~\cite{Keyes-P-measurement-PR1958,Akahama-P-hall-measure-JPSJ1983}. This is due to a drawback of the generalized gradient approximation to the exchange-correlation functional~\cite{PBEGGA} (see Appendix \ref{sec:abinitio-detail} for detailed conditions of the calculation) and supposed to be remedied by including the exchange effect more accurately~\cite{Tran-Yang-P-calc-PRB2014}.}
  \label{fig:phosphorene-sum}
 \end{center}
\end{figure*}

We also give a similar discussion for the Bernal stacking. Here, with only the nearest-neighbor hybridizations, a connected channel through the crystal is formed [Fig.~\ref{fig:graphite-sum}(e)]. For the second-nearest-neighbor layers, the relative shift is zero and therefore the rule depicted in Fig.~\ref{fig:graphite-rule}(d) applies. The remaining disconnected states become connected by the second-nearest-neighbor hybridizations and form channels through the crystal. Hence, there should emerge two types of bands, the former of which has larger bandwidth dominated by the nearest-neighbor hybridizations and the latter of which is flatter and dominated by the next-nearest hybridizations. The result of first-principles calculation is shown in Fig.~\ref{fig:graphite-sum}(f).

Finally, we consider the AA case. According to the rule of Fig.~\ref{fig:graphite-rule}(d), all the interlayer hybridizations are allowed among the C$_{A}$ and C$_{B}$ states, respectively. This yields two distinct channels connected by the nearest-neighbor hybridizations [Fig.~\ref{fig:graphite-sum}(h)]. These channels represent emergence of two-fold degenerate bands dominated by the nearest-neighbor hybridizations, which is consistent with the first-principles calculation presented in Fig.~\ref{fig:graphite-sum}(i).

Comparing the band structures, one would notice similarities in the energy scales. The more dispersive bands in the Bernal stacking has almost the same bandwidth as that of the band in the AA. The splitting of the rhombohedral bands is approximately half the bandwidth of the AA bands. On the other hand, the widths of the bands in the rhombohedral stacking and the flatter band in the Bernal stacking are almost the same. It is now clear that the former and latter features are dominated by the nearest-neighbor and next-nearest neighbor interlayer hybridizations, respectively.

\subsection{Black phosphorus}
\label{sec:phosphorus}
The recent successful exfoliation of black phosphorus monolayer (black phosphorene) has revealed its remarkable property as a two-dimensional semiconductor~\cite{Li-Zhang-phosphorene-transistor-nnano, Liu-Tomanek-phosphorene-mobility-acsnano}, which has added a new member to the possible building blocks of two-dimensional semiconductor devices. Numerous studies on its electronic structures in the monolayer and multilayer forms have been reported so far (for a review, see Refs.~\onlinecite{Morita-review1986,Liu-Ye-review2015,Ling-Dresselhaus-review2015,Carvalho-CastroNeto-review2016}). We reanalyze them from the viewpoint of the present theory. The black phosphorene forms in rectangular-net structure (Fig.~\ref{fig:phosphorene-sum}(a)), which is a marked feature compared with the systems addressed in the previous subsections, and therefore this should give an example if our theory on the interlayer hybridization is valid in a net other than the hexagonal one. 

As revealed in previous studies, the black phosphorene exhibits a direct gap at the $\Gamma$ point, whereas at the Brillouin-zone edge, where $\mathcal{K}_{\rm in}$ is located (Fig.~\ref{fig:phosphorene-sum}(b)), all the bands are doubly degenerate~(Fig.~\ref{fig:phosphorene-sum}(c)), which are forced by the presence of the two-fold screw and glide mirror symmetries (Fig.~\ref{fig:phosphorene-sum}(a))~\cite{Herring-nonsymmolph,Heine-textbook}. As in the case of graphene, for such degenerate bands, we cannot derive general selection rules for the interlayer hybridization only from the stacking shift. Nevertheless, we find below that the two statements derived from our theory are valid: (i) The interlayer hybridization of the states within the manifold $\mathcal{K}_{\rm in}(S;{\bm \tau})$ tends to be much suppressed (Sec.~\ref{sec:multiband-degene}), and (ii) the stacking-shift dependence of the electronic band structure is especially appreciable around $\mathcal{K}_{\rm in}(S;{\bm \tau})$ (Sec.~\ref{sec:stack-dependence-comment}). 

We calculated the band structure of the bulk black phosphorus; namely, black phosphorene in layered bulk forms. The familiar type of bulk black phosphorus [AB stacking (Fig.~\ref{fig:phosphorene-sum}(d))] has a structure where the inequivalent mirror planes are shared with the neighboring layers. In this case, the interlayer shift vector ${\bm \tau}={\bf a}_{2}/2$ and $\mathcal{K}_{\rm in}(m;$${\bf a}_{2}/2)$ (wavy line in Fig.~\ref{fig:phosphorene-sum}(b)) can therefore be employed. The calculations were then carried out for the AB and pathological AA stacking cases [${\bm \tau}$$=$$0$; Fig.~\ref{fig:phosphorene-sum}(e)]. The band structures for the Z--$\Gamma$--Y--T and U--X--S--R paths (Fig.~\ref{fig:phosphorene-sum}(f)) are shown in Fig.~\ref{fig:phosphorene-sum}(g). Remarkably, the bands along the paths $\Gamma$--Z and X--U are quite similar. On the other hand, in the paths Y--T and S--R, which are across $\mathcal{K}_{\rm in}(m;{\bf a}_{2}/2)$, the similarity is apparently lower and the dispersions tend to be smaller in the AB case. This feature represents the marked contrast in the hybridization properties in the two types of stacking and provides another example demonstrating the general applicability of the present theory.


\section{Application to general three-dimensional crystals}
\label{sec:three-dim}
In the previous sections, we have considered what happens on the electronic state when multiple layers are stacked, in a unified view of the interference effect on the hybridization. We have demonstrated that our theory enables us analysis based only on the stacking geometry for ${\bf k}$ points within $\mathcal{K}_{\rm in}(S; {\bm \tau})$. Below, we address an {\it inverse} problem: What can be derived by considering decomposition of three-dimensional crystals into stacked layers? Generally, three-dimensional periodic crystals are classified by the 14 Bravais lattices. Some of the Bravais lattices are decomposed into layers stacked so that the in-plane shift vector ${\bm \tau}$ for the neighboring layers makes $\mathcal{K}_{\rm in}(S; {\bm \tau})$ non-empty. Consequently, we obtain maps of ${\bf k}$-point paths for respective Bravais lattices, where the electronic band dispersions tend to be anomalously flat. This is due to the cancellation of the hybridization between the neighboring layers originating from the Bloch phase. We thereby propose a concept of the Bloch-phase induced flat-band paths (BIFP) for each Bravais lattice. Our theory reveals hidden ${\bf k}$-dependent anisotropy of the Bloch states in the three-dimensional lattice structure.

First, we consider a single-band tight-binding model on the FCC lattice [Fig.~\ref{fig:tb-FCC}(a)]. We demonstrate decomposition of the lattice into the layers with shift and derive the $k$-point paths where the band has little dispersion. This prediction is confirmed with first-principle calculations of the band structures in the NaCl-type semiconductors, whose structures belong to FCC.
\subsection{FCC tight-binding model}
\label{sec:tb-FCC}
\begin{figure*}[t]
 \begin{center}
  \includegraphics[scale=.50]{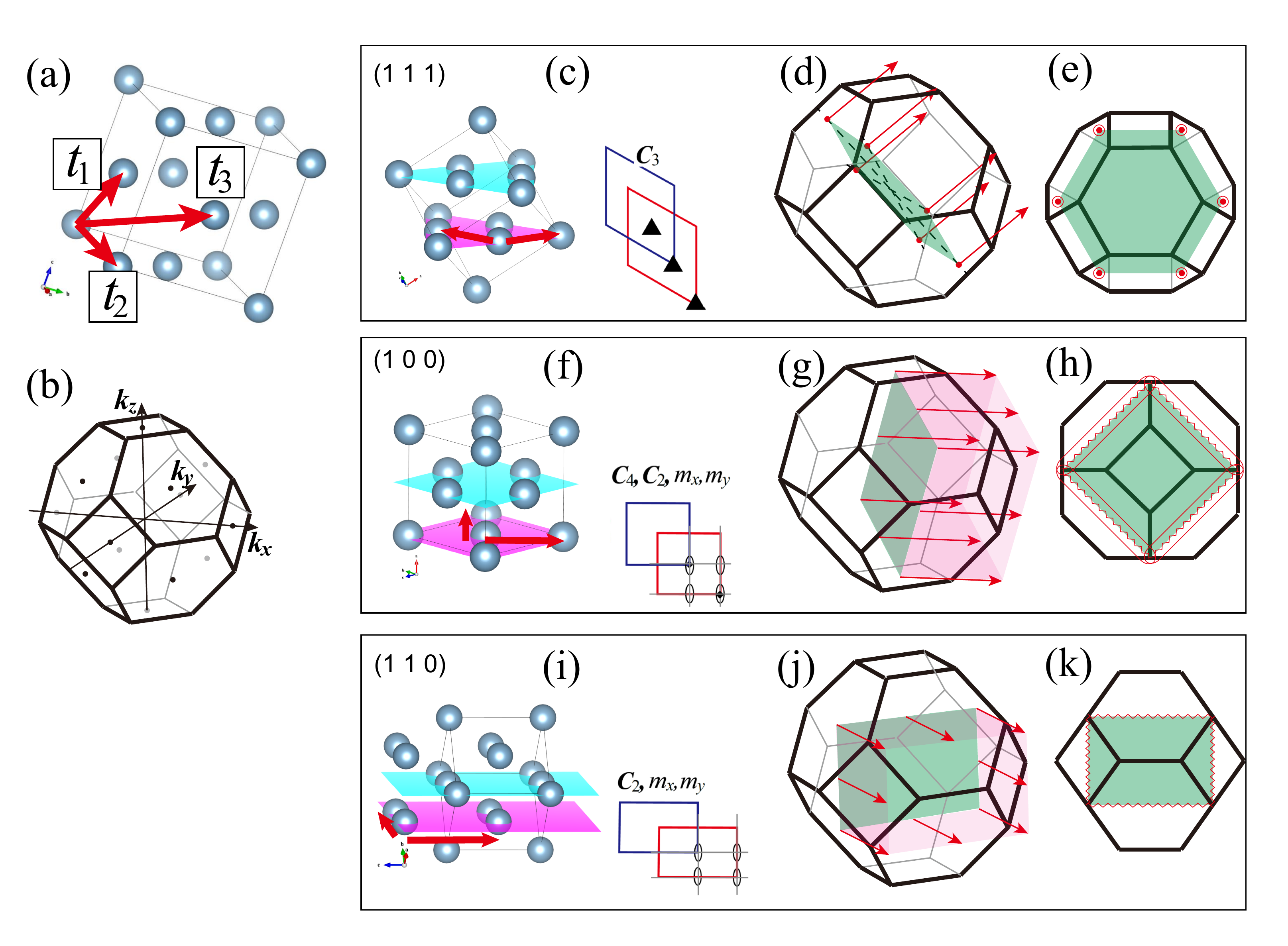}
  \caption{(a) the tight-binding model on the FCC lattice. Up to third-nearest neighbor inter-site hoppings are represented with arrows. (b) FCC BZ. (c) (f) and (i) decomposition of the FCC lattice into nets with shift perpendicular to various axes. (d) (g) (j) the two-dimensional BZ corresponding to the auxiliary lattice vectors depicted in (c) (f) and (i) respectively. $\mathcal{K}_{\rm in}(S; {\bm \tau})$ and the paths perpendicular to the net through $\mathcal{K}_{\rm in}(S; {\bm \tau})$ are depicted by arrows. (e), (h) and (k) top views of the paths.}
  \label{fig:tb-FCC}
 \end{center}
\end{figure*}

We here consider the single-band tight-binding model on the FCC lattice
\begin{eqnarray}
H
&=&H_{1}+H_{2}+H_{3}+\cdots
\nonumber \\
&=&t_{1}\sum_{\langle ij\rangle_{1}}c^{\dagger}_{i}c_{j}+t_{2}\sum_{\langle ij\rangle_{2}}c^{\dagger}_{i}c_{j}+t_{3}\sum_{\langle ij\rangle_{3}}c^{\dagger}_{i}c_{j}+\dots
\nonumber \\
&=&\sum_{{\bf k}}[\epsilon_{1}({\bf k})+\epsilon_{2}({\bf k})+\epsilon_{3}({\bf k})+\dots]\tilde{c}^{\dagger}_{{\bf k}}\tilde{c}_{{\bf k}}
,
\end{eqnarray}
where $H_{l} (l=1,2,3\cdots)$ denotes the $l$th nearest-neighbor intersite hopping terms and $\sum_{\langle ij \rangle_{l}}$ denotes the summation over the $l$th nearest pairs of sites. For example, the distances between the nearest-neighbor sites, next-nearest-neighbor sites, third nearest-neighbor sites $\dots$ are $\sqrt{2}a/2$, $a$, $\sqrt{6}a/2$, $\dots$, respectively, with $a$ being the lattice parameter~[Fig.~\ref{fig:tb-FCC}(a)]. We define $c^{\dagger}_{i}$ ($c_{i}$) as creation (annihilation) operator for the local orbital belonging to a one-dimensional representation of the point group $O_{h}$, which corresponds to the highest symmetry compatible with the FCC lattice. $t_{l}$ denotes the $l$th nearest neighbor hopping amplitude. Regardless of the specific values of $t_{1}, t_{2}, \dots$, the Hamiltonian is diagonalized with respect to the crystal wavenumber ${\bf k}$ defined within the FCC BZ [Fig.~\ref{fig:tb-FCC}(b)]. $\epsilon_{l}({\bf k})$ denotes the energy dispersion originating from $t_{l}$.

Seen from the (1 1 1) direction, the FCC lattice structure can be regarded as hexagonal nets stacked with shift, where the common global $C_{3}$ axis corresponds to the inequivalent axes for the neighboring layers [Fig.~\ref{fig:tb-FCC}(c)]. The ``intralayer" primitive lattice vectors are introduced as depicted in solid arrows in panel (c), and the corresponding two-dimensional BZ for the individual nets is determined to be a green plane in panel (d). Then, let us construct partial Bloch sums for each layer $|\Psi_{L, {\bf k}_{\rm 2D}} \rangle$ by summing up the local orbitals within each layer. Here, $L$ labels the index of the layer and ${\bf k}_{\rm 2D}$ is the wavenumber defined for the two-dimensional BZ. The partial Bloch sums can be hybridized between the layers via inter-site hopping terms connecting the sites in different layers. The resulting eigenstates are labeled by the $(1 1 1)$ component of the wavenumber and show energy dispersion along this direction. However, between any pairs of layers shifted by ${\bm \tau}$ [Fig.~\ref{fig:tb-FCC}(c)], hybridizations are perfectly canceled for ${\bf k}_{\rm 2D} \in \mathcal{K}_{\rm in}(C_{3}; {\bm \tau})$ because of Eq.~(\ref{eq:summary-singleband}). This occurs when ${\bf k}_{\rm 2D}$ is at the ``$K_{\rm 2D}$ or $K'_{\rm 2D}$ points" (namely, $K$ and $K'$ points of the two-dimensional hexagonal BZ). Namely, we get 
\begin{eqnarray}
&&
\langle \Psi_{L,K_{\rm 2D}}|H| \Psi_{L',K_{\rm 2D}}\rangle
=0
\\
&&(L'\!-\!L=\!3n\!+\!1, 3n\!+\!2; n=0, \pm1, \pm2, \dots)
.
\label{eq:hop-FCC}
\end{eqnarray}
Furthermore, it is easily found in the FCC crystal structure that up to the fifth-nearest hopping terms cannot connect the sites in the third- or further-nearest neighbor layers. In conjunction with Eq.~(\ref{eq:hop-FCC}), this fact yield 
\begin{eqnarray}
\langle \Psi_{L,K_{\rm 2D}}|H_{l}| \Psi_{L',K_{\rm 2D}}\rangle
=0
\hspace{5pt} (^{\forall} L\neq L'; l=1,2,3,4,5)
.
\end{eqnarray}
This means that the energy dispersions along $(1 1 1)$ through the $K_{\rm 2D}$ and $K'_{\rm 2D}$ points are not affected by these short-range hopping terms and dominated by $t_{6}$. Namely,
\begin{eqnarray}
\epsilon_{l}({\bf k})=const. (l=1,2,3,4,5)
\end{eqnarray}
for ${\bf k}=K_{\rm 2D}+s(1 1 1) {\rm or} K'_{\rm 2D}+s(1 1 1) (s:{\rm real})$. Thus, the band dispersion along these paths is anomalously flat even if the near-site hopping amplitudes are large. Discussions for the symmetrically equivalent directions such as (1 1 -1) also yield different paths where the band is anomalously flat.

\begin{figure*}[t]
 \begin{center}
  \includegraphics[scale=.55]{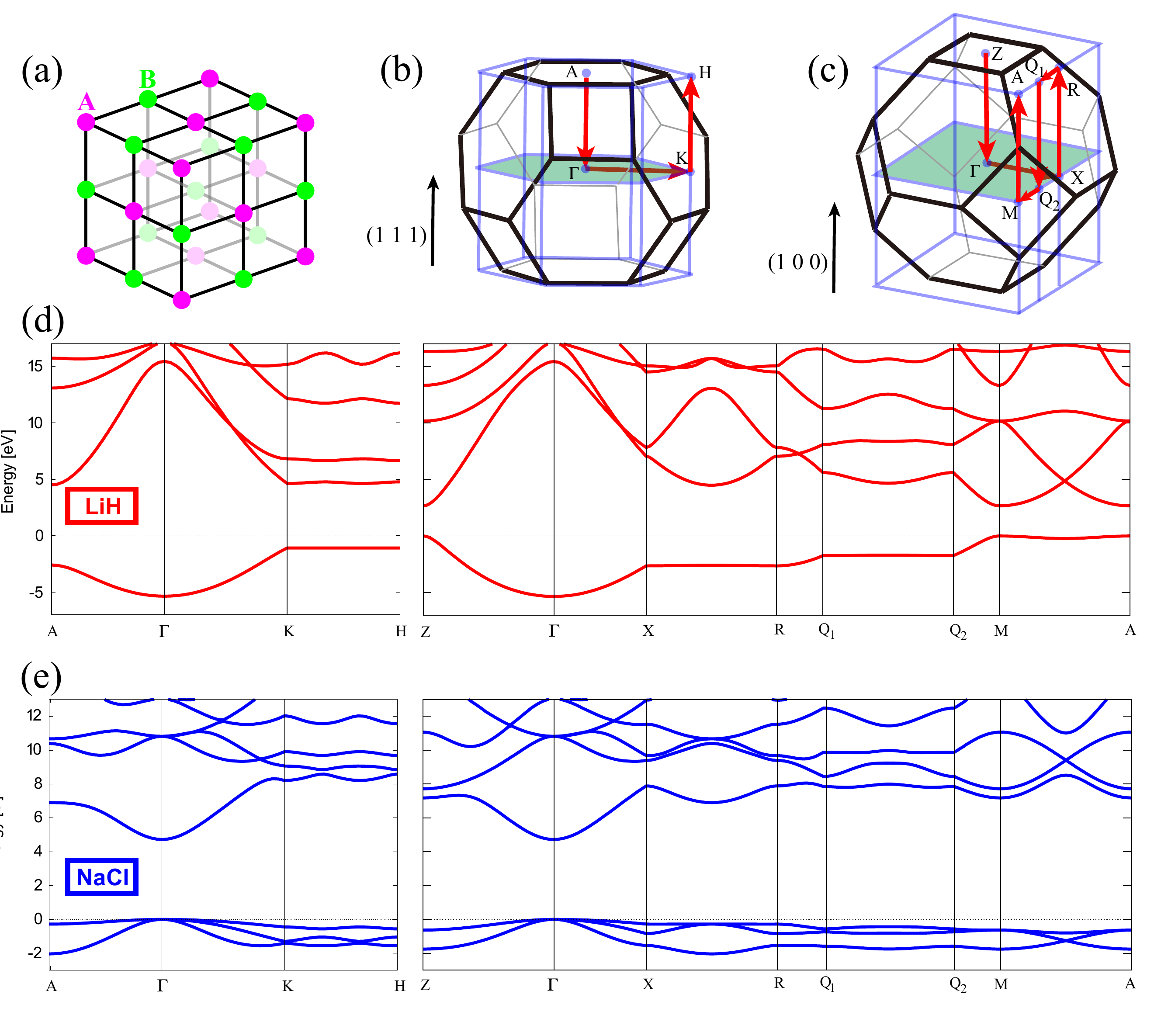}
  \caption{(a) the NaCl structure for $AB$ binary compounds, which belongs to the FCC lattice. (b) (c) auxiliary hexagonal and tetragonal BZs reflecting the layer decompositions in Figs.~\ref{fig:tb-FCC}(c) and \ref{fig:tb-FCC}(i). Note that the $Q_{1}$ and $Q_{2}$ points are introduced for convenience, not being special points in the tetragonal BZ. (d) [(e)] band structures in LiH (NaCl) calculated from first principles. The $k$-point paths are indicated in panels (b) and (c) respectively.}
  \label{fig:LiH-NaCl-bands}
 \end{center}
\end{figure*}

Next, the FCC lattice can be decomposed into the stacked square nets if seen from the (1 0 0) direction. With this view, inequivalent $C_{4}$ axes, $C_{2}$ axes, $m_{x}$ and $m_{y}$ planes of the odd-numbered and even-numbered layers coincide [Fig.~\ref{fig:tb-FCC}(f)]. The intralayer primitive lattice vectors and the corresponding BZ are defined as depicted in panels (f) and (g). Interlayer hybridizations are then prohibited for ${\bf k}\equiv{\bf k}_{\rm 2D}\in \mathcal{K}_{\rm in}(S; {\bm \tau})$ with $S=C_{4}, C_{2}, m_{x}, m_{y}$ and ${\bm \tau}={\bf a}_{1}/2+{\bf a}_{2}/2$, which represent all the points at the BZ edge (red or dark solid lines in panels (g) and (h); see Appendix~\ref{sec:list-net-map}). This prohibition yields 
\begin{eqnarray}
\epsilon_{1}({\bf k})=const.
\end{eqnarray}
for ${\bf k}={\bf k}_{\rm 2D}+s(1 0 0) (s:{\rm real})$. 

Seen from the (1 1 0) direction, on the other hand, the lattice can be decomposed into the stacked rectangular nets (panel (i)). With this stacking, the global $C_{2}$ axis and mirror planes $m_{x}$ and $m_{y}$ correspond to inequivalent symmetry reference axis/planes for the neighboring layers, respectively. The intralayer primitive lattice vectors and the corresponding BZ are defined as depicted in panels (i) and (j). According to the results in Fig.~\ref{fig:rectangle-sum}, interlayer hybridizations are prohibited for all the Bloch states at the BZ edge (panels (j) and (k)). In this case, however, even $H_{1}$ can generate the hopping between the second-nearest-neighbor layers, which are not shifted from each other. Then, $\epsilon_{1}(k)$ cannot be constant along the paths through the BZ edges toward the (1 1 0) direction.

Applying the present theory to the tight-binding model on the FCC lattice, we found ${\bf k}$-point paths where some near-site hopping terms does not contribute to the band dispersion. Significantly, in the paths along the (1 1 1) and (1 0 0) direction, the nearest-neighbor hopping terms $H_{1}$ do not give any dispersion. Here we note two key points in the above discussions: They respect only the kind of Bravais lattice and the symmetry compatible to it; second, although we have addressed a single-band model, the cancellation of the interlayer hybridization at ${\bf k}_{\rm 2D}\in\mathcal{K}_{\rm in}(S; {\bm \tau})$ is obviously valid in multiband systems, at least for intra-orbital ones~(Sec.~\ref{sec:multiband-degene}). Anomalously flat dispersions are therefore expected to be frequently realized along the same paths regardless of the atomic configuration in the unit cell. We hence define the BIFP as these paths for the FCC lattice. The general emergence of flat bands along the BIFP is later exemplified by the first-principles calculations for solid materials with the NaCl-type structure. 

\subsection{NaCl-type crystalline materials}
\label{sec:NaCl-type}
We here investigate the band structures of NaCl-type semiconductors: LiH and NaCl. Their structure does not show clear layered character as depicted in Fig.~\ref{fig:LiH-NaCl-bands}(a). Nevertheless, below we observe many anomalously flat dispersions along the BIFP. This represent that the corresponding Bloch states have very anisotropic effective masses along the BIFP despite the three-dimensional crystal structure, which is due to the interference of the Bloch phase induced by the stacking with shift.

We calculated the band structures of LiH and NaCl from the first principles for the paths drawn in Figs.~\ref{fig:LiH-NaCl-bands}(b) and (c). The detailed conditions for the first-principles calculations are available in Appendix ~\ref{sec:abinitio-detail}. Since the NaCl structure is $C_{3}$-symmetric seen from (1 1 1) and $C_{4}$-, $C_{2}$-, $m_{y}$- and $m_{z}$-symmetric seen from (1 0 0), the $K$--$H$, $X$--$R$, and $M$--$A$ paths belong to the BIFP, where the band structure is expected to show little dispersions. Also, any paths along the (1 0 0) direction through the edge of the 2D square BZ belong to the BIFP. In order to examine this, we introduce the $Q_{1}$ and $Q_{2}$ points at the halfway in the $R$--$A$ and $X$--$M$ paths, respectively, and take the $Q_{1}$--$Q_{2}$ path. 

For LiH [Fig.~\ref{fig:LiH-NaCl-bands}(d)], we see that the dispersion of the valence band formed by the H-$1s$ orbitals is much less in the $K$--$H$, $X$--$R$, $Q_{1}$--$Q_{2}$ and $M$--$A$ paths than in the $A$--$\Gamma$ and $Z$--$\Gamma$ paths, which verifies our expectations. On the other hand, for the conduction bands, although we see relatively flat dispersions for some of the bands, the dispersions are generally appreciable even in the former paths. 

A possible origin of the dispersions is obviously the hybridization between distant layers because the wave functions of the conduction states are more delocalized than those of the valence states. Another origin is the inter-orbital interlayer hybridizations. The conduction bands are formed by Li-$2s$ and $2p$ orbitals, which are approximately degenerate in energy. Since the in-plane Bloch sums formed by them corresponds to various representations of the in-plane symmetry, many types of hybridizations are possible even between the neighboring layers, though we do not go further about this point. Note that for the (1 0 0) direction, bands are flatter in the $Q_{1}$--$Q_{2}$ path compared with in the $X$--$R$ and $M$--$A$ paths. This is probably due to the energy splitting between the Li conduction states: In the isolated monolayer form, they well split at the $Q_{2}$ point because of the low symmetry, which reduce the impact of the inter-band hybridizations in the multilayer form (See Appendix~\ref{sec:multi-etc}).

For NaCl [Fig.~\ref{fig:LiH-NaCl-bands}(e)], we also see a number of bands with small dispersion in the BIFP. However, the dispersions of the valence bands are more significant compared with the LiH case. This is due to the degeneracy of the Cl-$3p$ states forming the valence bands. Note that we again find that the dispersions tend to be flat especially in the $Q_{1}$--$Q_{2}$ path for the both valence and conduction bands. Similarly to the LiH case, this is probably due to the lifted degeneracy of the Cl-$3p$ states induced by the low symmetry.

We have thus demonstrated that band dispersions indeed tend to be small in the BIFP. Although the magnitude of the flattening is essentially affected by the band degeneracy, we assert that similar flat dispersions should appear in the same paths for general FCC crystals.

\begin{table}[b]
\caption[t]
{Complete list of the Bravais lattices which can be decomposed into layers stacked with in-plane shift ${\bm \tau}$ so that $\mathcal{K}_{\rm in}(S; {\bm \tau})$ for the individual layers can be non-empty.}
\begin{center}
\label{tab:decompose}
\tabcolsep = 1mm
\begin{tabular}{|l |c|c|} \hline
 Crystal system& Centering& Decomposability \\ \hline
Triclinic & Primitive &  \\ \hline
Monoclinic & Primitive &  \\
 & Base-centered & yes \\ \hline
Orthorhombic & Primitive &  \\
 & Base-centered& yes \\
 & Face-centered& yes\\
 & Body-centered& yes\\ \hline
Tetragonal & Primitive& yes\\
 & Body-centered & yes\\ \hline
Rhombohedral &Primitive& yes\\ \hline
Hexagonal &Primitive& yes\\ \hline
Cubic & Primitive& yes\\ 
 & Face-centered& yes\\ 
 &Body-centered&yes \\
\hline
\end{tabular}
\end{center}
\end{table}

\subsection{General three-dimensional lattices}
\label{sec:general-3D}
It is worth mentioning again that the above discussion for the FCC system is {\it always} possible regardless of the structure within the unit cell. The key property of the FCC lattice is that it can be decomposed into layers so that their in-plane shift vector ${\bm \tau}$ makes $\mathcal{K}_{\rm in}(S; {\bm \tau})$ nonzero in the 2D BZ for these layers. As a matter of fact, 11 of the 14 Bravais lattices have this property (Table~\ref{tab:decompose}). Since the Brillouin-Zone shape is determined by the Bravais lattice, for these lattices, we can also define the BIFP and formulate one-to-one correspondences for the BZ and the BIFP. We provide in Appendix~\ref{sec:list-BZ} the representative BIFP derived from the decomposition. Note that we cannot give the complete list of BIFP for some Bravais lattices (see Appendix~\ref{sec:infinite-decompose}).

In Sec.~\ref{sec:tb-FCC}, we have seen that the nearest-neighbor hopping {\it does} affect the band structure along the BIFP with respect to the (1 1 0) direction. This represents that the probability of observing the flat bands along each BIFP depends on the lattice geometry. We can characterize this probability by referring to the distance between the neighboring layers $c_{\rm layer}$ in the decomposition scheme. For the BIFP yielded by the in-plane symmetry $S$, we define the threshold distance $d_{\rm th}$ as
\begin{eqnarray}
d_{\rm th}\equiv d_{\rm th}(S)=
\left \{
\begin{array}{c}
2c_{\rm layer} \ \ (S\neq C_{3}) \\
3c_{\rm layer} \ \ (S = C_{3}) 
\end{array}
\right.
.
\label{eq:th-dist}
\end{eqnarray}
Namely, $d_{\rm th}(S)$ corresponds to the shortest distance between the ``unshifted" layers---between which interlayer hybridization is not canceled. The following statement is consequently valid. Take an intra-orbital inter-site hopping between a certain pair of sites: If the distance between the sites is shorter than $d_{\rm th}$, such hopping does not contribute to the band dispersion along the BIFP. Equantion (\ref{eq:th-dist}) clarifies the advantage of the BIFPs yielded by $C_{3}$, where we should find flat dispersions especially frequently. The $K$--$H$ paths for bulk $r$-BN, 3R-MoS$_{2}$ and  $r$-graphite in Sec.~\ref{sec:layered-semicon} and those along the (1 1 1)-BIFP in the NaCl-type crystals in Sec.~\ref{sec:NaCl-type} correspond to this case. 

We occasionally encounter the cases that, in addition to the BIFP related to the Bravais lattice, specific atomic configurations within the unit cell and lattice parameters result in extra ${\bf k}$-point paths where near-site hoppings do not affect the band dispersions. We have in fact seen a typical example in Sec.~\ref{sec:BN-bulk} as $h$-BN. Although the primitive hexagonal lattice cannot generally yield the BIFP along the (0 0 1) direction (see Appendix~\ref{sec:list-BZ}), the hopping terms between the nearest-neighbor layers do not affect the dispersion along the $K$--$H$ path. Construction of the BIFP for such systems can be done by introducing the site-specific nets as demonstrated in Sec.~\ref{sec:layered-semicon}. The one-to-one correspondences between each space group (230 in total) and the BIFP could be formulated, though we do not address this issue in this paper.


\section{Discussions}
\label{sec:discussion}

\subsection{Relevance to experiments}
In the previous sections, we have demonstrated that there are $k$-points where the interlayer hybridizations are prohibited depending on the stacking shift and that much affects the electronic structures in multilayered systems. This characteristic is especially utilizable for van der Waals-coupled multilayer materials because they can adopt various stacking patterns thanks to the weak interlayer binding. In particular, when the valence-band top and/or conduction-band bottom are located at any of $\mathcal{K}_{\rm in}(S;{\bm \tau})$ in the monolayer form, one can demonstrate stacking-dependent characteristics of electronic states with low-energy probes. It is expected to be often the case because $\mathcal{K}_{\rm in}(S;{\bm \tau})$ includes some of the high-symmetry special points, where the band extrema are located. In fact, some of the authors have shown that the dimensionality of the excitons in MoS$_{2}$ is controlled by the stacking geometry through the reflectivity measurement~\cite{Akashi2015}. Since there has recently been a notable progress in the experimental technique to control the stacking shift~\cite{Kim-graphene-pret-a-porte-PRB, Jiang-Wu-MoS2-fold-NNano}, our theory should have a broad potential applicability. 

Even in the general three-dimensional crystals, the Bloch-phase interference can serve significant effects if the stacked-bilayer structure is formed locally. For example, BiS$_{2}$-based superconductors~\cite{Mizuguchi-Bi4O4S3-PRBR2012, Mizuguchi-LaOFBiS2-JPSJ2012} having BiS bilayers show abrupt increase of the superconducting transition temperature $T_{\rm c}$~\cite{Wolowiec-Yazici-pressure-PRB2013} concomitantly with the pressure-induced structural phase transition from the tetragonal to monoclinic structure~\cite{Tomita-Mizuguchi-pressure-structure-JPSJ2014, Guo-Yuan-EuBiS2F-pressure-structure-PRB2015}. It has been pointed out that the strong interlayer hybridization at the conduction-band bottom is switched on by this transition~\cite{Ochi-BiS2-JPSJ}, which could be crucial for enhancing the $T_{\rm c}$. This switching is due to the change of the interlayer shift, which breaks the interference condition. For the low-dimensional Bloch states emerging along the BIFP, however, it will be difficult to observe. Such states usually emerge in the middle of the valence/conduction bands because the BIFP usually do not pass through the special points of the {\it three-dimensional} BZ. In order to observe such high energy states, heavy doping or a wavenumber- and energy-resolved electronic probe is necessary. Still, we believe that our maps of the BIFP are useful for electronic materials search and design because of their general applicability. With a broad range of candidate materials, it could be possible to find suitable ones where we can detect the low-dimensional Bloch states. In fact, in LiH, the energy of the flat band in the $M$--$A$ path is near the Fermi energy, where holes could be injected with chemical or field-effect doping.

\subsection{Remark for tight-binding modeling}
A general remark on the development of tight-binding models for layered materials is derived from our theory. When one theoretically studies a layered system, only a small number of inter-site hopping matrix elements are usually retained in the model, particularly for those between the sites of different layers. However, when the layers are shifted from each other, such modeling will suffer from underestimation of interlayer transfer of electrons in the states within the SAIM, unless the hopping terms between ``unshifted" layers are included. Take for example the case of the graphite with the Bernal or rhombohedral stackings (Sec.~\ref{sec:graphite}). If only the hopping matrix elements up to the nearest-neighbor layers are retained in the modeling, the electrons in the $K$-point states never propagate across the whole layers (See Fig.~\ref{fig:graphite-sum} (e)). To avoid such too idealized situations, the inter-site hopping between the second-nearest neighbor layers ($\gamma'_{2}$ in the Partoens-Peeters modeling~\cite{Partoens-multilayer2006}) is needed for the both stackings.



\subsection{Possible extensions}
In this work, we have addressed only the interlayer transitions of electrons induced by the intrinsic Hamiltonian. The important fact is that the interlayer transitions concerning electronic states in the SAIM is subject to the stacking geometry. This is also extended to more general interlayer transitions via external perturbations such as incident light. The only difference is that the operator does not always transform according to the identity representation of the space group as does the Hamiltonian. Reconsideration of optical absorption spectra for few-layer graphene~\cite{Mak-Heinz-fewlayer-graphene-PRL2010} from the present viewpoint could give a new insight. Also, an extension to phononic systems is possible considering that the displacement vector of the phonon mode obeys the Bloch theorem. By expanding the total Born-Oppenheimer energy with respect to the mode displacement vectors for the isolated layers, we can derive the rule for which the second derivatives become zero. Our maps for the SAIM and BIFP will be useful when one considers such extensions.

\section{Summary and Conclusion}
\label{sec:summary}
We have addressed the Bloch-phase interference on the electronic motion in layered crystals and presented a general theory on the interlayer hybridizations in view of the stacking geometry. The key result is the development of the SAIM in the two-dimensional BZ, displaying which Bloch states are prevented from the interlayer hybridization depending on the stacking shift. We have demonstrated that our theory is useful for studying the electronic states in multilayer systems with applications to BN, graphite, TMD and black phosphorus. Also, we have shown that the prevention of the interlayer hybridization is robust even in general three-dimensional crystals and it induces strongly anisotropic electronic bands where effects of some inter-site hopping amplitudes are canceled. The theory is based on the classification of the periodic lattice structure and therefore in principle applicable to every material with periodic structure, which gives us a simple view on understanding and controlling the electronic states. 

\begin{acknowledgments}
This work was supported by MEXT Element Strategy Initiative to Form Core Research Center in Japan and JSPS KAKENHI Grant Number 15K20940 (R. A.) from Japan Society for the Promotion of Science (JSPS). We thank Ryotaro Arita, Shinji Tsuneyuki, Masayuki Ochi, Peter Maksym, Susumu Saito, and Ryuji Suzuki for careful reading of the manuscript and valuable comments.
\end{acknowledgments}

\begin{appendix}
\section{Derivation of Eqs.~(\ref{eq:trans-pw}) and (\ref{eq:trans-cell})}
\label{sec:derivation}
We first derive Eq.~(\ref{eq:trans-pw}). Suppose the origin of ${\bf r}$ is located on the reference point of $S_{1}$. We then trivially get 
\begin{eqnarray}
S_{1}e^{i{\bf k}\cdot{\bf r}}
=
e^{i{\bf k}\cdot{S^{-1}{\bf r}}}
=
e^{i(S{\bf k})\cdot{{\bf r}}}
\end{eqnarray}
and
\begin{eqnarray}
\varphi^{\rm pw}_{1}({\bf r})
=
e^{i(S{\bf k}-{\bf k})\cdot{{\bf r}}}
.
\end{eqnarray}
Equation~(\ref{eq:trans-pw}) is also obtained straightforwardly:
\begin{eqnarray}
S_{j}e^{i{\bf k}\cdot{\bf r}}
&\equiv&
T_{{\bm \tau}_{j-1}}
S_{1}
T^{-1}_{{\bm \tau}_{j-1}}
e^{i{\bf k}\cdot{\bf r}}
\nonumber \\
&=&
e^{-i(S{\bf k}-{\bf k})\cdot{\bm \tau}_{j-1}}
e^{i(S{\bf k})\cdot{\bf r}}
,
\end{eqnarray}
where we used $T_{{\bm \tau}}f({\bf r})=f({\bf r}-{\bm \tau})$ for general function $f({\bf r})$. This form is translated to Eq.~(\ref{eq:trans-pw}).

Before going to Eq.~(\ref{eq:trans-cell}), we introduce the plane-wave expansion of the Bloch function
\begin{eqnarray}
\psi_{{\bf k}}({\bf r})
=
e^{i{\bf k}\cdot{\bf r}}
\left[
\sum_{{\bf G}}
e^{i{\bf G}\cdot{\bf r}}
c_{{\bf k}, {\bf G}}
\right]
,
\end{eqnarray}
where the factor in the bracket corresponds to the cell-periodic part $u_{{\bf k}}({\bf r})$ and $c_{{\bf k}, {\bf G}}$ is the expansion coefficient. ${\bf G}$ is the reciprocal lattice vector. To obtain Eq.~(\ref{eq:trans-cell}), we here derive constraints between $c_{{\bf k}, {\bf G}}$, ${\bf G}$ and ${\bm \tau}_{j-1}$ imposed by the symmetry. Since $\psi_{{\bf k}}({\bf r})$ is an eigenfunction of $S$, the following expression holds: $S_{1}\psi_{{\bf k}}({\bf r})={\rm exp}(i\varphi_{1})\psi_{{\bf k}}({\bf r})$. Transforming this equation into the plane-wave expansion form,
\begin{eqnarray}
\sum_{{\bf G}}
e^{i(S{\bf G}+S{\bf k})\cdot{\bf r}}
c_{{\bf k}, {\bf G}}
=
\sum_{{\bf G}}
e^{i({\bf G}+{\bf k})\cdot{\bf r}}
e^{i\varphi_{1}}
c_{{\bf k}, {\bf G}}
.
\\
\end{eqnarray}
Dividing the both sides by ${\rm exp}(i{\bf k}\cdot{\bf r})$ and redefining the vector ${\bf G}$, we get
\begin{eqnarray}
\sum_{{\bf G}}
e^{i{\bf G}\cdot{\bf r}}
c_{{\bf k}, S^{-1}({\bf G}-{\bf G}_{0})}
=
\sum_{{\bf G}}
e^{i{\bf G}\cdot{\bf r}}
e^{i\varphi_{1}}
c_{{\bf k}, {\bf G}}
.
\end{eqnarray}
Here, a reciprocal lattice vector ${\bf G}_{0}$ is given by ${\bf G}_{0}=S{\bf k}-{\bf k}$, which is required for $\psi_{\bf k}({\bf r})$ to be an eigenfunction of $S$. The uniqueness of the Fourier expansion yields
\begin{eqnarray}
e^{i\varphi_{1}}
=
\frac{c_{{\bf k},S^{-1}({\bf G}-{\bf G}_{0})}}{c_{{\bf k},{\bf G}}}
\ \ 
(^{\forall} {\bf G})
.
\label{eq:pw-constraint}
\end{eqnarray}
Applying the similar discussion to $S_{j}\psi_{{\bf k}}({\bf r})={\rm exp}(i\varphi_{j})\psi_{{\bf k}}({\bf r})$,
\begin{eqnarray}
&&
e^{i\varphi_{j}}
\frac{c_{{\bf k},{\bf G}}}{c_{{\bf k},S^{-1}({\bf G}-{\bf G}_{0})}}
e^{i(S^{-1}{\bf G}_{0})\cdot {\bm \tau}_{j-1}}
\nonumber \\
&&\hspace{20pt}
=
e^{-i({\bf G}-S^{-1}{\bf G})\cdot {\bm \tau}_{j-1}}
\ \ 
(^{\forall} {\bf G})
.
\end{eqnarray}
Since the left-hand side is independent of ${\bf G}$ because of Eq.~(\ref{eq:pw-constraint}), the right-hand side must also be independent of ${\bf G}$; namely, 
\begin{eqnarray}
{\rm exp}[-i({\bf G}-S^{-1}{\bf G})\cdot {\bm \tau}_{j-1}]
= {\rm const.}
=1
.
\label{eq:pw-constraint2}
\end{eqnarray}
The second equation above is yielded by substituting ${\bf G}=0$.

Finally, let us move on to Eq.~(\ref{eq:trans-cell}). Operating $S_{1}$ on $u_{{\bf k}}({\bf r})$, we get
\begin{eqnarray}
S_{1}u_{{\bf k}}({\bf r})
&=&
\sum_{{\bf G}}
e^{i(S{\bf G})\cdot{\bf r}}
c_{{\bf k}, {\bf G}}
\nonumber \\
&=&
\sum_{{\bf G}}
e^{i{\bf G}\cdot{\bf r}}
c_{{\bf k}, S^{-1}{\bf G}}
.
\label{eq:S1-cell}
\end{eqnarray}
Similarly, 
\begin{eqnarray}
S_{j}u_{{\bf k}}({\bf r})
&\equiv&
T_{{\bm \tau}_{j-1}}
S_{1}
T^{-1}_{{\bm \tau}_{j-1}}
\sum_{{\bf G}}
e^{i{\bf G}\cdot{\bf r}}
c_{{\bf k}, {\bf G}}
\nonumber \\
&=&
\sum_{{\bf G}}
e^{i(S{\bf G})\cdot{\bf r}}
e^{-i(S{\bf G}-{\bf G})\cdot{\bm \tau}_{j-1}}
c_{{\bf k}, {\bf G}}
\nonumber \\
&=&
\sum_{{\bf G}}
e^{i{\bf G}\cdot{\bf r}}
e^{-i({\bf G}-S^{-1}{\bf G})\cdot{\bm \tau}_{j-1}}
c_{{\bf k}, S^{-1}{\bf G}}
.
\end{eqnarray}
Comparing this equation with Eq.~(\ref{eq:S1-cell}) and using Eq.~(\ref{eq:pw-constraint2}), we finally obtain Eq.~(\ref{eq:trans-cell}).

\section{Summary on stacking-adapted interference manifold (SAIM)}
\label{sec:list-net-map}
\subsection{Definition of the nets}
\label{sec:def-nets}
Nets are defined by the lengths of the two lattice translation vectors (${\bf a}_{1}$ and ${\bf a}_{2}$) and their relative angle ($\theta$) as follows:
\begin{eqnarray}
&&{\rm Oblique\  net:} |{\rm a}_{1}| \neq |{\rm a}_{2}|\ \  {\rm and} \ \ \theta \neq 90^{\circ},
\\
&&{\rm Rectangular\  net:} |{\rm a}_{1}| \neq |{\rm a}_{2}|\ \ {\rm and} \ \ \theta = 90^{\circ},
\\
&&{\rm Diamond\  net:} |{\rm a}_{1}| = |{\rm a}_{2}| \ \  {\rm and} \ \ \theta \neq 60^{\circ}, 90^{\circ}, 120^{\circ},
\\
&&{\rm Hexagonal\  net:} |{\rm a}_{1}| = |{\rm a}_{2}| \ \  {\rm and} \ \ \theta = 60^{\circ}, 120^{\circ},
\\
&&{\rm Square\  net:} |{\rm a}_{1}| = |{\rm a}_{2}| \ \  {\rm and} \ \ \theta = 90^{\circ}.
\end{eqnarray}

\subsection{Compatibility relation}
\begin{figure}[htbp]
 \begin{center}
  \includegraphics[scale=.48]{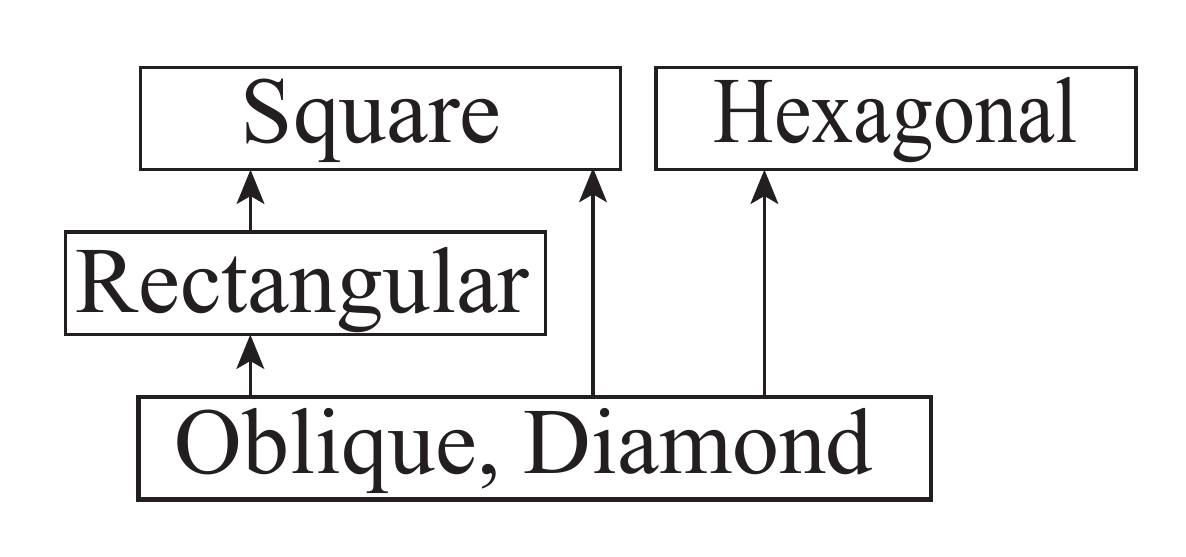}
  \caption{Compatibility relations for SAIM.}
  \label{fig:net-category}
 \end{center}
\end{figure}

By imposing some constraints on the lattice vectors defining a certain net, another net with higher symmetry is generated. Accordingly, the SAIM $\mathcal{K}_{\rm in}(S;{\bm \tau})$ for the former net also applies to those of the latter. We illustrate this relation in Fig.~\ref{fig:net-category} so that $\mathcal{K}_{\rm in}(S;{\bm \tau})$ for a net applies to those indicated by arrows. Note that the same set of SAIM is applicable to oblique and diamond nets.

\subsection{$\mathcal{K}_{\rm in}(S; {\bm \tau})$ for five nets}
\begin{figure}[htbp]
 \begin{center}
  \includegraphics[scale=.48]{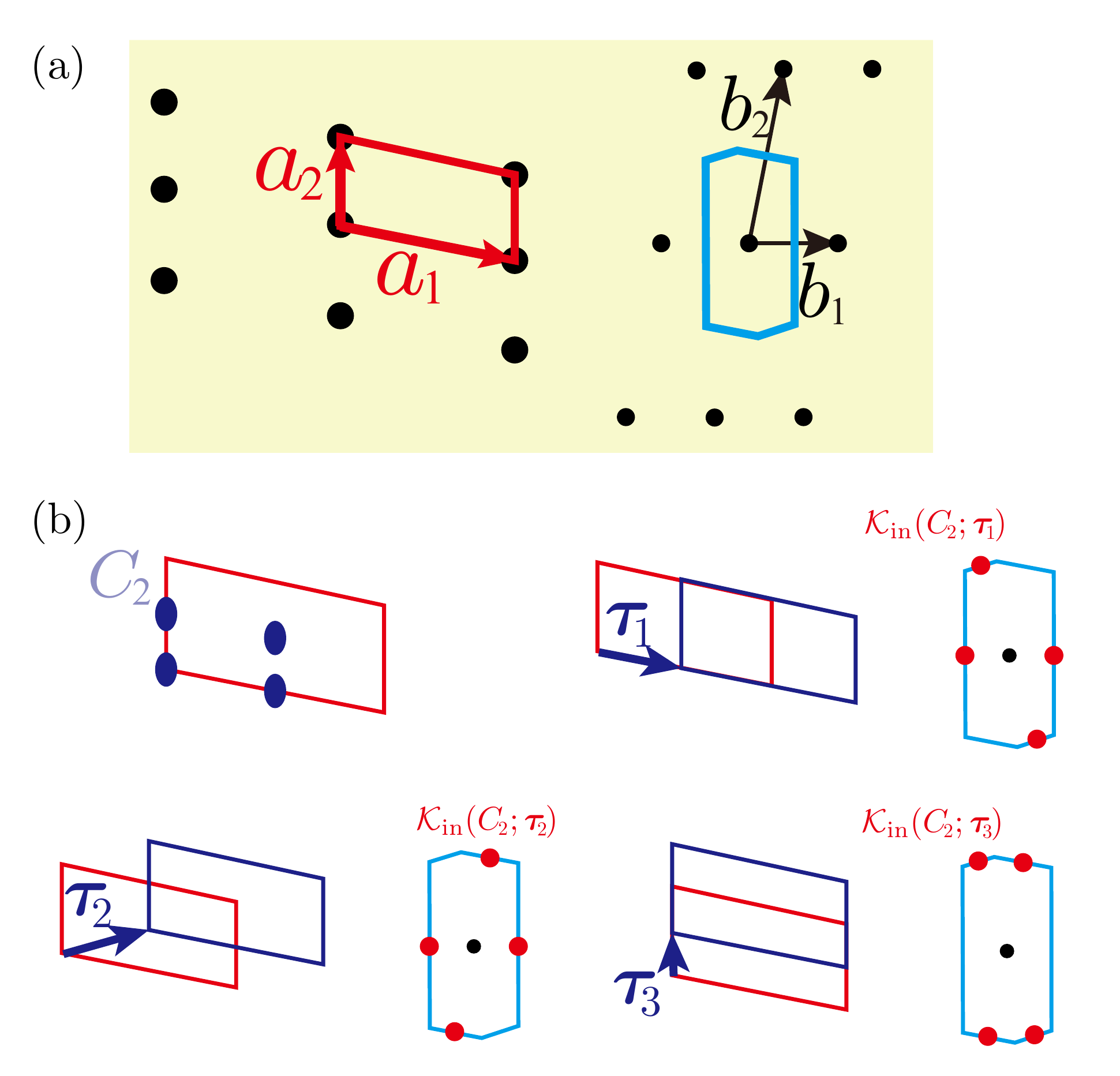}
  \caption{SAIM for the oblique net. (a) the primitive translation vectors ${\bf a}_{1}$ and ${\bf a}_{2}$, the corresponding reciprocal vectors ${\bf b}_{1}$ and ${\bf b}_{2}$, and the BZ. (b), possible lists of SAIM $\mathcal{K}_{\rm in}(S; {\bm \tau})$ when the crystal structure is $C_{2}$-symmetric.}
  \label{fig:oblique-sum}
 \end{center}
\end{figure}

\begin{figure}[htbp]
 \begin{center}
  \includegraphics[scale=.48]{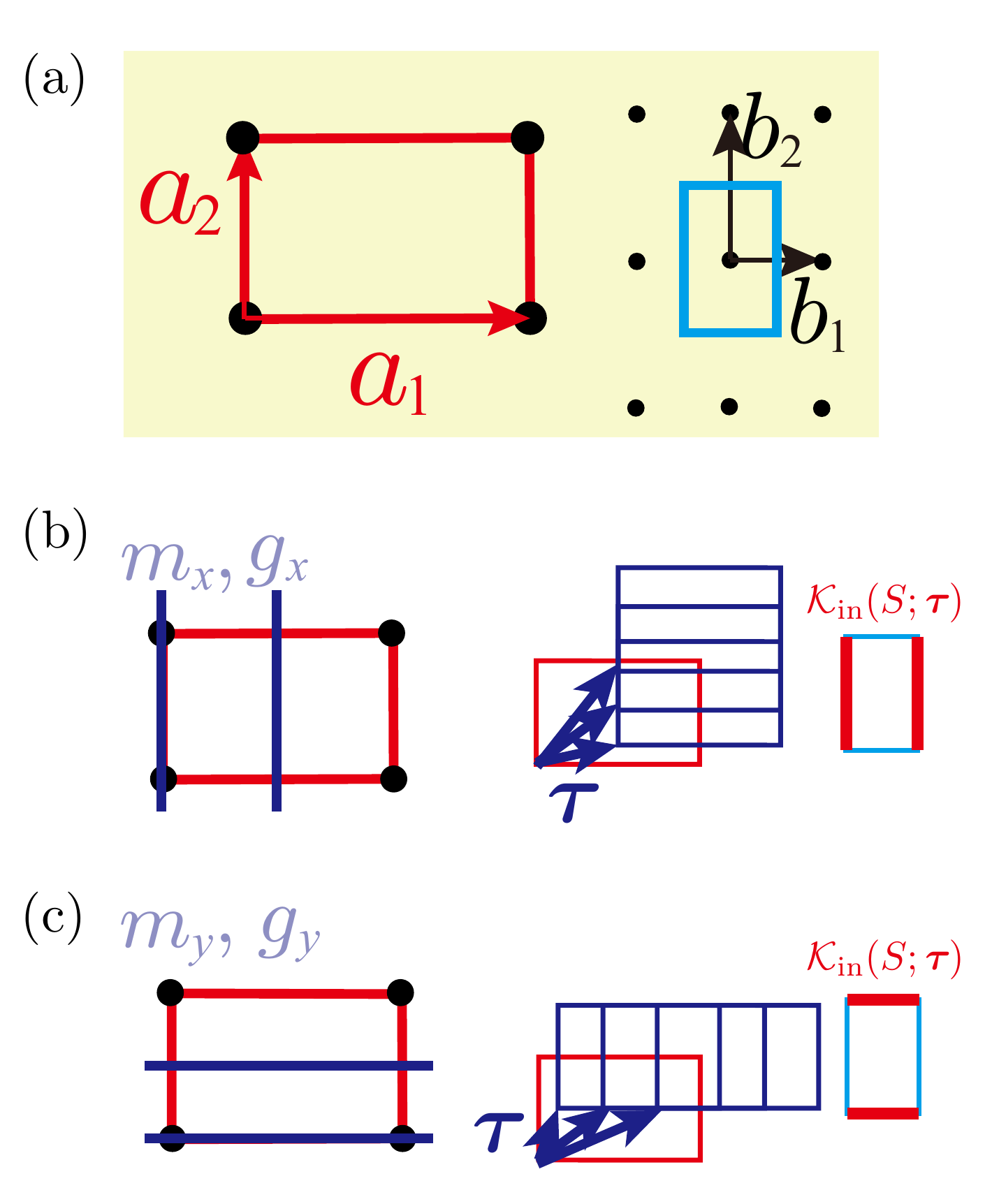}
  \caption{SAIM for the rectangular net. (a) the primitive translation vectors ${\bf a}_{1}$ and ${\bf a}_{2}$, the corresponding reciprocal vectors ${\bf b}_{1}$ and ${\bf b}_{2}$, and the BZ. (b) [(c)] $\mathcal{K}_{\rm in}(S; {\bm \tau})$ when the structure is $m_{x}$- or $g_{x}$-symmetric ($m_{y}$- or $g_{y}$-symmetric).}
  \label{fig:rectangular-sum-app}
 \end{center}
\end{figure}

\begin{figure}[htbp]
 \begin{center}
  \includegraphics[scale=.48]{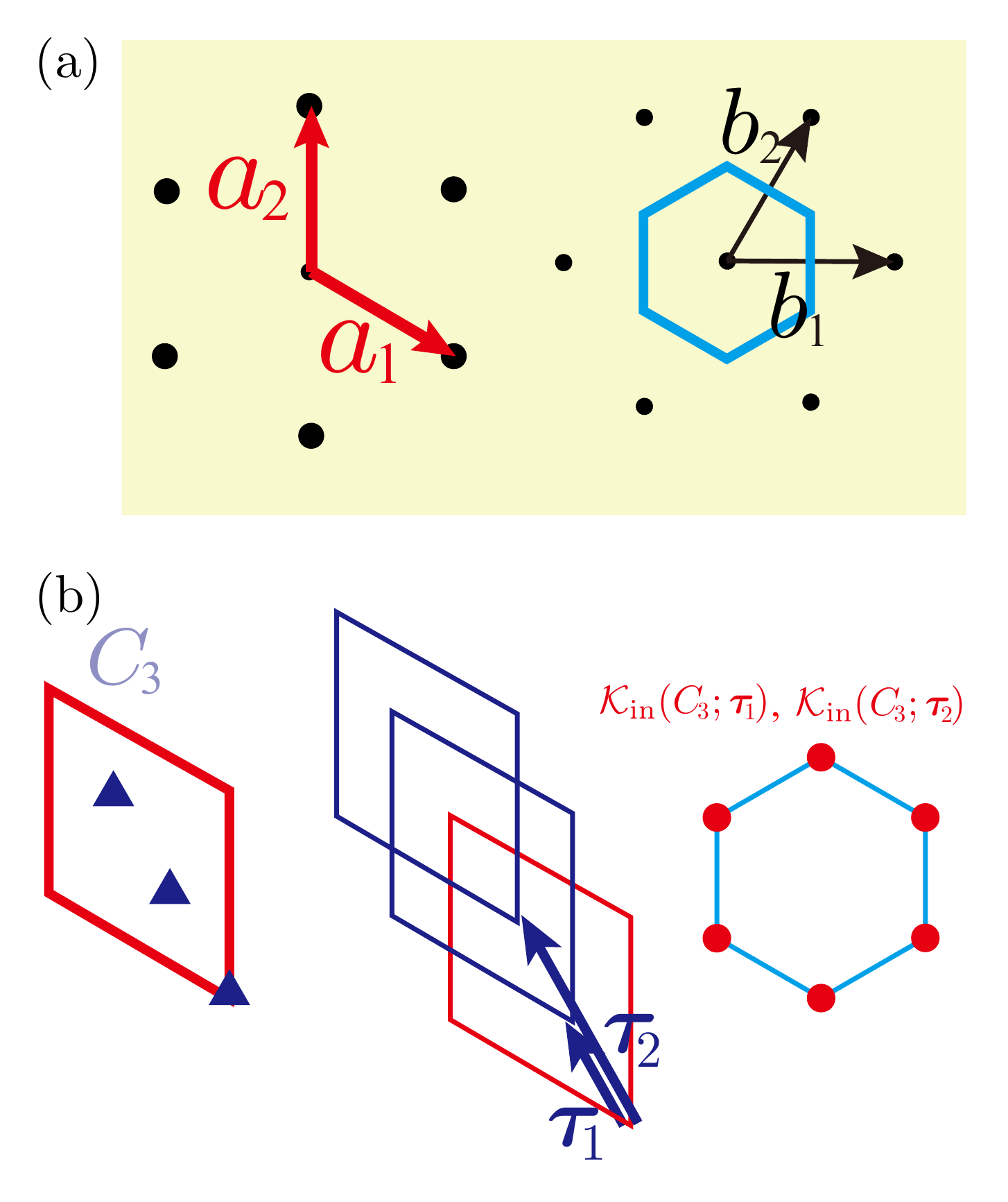}
  \caption{SAIM for the hexagonal net. (a) the primitive translation vectors ${\bf a}_{1}$ and ${\bf a}_{2}$, the corresponding reciprocal vectors ${\bf b}_{1}$ and ${\bf b}_{2}$, and the BZ. (b) $\mathcal{K}_{\rm in}(S; {\bm \tau})$ when the structure is $C_{3}$-symmetric.}
  \label{fig:hexagonal-sum}
 \end{center}
\end{figure}

\begin{figure}[htbp]
 \begin{center}
  \includegraphics[scale=.48]{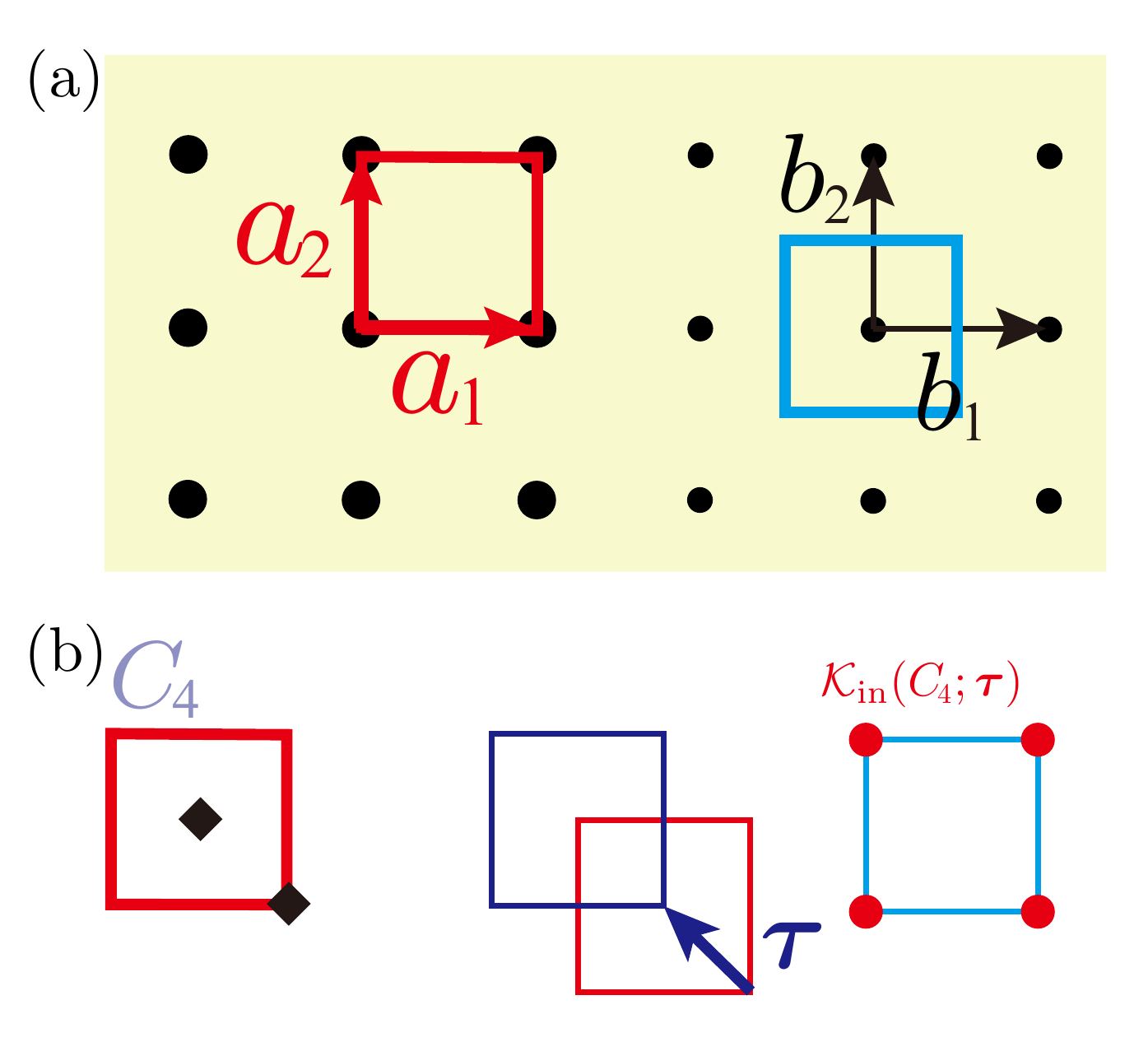}
  \caption{SAIM for the square net. (a) The primitive translation vectors ${\bf a}_{1}$ and ${\bf a}_{2}$, the corresponding reciprocal vectors ${\bf b}_{1}$ and ${\bf b}_{2}$, and the BZ. (b) $\mathcal{K}_{\rm in}(S; {\bm \tau})$ when the structure is $C_{4}$-symmetric.}
  \label{fig:square-sum}
 \end{center}
\end{figure}

Below, we show a complete list of non-empty $\mathcal{K}_{\rm in}$ for the five nets. For each net, we summarize all $\mathcal{K}_{\rm in}$ in the text for convenience, though in Figs.~\ref{fig:oblique-sum}--\ref{fig:square-sum} we depict only those which cannot be applicable to the nets with lower symmetries according to the compatibility relations in Fig.~\ref{fig:net-category}.

\subsubsection{Oblique and diamond nets}

\begin{eqnarray}
\mathcal{K}_{\rm in}(C_{2};{\bm \tau}_{1})
=
\{
{\bf k}|
{\bf k}=\frac{1}{2}{\bf b}_{1}, \ \frac{1}{2}{\bf b}_{1}+\frac{1}{2}{\bf b}_{2}
\}
,\\
\mathcal{K}_{\rm in}(C_{2};{\bm \tau}_{2})
=
\{
{\bf k}|
{\bf k}=\frac{1}{2}{\bf b}_{1}, \ \frac{1}{2}{\bf b}_{2}
\}
,\\
\mathcal{K}_{\rm in}(C_{2};{\bm \tau}_{3})
=
\{
{\bf k}|
{\bf k}=\frac{1}{2}{\bf b}_{2}, \ \frac{1}{2}{\bf b}_{1}+\frac{1}{2}{\bf b}_{2}
\}
,
\end{eqnarray}
with
\begin{eqnarray}
{\bm \tau}_{1}=\frac{1}{2}{\bf a}_{1}
, 
\\ 
{\bm \tau}_{2}=\frac{1}{2}{\bf a}_{1}+\frac{1}{2}{\bf a}_{2}
, 
\\
{\bm \tau}_{3}=\frac{1}{2}{\bf a}_{2}
,
\end{eqnarray}
respectively.

\subsubsection{Rectangular net}
(i)$S=C_{2}$
\begin{eqnarray}
\mathcal{K}_{\rm in}(C_{2};{\bm \tau}_{1})
=
\{
{\bf k}|
{\bf k}=\frac{1}{2}{\bf b}_{1}, \ \frac{1}{2}{\bf b}_{1}+\frac{1}{2}{\bf b}_{2}
\}
,\\
\mathcal{K}_{\rm in}(C_{2};{\bm \tau}_{2})
=
\{
{\bf k}|
{\bf k}=\frac{1}{2}{\bf b}_{1}, \ \frac{1}{2}{\bf b}_{2}
\}
,\\
\mathcal{K}_{\rm in}(C_{2};{\bm \tau}_{3})
=
\{
{\bf k}|
{\bf k}=\frac{1}{2}{\bf b}_{2}, \ \frac{1}{2}{\bf b}_{1}+\frac{1}{2}{\bf b}_{2}
\}
,
\end{eqnarray}
with
\begin{eqnarray}
{\bm \tau}_{1}=\frac{1}{2}{\bf a}_{1}
, 
\\ 
{\bm \tau}_{2}=\frac{1}{2}{\bf a}_{1}+\frac{1}{2}{\bf a}_{2}
, 
\\
{\bm \tau}_{3}=\frac{1}{2}{\bf a}_{2}
,
\end{eqnarray}
respectively.

(ii)$S=m_{x}, g_{x}$
\begin{eqnarray}
\mathcal{K}_{\rm in}(S;{\bm \tau}(t))
=
\{
{\bf k}|
{\bf k}=\frac{1}{2}{\bf b}_{1}+s{\bf b}_{2}
\}
(^{\forall} s)
\end{eqnarray}
with
\begin{eqnarray}
{\bm \tau}(t)=\frac{1}{2}{\bf a}_{1}+t{\bf a}_{2}
. 
(^{\forall} t)
\end{eqnarray}

(iii)$S=m_{y}, g_{y}$
\begin{eqnarray}
\mathcal{K}_{\rm in}(S;{\bm \tau}(t))
=
\{
{\bf k}|
{\bf k}=\frac{1}{2}{\bf b}_{2}+s{\bf b}_{1}
\}
(^{\forall} s)
\end{eqnarray}
with
\begin{eqnarray}
{\bm \tau}(t)=\frac{1}{2}{\bf a}_{2}+t{\bf a}_{1}
. 
(^{\forall} t)
\end{eqnarray}

\subsubsection{Hexagonal net}

(i)$S=C_{2}$
\begin{eqnarray}
\mathcal{K}_{\rm in}(C_{2};{\bm \tau}_{1})
=
\{
{\bf k}|
{\bf k}=\frac{1}{2}{\bf b}_{1}, \ \frac{1}{2}{\bf b}_{1}+\frac{1}{2}{\bf b}_{2}
\}
,\\
\mathcal{K}_{\rm in}(C_{2};{\bm \tau}_{2})
=
\{
{\bf k}|
{\bf k}=\frac{1}{2}{\bf b}_{1}, \ \frac{1}{2}{\bf b}_{2}
\}
,\\
\mathcal{K}_{\rm in}(C_{2};{\bm \tau}_{3})
=
\{
{\bf k}|
{\bf k}=\frac{1}{2}{\bf b}_{2}, \ \frac{1}{2}{\bf b}_{1}+\frac{1}{2}{\bf b}_{2}
\}
,
\end{eqnarray}
with
\begin{eqnarray}
{\bm \tau}_{1}=\frac{1}{2}{\bf a}_{1}
, 
\\ 
{\bm \tau}_{2}=\frac{1}{2}{\bf a}_{1}+\frac{1}{2}{\bf a}_{2}
, 
\\
{\bm \tau}_{3}=\frac{1}{2}{\bf a}_{2}
,
\end{eqnarray}
respectively.

(ii)$S=C_{3}$
\begin{eqnarray}
\mathcal{K}_{\rm in}(C_{3};{\bm \tau}_{1})
&=&
\left\{
{\bf k}|
{\bf k}=\frac{1}{3}{\bf b}_{1}+\frac{1}{3}{\bf b}_{2}, -\frac{1}{3}{\bf b}_{1}+\frac{2}{3}{\bf b}_{2}
\right\}
,
\\
\mathcal{K}_{\rm in}(C_{3};{\bm \tau}_{2})
&=&
\left\{
{\bf k}|
{\bf k}=\frac{1}{3}{\bf b}_{1}+\frac{1}{3}{\bf b}_{2},-\frac{1}{3}{\bf b}_{1}+\frac{2}{3}{\bf b}_{2}
\right\}
,
\end{eqnarray}
with
\begin{eqnarray}
{\bm \tau}=\frac{2}{3}{\bf a}_{1}+\frac{1}{3}{\bf a}_{2}
, 
\frac{1}{3}{\bf a}_{1}+\frac{2}{3}{\bf a}_{2}
.
\end{eqnarray}

\subsubsection{Square net}

(i)$S=C_{2}$
\begin{eqnarray}
\mathcal{K}_{\rm in}(C_{2};{\bm \tau}_{1})
=
\{
{\bf k}|
{\bf k}=\frac{1}{2}{\bf b}_{1}, \ \frac{1}{2}{\bf b}_{1}+\frac{1}{2}{\bf b}_{2}
\}
,\\
\mathcal{K}_{\rm in}(C_{2};{\bm \tau}_{2})
=
\{
{\bf k}|
{\bf k}=\frac{1}{2}{\bf b}_{1}, \ \frac{1}{2}{\bf b}_{2}
\}
,\\
\mathcal{K}_{\rm in}(C_{2};{\bm \tau}_{3})
=
\{
{\bf k}|
{\bf k}=\frac{1}{2}{\bf b}_{2}, \ \frac{1}{2}{\bf b}_{1}+\frac{1}{2}{\bf b}_{2}
\}
,
\end{eqnarray}
with
\begin{eqnarray}
{\bm \tau}_{1}=\frac{1}{2}{\bf a}_{1}
, 
\\ 
{\bm \tau}_{2}=\frac{1}{2}{\bf a}_{1}+\frac{1}{2}{\bf a}_{2}
, 
\\
{\bm \tau}_{3}=\frac{1}{2}{\bf a}_{2}
,
\end{eqnarray}
respectively.

(ii)$S=m_{x}, g_{x}$
\begin{eqnarray}
\mathcal{K}_{\rm in}(S;{\bm \tau}(t))
=
\{
{\bf k}|
{\bf k}=\frac{1}{2}{\bf b}_{1}+s{\bf b}_{2}
\}
(^{\forall} s)
\end{eqnarray}
with
\begin{eqnarray}
{\bm \tau}(t)=\frac{1}{2}{\bf a}_{1}+t{\bf a}_{2}
. 
(^{\forall} t)
\end{eqnarray}

(iii)$S=m_{y}, g_{y}$
\begin{eqnarray}
\mathcal{K}_{\rm in}(S;{\bm \tau}(t))
=
\{
{\bf k}|
{\bf k}=\frac{1}{2}{\bf b}_{2}+s{\bf b}_{1}
\}
(^{\forall} s)
\end{eqnarray}
with
\begin{eqnarray}
{\bm \tau}(t)=\frac{1}{2}{\bf a}_{2}+t{\bf a}_{1}
. 
(^{\forall} t)
\end{eqnarray}

(iv) $S=C_{4}$
\begin{eqnarray}
\mathcal{K}_{\rm in}(C_{4};{\bm \tau})
=
\left\{
\begin{array}{l}
{\bf k}|
{\bf k}=\frac{1}{2}{\bf b}_{1}+\frac{1}{2}{\bf b}_{2}
\end{array}
\right\}
\end{eqnarray}
with
\begin{eqnarray}
{\bm \tau}=\frac{1}{2}{\bf a}_{1}+\frac{1}{2}{\bf a}_{2}
.
\end{eqnarray}

\section{More on multiband and degeneracy}
\label{sec:multi-etc}
The advantage of the single-band case is that we can derive the selection rule without referring to the specific character of the cell-periodic part of the Bloch wave function because Eq.~(\ref{eq:trans-cell}) trivially holds regardless of its detail. In multiband cases, where the Bloch wave function has a band index, we have to at least care for interband interlayer hybridizations with an extended formula of Eq.~(\ref{eq:theorem-proof}) as 
\begin{eqnarray}
&&\langle 1,m,{\bf k}| H({\bm \tau}_{j}) |2,n,{\bf k} \rangle
\nonumber \\
&&
\hspace{10pt}=
e^{-{\rm i}\varphi_{mn;j}}
 \langle 1,m,{\bf k}| H({\bm \tau}_{j}) |2,n,{\bf k} \rangle
,
\label{eq:theorem-proof-multi}
\end{eqnarray}
with
\begin{eqnarray}
\varphi_{mn;j}
=({\bf k}-S{\bf k})\cdot {\bm \tau}_{j}+\varphi^{\rm cell}_{mn},
\label{eq:phases-mn}
\\
\varphi^{\rm cell}_{mn} =\varphi^{\rm cell}_{m;j}({\bf r})-\varphi^{\rm cell}_{n;1}({\bf r})
.
\end{eqnarray}
The state $|l,n,{\bf k} \rangle$ denotes the Bloch state of the $n$th band with wavenumber ${\bf k}$ $ | n, {\bf k}\rangle$ for the $l$th layer. Note that $|l,n,{\bf k} \rangle$ is an eigenstate of the symmetry operation $S$. $\varphi^{\rm cell}_{m,j}({\bf r})$, reflecting the character of the cell-periodic part of the band $m$, is defined by the operation on the cell-periodic part of the wave function as $S_{j}u_{m{\bf k}}({\bf r})={\rm exp}[i\varphi^{\rm cell}_{m,j}({\bf r})]u_{m{\bf k}}({\bf r})$. In Eq.~(\ref{eq:phases-mn}), $\varphi^{\rm cell}_{mn}$ is independent of ${\bf r}$ and $j$. The former independence is because of the independence of $\varphi^{\rm pw}_{j}({\bf r})$ from the band index and condition $\varphi^{\rm pw}_{j}({\bf r})+\varphi^{\rm cell}_{m;j}({\bf r})={\rm const.} (^{\forall} m,j)$; the latter comes from the fact that Eq.~(\ref{eq:pw-constraint2}) trivially holds regardless of the band index. Although the intraband interlayer hybridization ($m=n$) becomes zero for ${\bf k} \in \mathcal{K}_{\rm in}(S; {\bm \tau})$, the inter-band hybridization ($m\neq n$) can generally be nonzero due to $\varphi^{\rm cell}_{mn}$. Nevertheless, we can neglect the latter if the target band is energetically isolated from the other bands. The mixing ratio of the wave functions and the energy shift due to the hybridization are scaled by $O(t_{12;mm'}/\Delta E)$ and $O((t_{12;mm'}/\Delta E)^{2})$ for $t_{12;mm'}<<\Delta E$, respectively. Here, $\Delta E$ denotes the energy difference between the states $|1,m,k \rangle$ and $|2,n,k \rangle$ and $t_{12:mm'}$$\equiv$$\langle1,m,k |$$H({\bm \tau})$$|2,n,k \rangle$. Consequently, we can safely focus on only one band even when the unit cell includes more than one orbitals and therefore do not need information of $\varphi^{\rm cell}_{mn}$, if the target band is energetically well separated from other bands. When $\Delta E$ is rather small, however, the hybridization between the states $m\neq n$ becomes non-negligible and we need to consider specific characters of the Bloch states.

\section{Details of first-principles calculations}
\label{sec:abinitio-detail}
The first-principles calculations for the bilayer BN (Sec.~\ref{sec:BN-bi}) were done with the plane-wave pseudopotential method using {\it extended Tokyo Ab-initio Program Package} (xTAPP\cite{xTAPP}). The atomic potentials were approximated with the norm-conserving pseudopotentials~\cite{Troullier-Martins}. The band-structure calculations for the bulk systems [BN (Sec.~\ref{sec:BN-bulk}), TMD (Sec.~\ref{sec:TMD}), graphite~(Sec.~\ref{sec:graphite}), black phosphorus~(Sec.~\ref{sec:phosphorus}), LiH and NaCl (Sec.~\ref{sec:NaCl-type})] were performed with the full-potential linearized augmented plane-wave method as implemented in {\it wien2k}~\cite{WIEN2k}. The local-density approximation was employed for the exchange-correlation potentials~\cite{Ceperley-Alder,LDA-PZ} in the calculations for BN, graphite, LiH and NaCl. The results for MoS$_{2}$ and black phosphorus, the former of which were quoted from Ref.~\onlinecite{Akashi2015}, have been obtained with the generalized gradient approximation~\cite{PBEGGA}.

For bilayer BN, the lattice parameter $a$ and interlayer distance $c$ were set to theoretically optimized values for the bulk hexagonal structure~\cite{Kresse-graphite-BN-band1996}, whereas for bulk BN and graphite (black phosphorus), they were set to those for the bulk hexagonal (AB-stacked) structure determined experimentally~\cite{BN-Xray-exp,graphite-Xray-exp,P-As-Xray-exp}. The cubic lattice parameter for FCC-LiH and NaCl was also set to experimental values~\cite{Zintl-Harder-LiH, Nickels-Wallace-NaCl}.

The wave function energy cutoff for the plane-wave pseudopotential calculation was set to 64.0 Ry, whereas the cutoff parameter $RK_{\rm max}$ for the full-potential linearized augmented plane-wave calculations was set to $\geq 6.0$. Charge densities were converged using the equal {\bf k}-point meshes whose typical distance between the neighboring points are $\lesssim$$2\pi \times$ 0.05~\AA$^{-1}$.

\section{Bloch-phase induced flat-band paths (BIFPs) for the Bravais lattices}
\label{sec:list-BZ}
Among the 14 Bravais lattices, we can find the BIFP for 11 lattices regardless of the atomic configuration in the unit cell, which are depicted below in the corresponding BZs~\cite{BZ-list}. The primitive lattice vectors are defined in the Cartesian coordinate, whereas the BIFP are represented in the reciprocal-vector coordinate for convenience.  The probability of finding flat bands along the BIFP is characterized by the threshold distance $d_{\rm th}$ as discussed in Sec.~\ref{sec:general-3D}. The paths are represented by the variable $t$.
\begin{figure}[h!]
 \begin{center}
  \includegraphics[scale=.36]{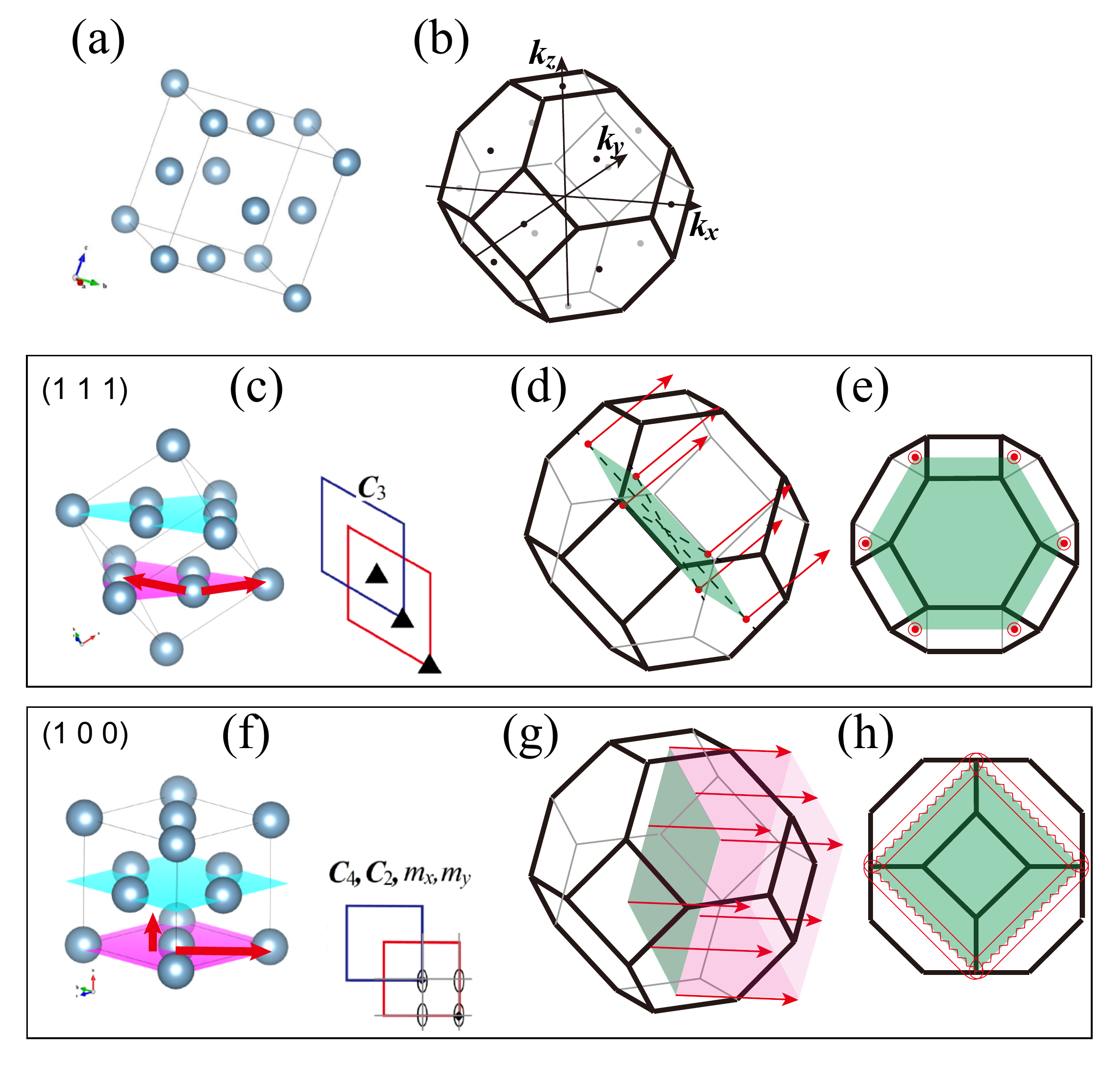}
  \caption{(a) Face-centered cubic lattice and (b) its corresponding BZ. (c) [(f)] Decomposition of the lattice into layers perpendicular to the (1 1 1) [(1 0 0)] direction. The primitive lattice vectors for each layer are represented by arrows. The interlayer shift and possible in-plane symmetry elements having two or more symmetry references are shown in the right. (d) (e) [(g) (h)] BIFP yielded by the decomposition of panel (c) [(f)]. Note that we show only the BIFP when the crystal structure has maximal symmetry compatible to the lattice and we do not show the BIFP whose $d_{\rm th}$ is not larger than the distance of the nearest-neighbor atoms. These points also apply to the later figures.}
  \label{fig:FCC-sum}
 \end{center}
\end{figure}

\subsection{Face centered cubic (FCC) lattice}
\begin{eqnarray}
{\bf a}_{1}=(a/2, a/2, 0)
\\
{\bf a}_{2}=(a/2, 0, a/2)
\\
{\bf a}_{3}=(0, a/2, a/2)
.
\end{eqnarray}
(i) (1 1 1)
\begin{eqnarray}
d_{\rm th}&=&\sqrt{3}a 
\\
 {\bf k}_{\rm BIFP}(t)&=&(\frac{1}{3}+t, -\frac{1}{3}+t, t) \ \ (C_{3})
\end{eqnarray}
(ii) (1 0 0)
\begin{eqnarray}
&& d_{\rm th}= a 
\\
&& {\bf k}_{\rm BIFP}(t)=\\
&& \hspace{5pt}
 \left\{
 \begin{array}{ll}
 (-\frac{1}{4}+t, -\frac{1}{4}+t, -\frac{1}{2}) &  (C_{2})\\
 (\frac{1}{4}+t, -\frac{1}{4}+t, 0)  &  (C_{2})
 \\
 (t, -\frac{1}{2}+t, -\frac{1}{2})  &  (C_{4})
  \\
 (-\frac{1}{4}+\frac{1}{2}s+t, -\frac{1}{4}-\frac{1}{2}s+t,-\frac{1}{2}) & (m_{x}) 
\\
 (\frac{1}{4}-\frac{1}{2}s+t, -\frac{1}{4}-\frac{1}{2}s+t,-s) & (m_{y}) 
 \end{array}
\right.
\end{eqnarray}

\begin{figure}[h!]
 \begin{center}
  \includegraphics[scale=.36]{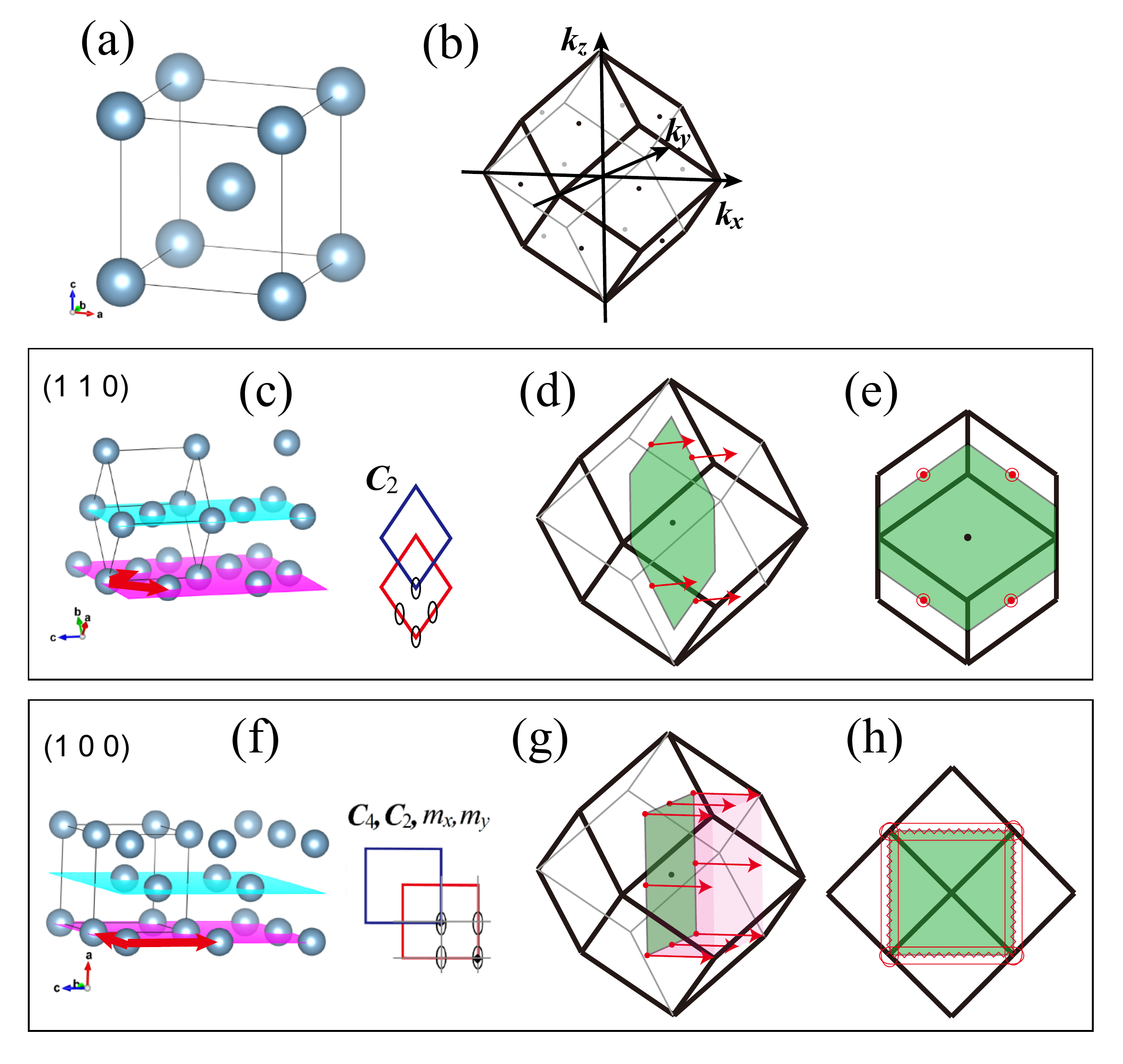}
  \caption{(a) Body-centered cubic lattice and (b) its corresponding BZ. (c) [(f)] Decomposition of the lattice into layers perpendicular to the (1 1 0) [(1 0 0)] direction. (d) (e) [(g) (h)] BIFP yielded by the decomposition of panel (c) [(f)].}
  \label{fig:BCC-sum}
 \end{center}
\end{figure}

\subsection{Body centered cubic (BCC) lattice}

\begin{eqnarray}
{\bf a}_{1}=(-a/2, a/2, a/2)
\\
{\bf a}_{2}=(a/2, -a/2, a/2)
\\
{\bf a}_{3}=(a/2, a/2, -a/2)
.
\end{eqnarray}

(i) (1 1 1)
\begin{eqnarray}
&& d_{\rm th}= \frac{\sqrt{3}}{2}a 
\\
&& {\bf k}_{\rm BIFP}(t)=\\
&& \hspace{5pt}
 \begin{array}{ll}
 (t, -\frac{1}{3}+t, \frac{1}{3}+t) &  (C_{3}) 
 \end{array}
\end{eqnarray}

(ii) (1 1 0)
\begin{eqnarray}
&& d_{\rm th}=\sqrt{2}a 
\\
&& {\bf k}_{\rm BIFP}(t)=\\
&& \hspace{5pt}
 \left\{
 \begin{array}{ll}
 (-\frac{1}{2}, 0, \frac{1}{4}+t) &  (C_{2})\\
 (0, -\frac{1}{2}, \frac{1}{4}+t) &  (C_{2})
 \end{array}
\right.
\end{eqnarray}

(iii) (1 0 0)
\begin{eqnarray}
&& d_{\rm th}=a 
\\
&& {\bf k}_{\rm BIFP}(t)=\\
&& \hspace{5pt}
 \left\{
 \begin{array}{ll}
 (-\frac{1}{4}-t, \frac{1}{4}+t, -\frac{1}{4}+t) &  (C_{2})\\
 (-\frac{1}{4}-t, -\frac{1}{4}+t, \frac{1}{4}+t) &  (C_{2})\\
 (-\frac{1}{2}-t, t, t) & (C_{4}) \\
 (-\frac{1}{4}+\frac{s}{2}-t,\frac{1}{4}-\frac{s}{2}+ t,-\frac{1}{4}+\frac{s}{2}+ t) & (m_{x}) \\
 (-\frac{1}{4}+\frac{s}{2}-t,-\frac{1}{4}+\frac{s}{2}+ t,\frac{1}{4}-\frac{s}{2}+ t) & (m_{y})
 \end{array}
\right.
\end{eqnarray}

\begin{figure}[h!]
 \begin{center}
  \includegraphics[scale=.36]{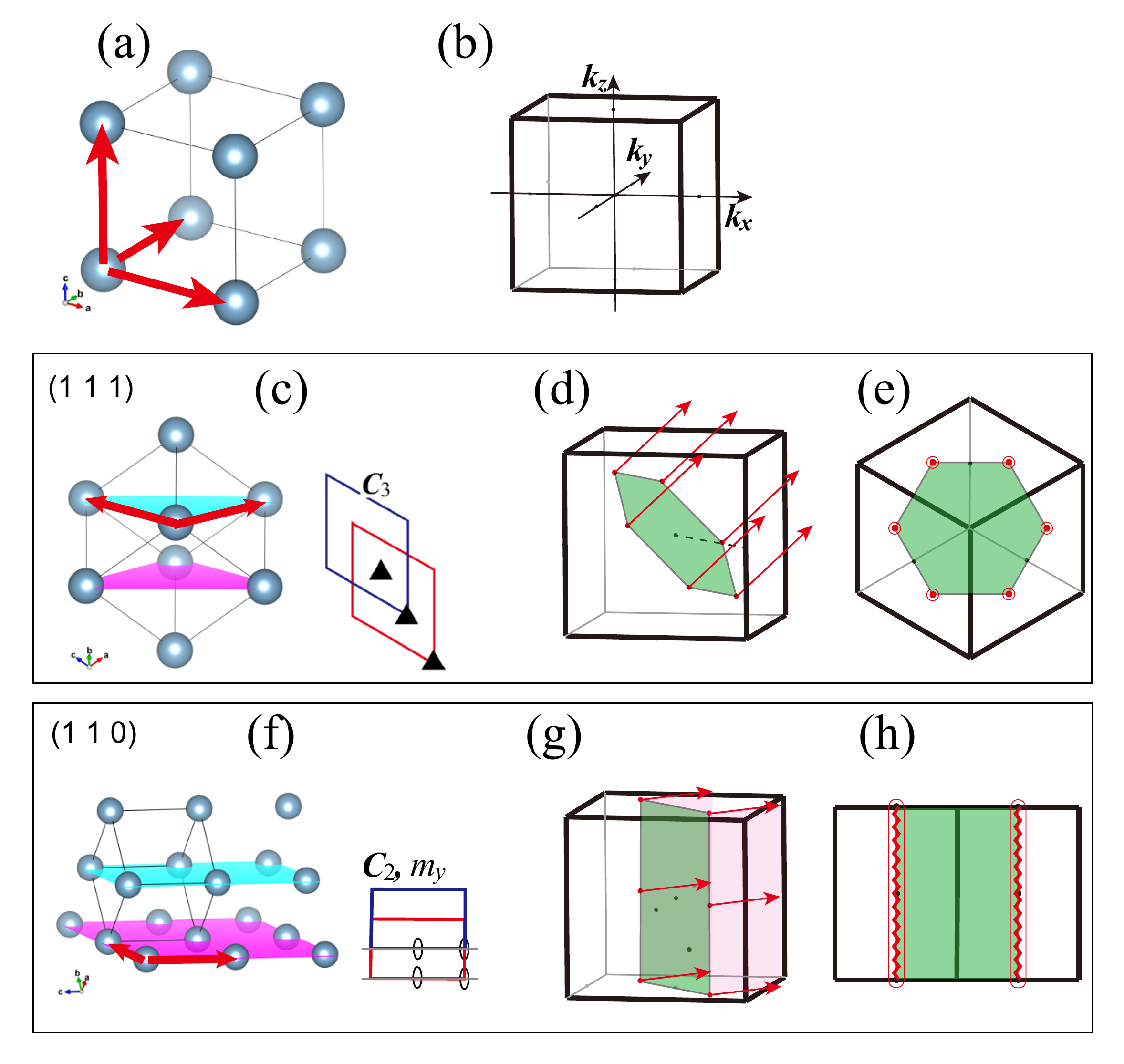}
  \caption{(a) Simple cubic lattice and (b) its corresponding BZ. (c) [(f)] Decomposition of the lattice into layers perpendicular to the (1 1 1) [(1 1 0)] direction. (d) (e) [(g) (h)] BIFP yielded by the decomposition of panel (c) [(f)].}
  \label{fig:PC-sum}
 \end{center}
\end{figure}

\subsection{Simple cubic (SC) lattice}
\begin{eqnarray}
{\bf a}_{1}=(a,0,0)
\\
{\bf a}_{2}=(0,a,0)
\\
{\bf a}_{3}=(0,0,a)
.
\end{eqnarray}

(i) (1 1 1)
\begin{eqnarray}
&& d_{\rm th}= \sqrt{3}a 
\\
&& {\bf k}_{\rm BIFP}(t)=\\
&& \hspace{5pt}
 \begin{array}{ll}
 (t, \frac{1}{3}+t, -\frac{1}{3}+t) &  (C_{3}) 
 \end{array}
\end{eqnarray}

(ii) (1 1 0)
\begin{eqnarray}
&& d_{\rm th}=a 
\\
&& {\bf k}_{\rm BIFP}(t)=\\
&& \hspace{5pt}
 \left\{
 \begin{array}{ll}
 (\frac{1}{4}+t, -\frac{1}{4}+t, 0) &  (C_{2})\\
 (\frac{1}{4}+t, -\frac{1}{4}+t, -\frac{1}{2}) &  (C_{2})\\
 (\frac{1}{4}+t, -\frac{1}{4}+t, -s) & (m_{x})
 \end{array}
\right.
\end{eqnarray}

\begin{figure}[h!]
 \begin{center}
  \includegraphics[scale=.36]{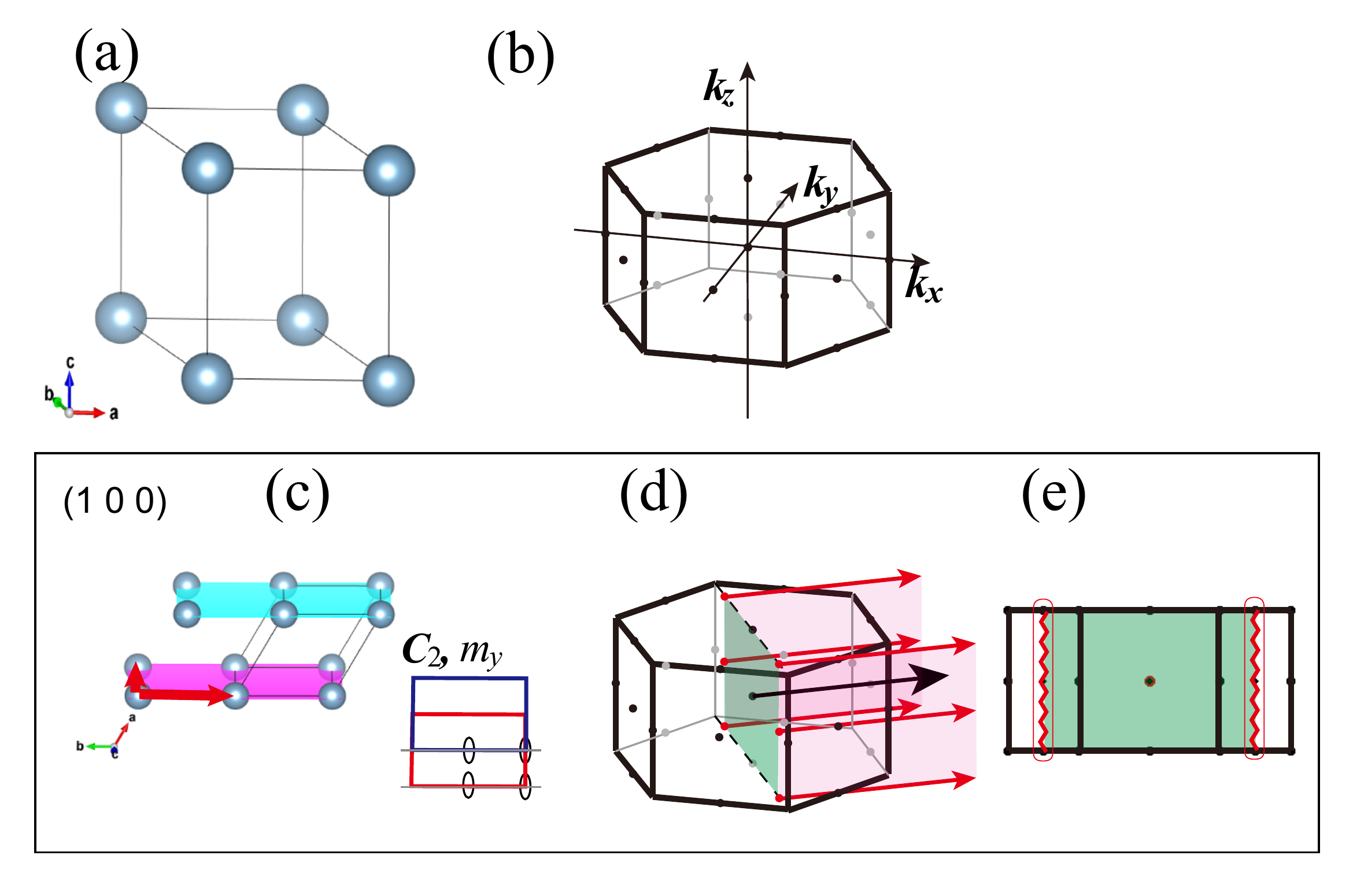}
  \caption{(a) Primitive hexagonal lattice and (b) its corresponding BZ. (c) Decomposition of the lattice into layers perpendicular to the (1 0 0) direction. (d) (e) BIFP yielded by the decomposition of panel (c).}
  \label{fig:PH-sum}
 \end{center}
\end{figure}

\subsection{Hexagonal (H) lattice}
\begin{eqnarray}
{\bf a}_{1}=(\sqrt{3}a/2,-a/2,0)
\\
{\bf a}_{2}=(0,a,0)
\\
{\bf a}_{3}=(0,0,c)
.
\end{eqnarray}

(i) (1 0 0)
\begin{eqnarray}
&& d_{\rm th}=\sqrt{3}a 
\\
&& {\bf k}_{\rm BIFP}(t)=\\
&& \hspace{5pt}
 \left\{
 \begin{array}{ll}
 (\frac{1}{4}+t, -\frac{1}{2}, 0) &  (C_{2})\\
 (\frac{1}{4}+t, -\frac{1}{2}, -\frac{1}{2}) &  (C_{2})\\
 (\frac{1}{4}+t, -\frac{1}{2}, -s) & (m_{x})
 \end{array}
\right.
\end{eqnarray}

(i) (1 -1 0)
\begin{eqnarray}
&& d_{\rm th}=a 
\\
&& {\bf k}_{\rm BIFP}(t)=\\
&& \hspace{5pt}
 \left\{
 \begin{array}{ll}
 (\frac{1}{4}-t, 2t, 0) &  (C_{2})\\
 (\frac{1}{4}-t, 2t, \frac{1}{2}) &  (C_{2})\\
 (\frac{1}{4}-t, 2t, s) & (m_{x})
 \end{array}
\right.
\end{eqnarray}

\begin{figure}[h!]
 \begin{center}
  \includegraphics[scale=.36]{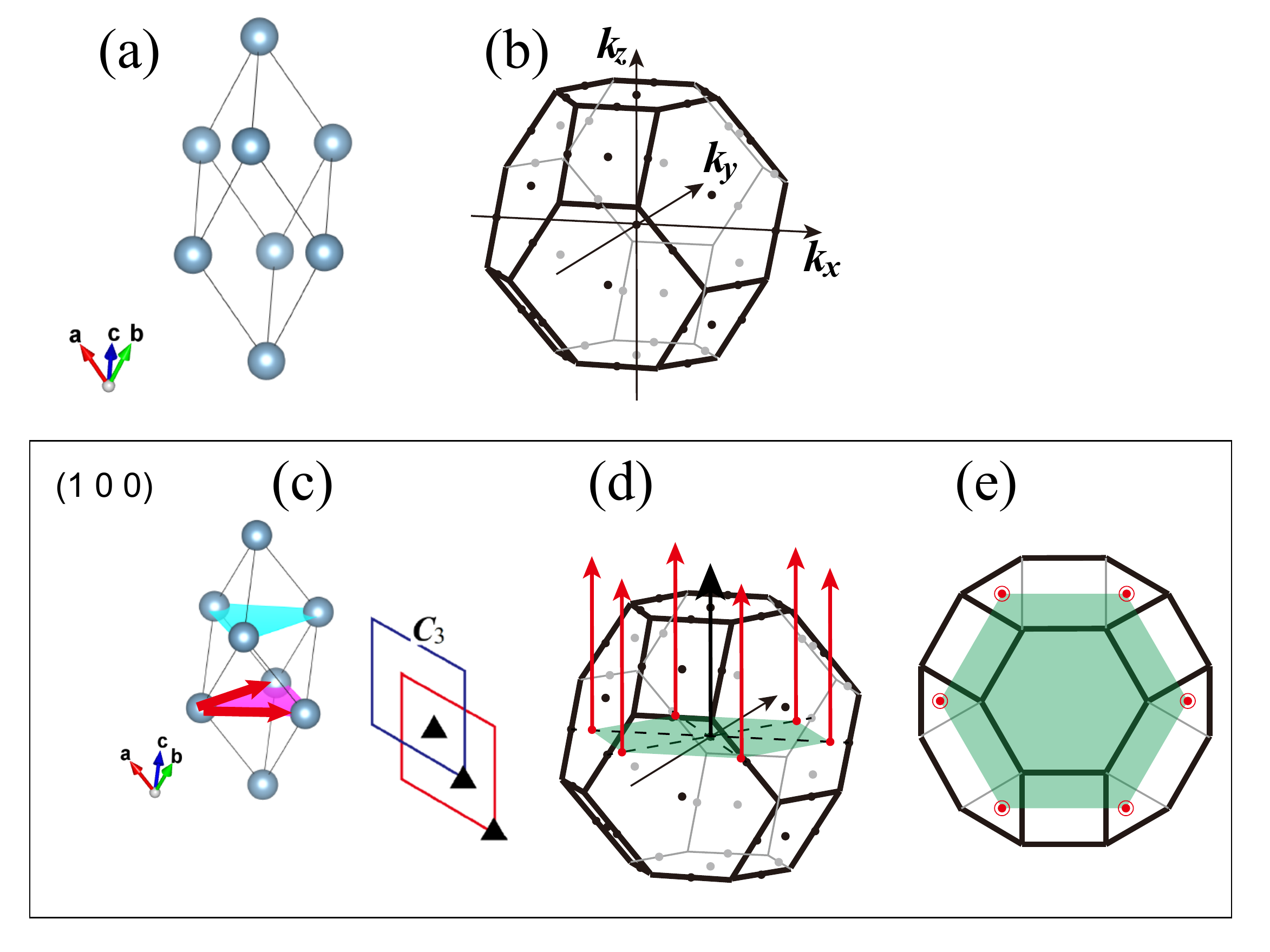}
  \caption{(a) Rhombohedral lattice and (b) its corresponding BZ. (c) [(f)] Decomposition of the lattice into layers perpendicular to the (0 0 1) direction. (d) (e) BIFP yielded by the decomposition of panel (c).}
  \label{fig:R-sum}
 \end{center}
\end{figure}

\subsection{Rhombohedral (Rh) lattice}
\begin{eqnarray}
{\bf a}_{1}\!=\!(\sqrt{\frac{1\!-\!\cos\alpha}{2}}a, -\sqrt{\frac{1\!-\!\cos\alpha}{6}}a, \sqrt{\frac{1\!+\!2\cos\alpha}{3}}a)
\nonumber \\
\\
{\bf a}_{2}\!=\!(0, 2\sqrt{\frac{1\!-\!\cos\alpha}{6}}a, \sqrt{\frac{1\!+\!2\cos\alpha}{3}}a)
\nonumber \\
\\
{\bf a}_{3}\!=\!(-\sqrt{\frac{1\!-\!\cos\alpha}{2}}a, -\sqrt{\frac{1\!-\!\cos\alpha}{6}}a, \sqrt{\frac{1\!+\!2\cos\alpha}{3}}a).
\nonumber \\
\end{eqnarray}

(i) (0 0 1)
\begin{eqnarray}
&& d_{\rm th}=\sqrt{3(1\!+\!2\cos\alpha)}a
\\
&& {\bf k}_{\rm BIFP}(t)=\\
&& \hspace{5pt}
 \left\{
 \begin{array}{ll}
 (-\frac{1}{3}+t, \frac{1}{3}+t, t) &  (C_{3})
 \end{array}
\right.
\end{eqnarray}

\begin{figure}[htbp]
 \begin{center}
  \includegraphics[scale=.36]{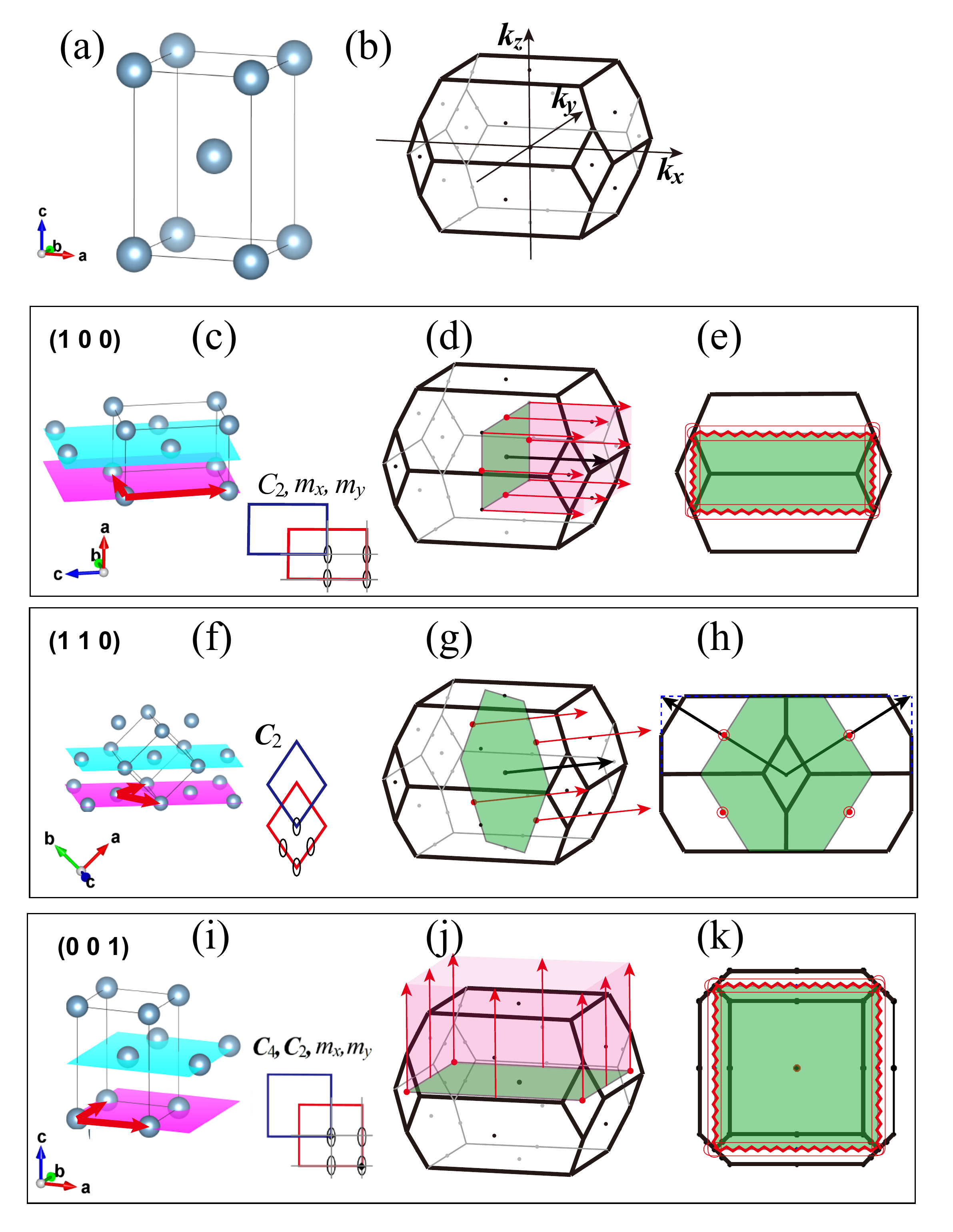}
  \caption{(a) Body-centered tetragonal lattice and (b) its corresponding BZ. (c) (f) (i) Decomposition of the lattice into layers perpendicular to the (1 0 0), (1 1 0), and (0 0 1) directions, respectively. (d) (e), (g) (h), and (j) (k) BIFP yielded by the decomposition of panels (c), (f), and (i), respectively.}
  \label{fig:ICT-sum}
 \end{center}
\end{figure}

\subsection{Body centered tetragonal (ICT) lattice}
\begin{eqnarray}
{\bf a}_{1}=(a/2,-a/2,c/2)
\\
{\bf a}_{2}=(a/2,a/2,c/2)
\\
{\bf a}_{3}=(-a/2,-a/2,c/2)
.
\end{eqnarray}

(i) (1 1 0)
\begin{eqnarray}
&& d_{\rm th}=\sqrt{2}a 
\\
&& {\bf k}_{\rm BIFP}(t)=\\
&& \hspace{5pt}
 \left\{
 \begin{array}{ll}
 (0, -\frac{1}{4}+t, -\frac{1}{4}-t) &  (C_{2})\\
  (-\frac{1}{2}, -\frac{1}{4}+t, -\frac{1}{4}-t) &  (C_{2})
 \end{array}
\right.
\end{eqnarray}
(ii) (1 0 0)
\begin{eqnarray}
&& d_{\rm th}=a 
\\
&& {\bf k}_{\rm BIFP}(t)=\\
&& \hspace{5pt}
 \left\{
 \begin{array}{ll}
 (\frac{1}{4}+t, -\frac{1}{4}+t, \frac{1}{4}-t) &  (C_{2})\\
 (-\frac{1}{4}+t, -\frac{1}{4}+t, -\frac{1}{4}-t) &  (C_{2})\\
  (\frac{1}{4}-\frac{s}{2}+t, -\frac{1}{4}-\frac{s}{2}+t, \frac{1}{4}-\frac{s}{2}-t) &  (m_{x})\\
 (-\frac{1}{4}+\frac{s}{2}+t, -\frac{1}{4}-\frac{s}{2}+t, -\frac{1}{4}+\frac{s}{2}-t) &  (m_{y})
 \end{array}
\right.
\end{eqnarray}
(iii) (0 0 1)
\begin{eqnarray}
&& d_{\rm th}=c 
\\
&& {\bf k}_{\rm BIFP}(t)=\\
&& \hspace{5pt}
 \left\{
 \begin{array}{ll}
 (\frac{1}{4}+t, \frac{1}{4}+t, -\frac{1}{4}+t) &  (C_{2})\\
 (-\frac{1}{4}+t, \frac{1}{4}+t, -\frac{1}{4}+t) &  (C_{2})\\
  (t, \frac{1}{2}+t, -\frac{1}{2}+t) &  (C_{4})\\
 (\frac{1}{4}-\frac{s}{2}+t, \frac{1}{4}+\frac{s}{2}+t, -\frac{1}{4}-\frac{s}{2}+t) &  (m_{x})\\
 (-\frac{1}{4}+\frac{s}{2}+t, \frac{1}{4}+\frac{s}{2}+t, -\frac{1}{4}-\frac{s}{2}+t) &  (m_{y})
 \end{array}
\right.
\end{eqnarray}

\begin{figure}[h!]
 \begin{center}
  \includegraphics[scale=.36]{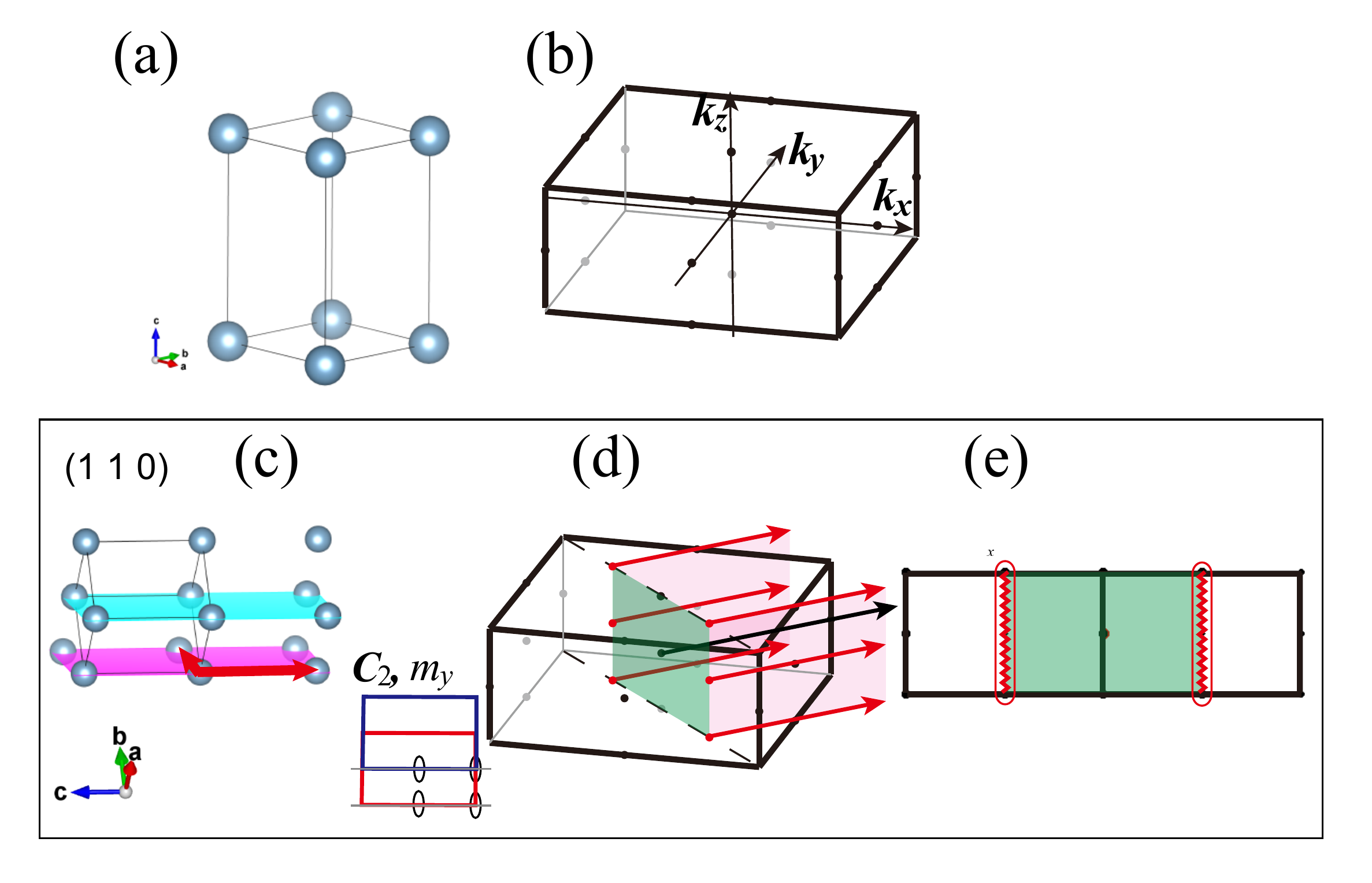}
  \caption{(a) Primitive tetragonal lattice and (b) its corresponding BZ. (c) Decomposition of the lattice into layers perpendicular to the (1 1 0) direction. (d) (e) BIFP yielded by the decomposition of panel (c) [(f)].}
  \label{fig:PT-sum}
 \end{center}
\end{figure}

\subsection{Primitive tetragonal (PT) lattice}
\begin{eqnarray}
{\bf a}_{1}=(a, 0, 0)
\\
{\bf a}_{2}=(0, a, 0)
\\
{\bf a}_{3}=(0, 0, c)
.
\end{eqnarray}

(i) (1 1 0)
\begin{eqnarray}
&& d_{\rm th}=\sqrt{2}a 
\\
&& {\bf k}_{\rm BIFP}(t)=\\
&& \hspace{5pt}
 \left\{
 \begin{array}{ll}
 (\frac{1}{4}+t, -\frac{1}{4}+t, 0) &  (C_{2})\\
 (\frac{1}{4}+t, -\frac{1}{4}+t, -\frac{1}{2}) &  (C_{2})\\
 (\frac{1}{4}+t, -\frac{1}{4}+t, -s) &  (m_{x})
 \end{array}
\right.
\end{eqnarray}

\begin{figure}[h!]
 \begin{center}
  \includegraphics[scale=.36]{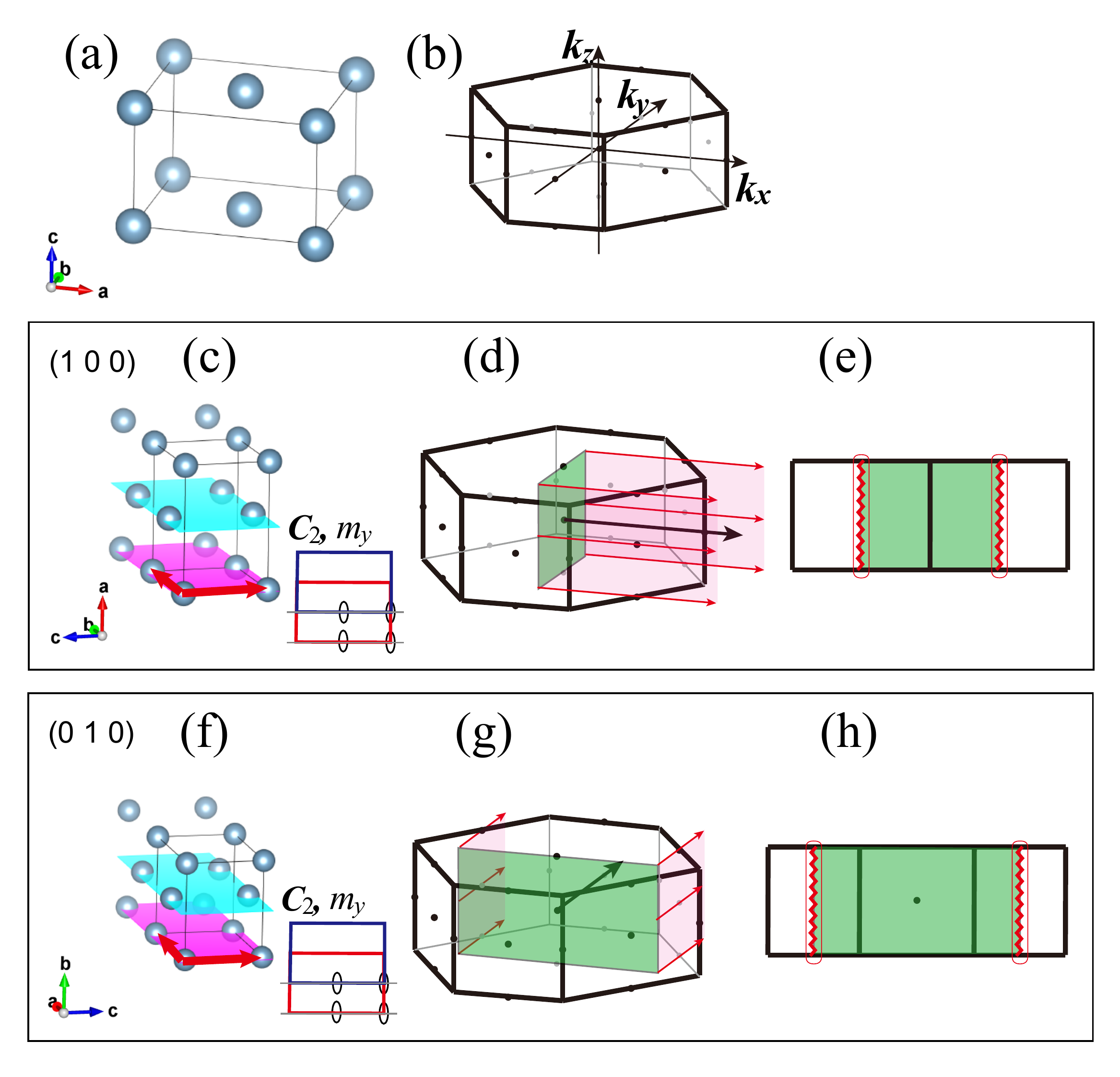}
  \caption{(a) Base-centered orthorhombic lattice and (b) its corresponding BZ. (c) [(f)] Decomposition of the lattice into layers perpendicular to the (1 0 0) [(0 1 0)] direction. The primitive lattice vectors for each layer are represented by arrows. The interlayer shift and possible in-plane symmetry elements having two or more symmetry references are shown in the right. (d) (e) [(g) (h)] BIFP yielded by the decomposition of panel (c) [(f)].}
  \label{fig:BCO-sum}
 \end{center}
\end{figure}

\subsection{Base centered orthorhombic (BCO) lattice}
\begin{eqnarray}
{\bf a}_{1}=(a/2,b/2,0)
\\
{\bf a}_{2}=(-a/2,b/2,0)
\\
{\bf a}_{3}=(0, 0, c)
.
\end{eqnarray}

(i) (0 1 0)
\begin{eqnarray}
&& d_{\rm th}=b 
\\
&& {\bf k}_{\rm BIFP}(t)=\\
&& \hspace{5pt}
 \left\{
 \begin{array}{ll}
 (\frac{1}{4}+t, -\frac{1}{4}+t, 0) &  (C_{2})\\
 (\frac{1}{4}+t, -\frac{1}{4}+t, -\frac{1}{2}) &  (C_{2})\\
 (\frac{1}{4}+t, -\frac{1}{4}+t, -s) &  (m_{x})
 \end{array}
\right.
\end{eqnarray}

(ii) (1 0 0)
\begin{eqnarray}
&& d_{\rm th}=a
\\
&& {\bf k}_{\rm BIFP}(t)=\\
&& \hspace{5pt}
 \left\{
 \begin{array}{ll}
 (\frac{1}{4}+t, \frac{1}{4}-t, 0) &  (C_{2})\\
 (\frac{1}{4}+t, \frac{1}{4}-t, \frac{1}{2}) &  (C_{2})\\
 (\frac{1}{4}+t, \frac{1}{4}-t, s) &  (m_{x})
 \end{array}
\right.
\end{eqnarray}

\begin{figure}[h!]
 \begin{center}
  \includegraphics[scale=.36]{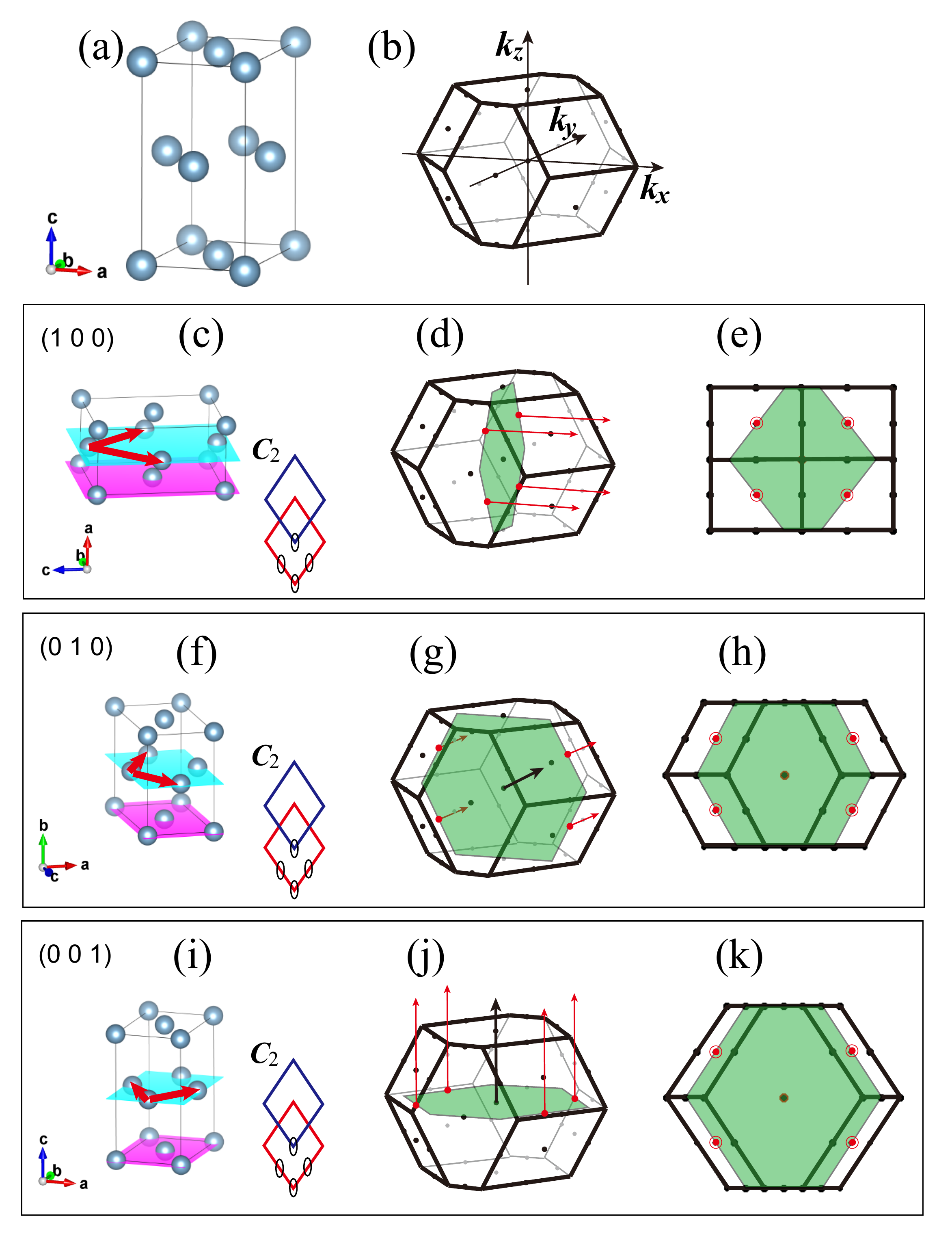}
  \caption{(a) Face-centered orthorhombic lattice and (b) its corresponding BZ. (c) (f) (i) Decomposition of the lattice into layers perpendicular to the (1 0 0), (0 1 0) and (0 0 1) directions, respectively. (d) (e), (g) (h), and (j) (k) BIFP yielded by the decomposition of panel (c), (f), and (i), respectively.}
  \label{fig:FCO-sum}
 \end{center}
\end{figure}

\subsection{Face centered orthorhombic (FCO) lattice}
\begin{eqnarray}
{\bf a}_{1}=(a/2,0,c/2)
\\
{\bf a}_{2}=(a/2,b/2,0)
\\
{\bf a}_{3}=(0, b/2, c/2)
.
\end{eqnarray}

(i) (0 0 1)
\begin{eqnarray}
&& d_{\rm th}=c
\\
&& {\bf k}_{\rm BIFP}(t)=\\
&& \hspace{5pt}
 \left\{
 \begin{array}{ll}
 (\frac{1}{4}+t, \frac{1}{2}, \frac{1}{4}+t) &  (C_{2})\\
 (-\frac{1}{4}+t, 0, \frac{1}{4}+t) &  (C_{2})
 \end{array}
\right.
\end{eqnarray}

(ii) (0 1 0)
\begin{eqnarray}
&& d_{\rm th}=b
\\
&& {\bf k}_{\rm BIFP}(t)=\\
&& \hspace{5pt}
 \left\{
 \begin{array}{ll}
 (0,\frac{1}{4}+t, -\frac{1}{4}+t) &  (C_{2})\\
 (-\frac{1}{2},-\frac{1}{4}+t, -\frac{1}{4}+t) &  (C_{2})
 \end{array}
\right.
\end{eqnarray}

(iii) (1 0 0)
\begin{eqnarray}
&& d_{\rm th}=a
\\
&& {\bf k}_{\rm BIFP}(t)=\\
&& \hspace{5pt}
 \left\{
 \begin{array}{ll}
 (-\frac{1}{4}+t, -\frac{1}{4}+t,-\frac{1}{2}) &  (C_{2})\\
 (-\frac{1}{4}+t,\frac{1}{4}+t, 0) &  (C_{2})
 \end{array}
\right.
\end{eqnarray}

\begin{figure}[h!]
 \begin{center}
  \includegraphics[scale=.36]{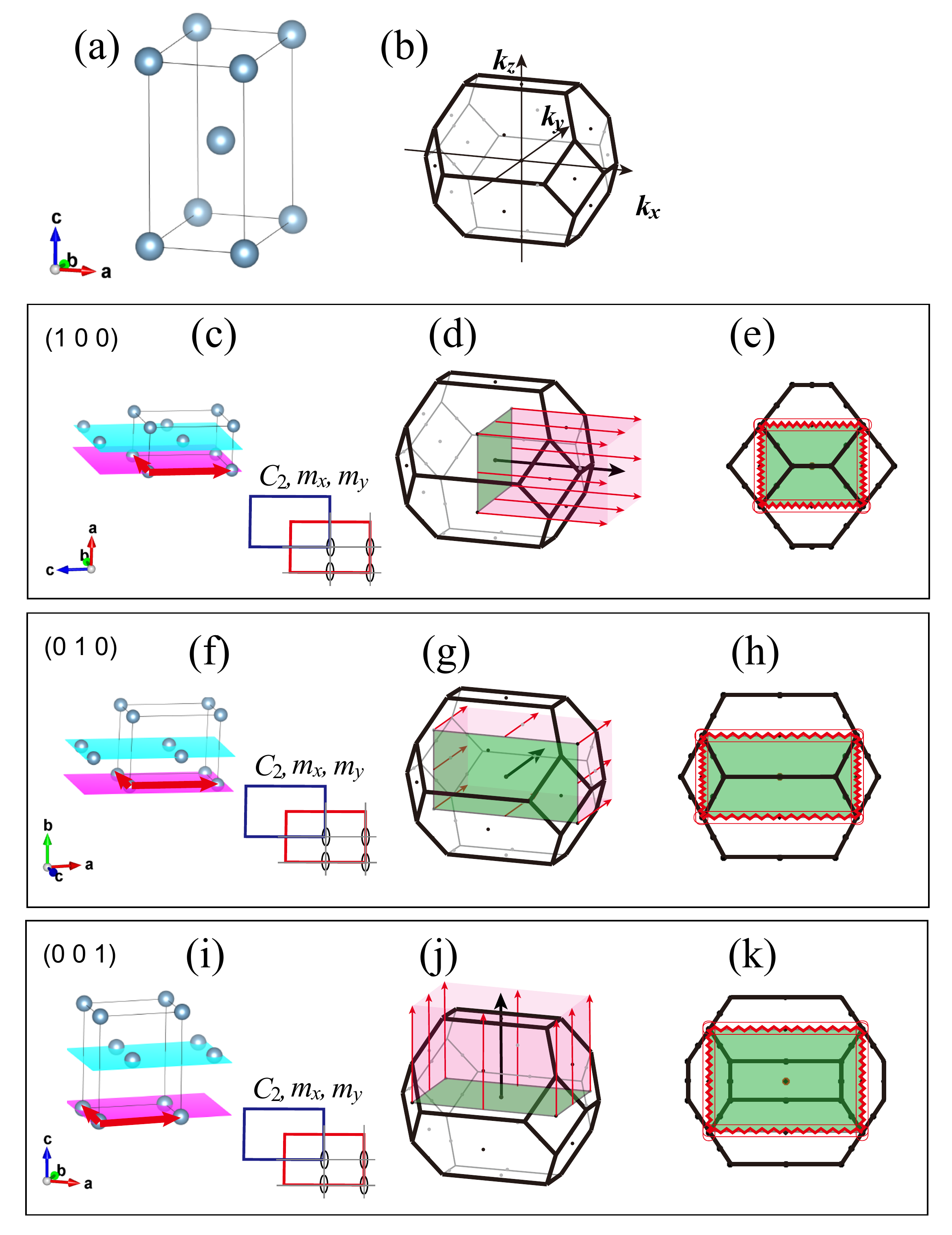}
  \caption{(a) Body-centered orthorhombic lattice and (b) its corresponding BZ. (c) (f) (i) Decomposition of the lattice into layers perpendicular to the (1 0 0), (0 1 0) and (0 0 1) direction. (d) (e), (g) (h), and (j) (k) BIFP yielded by the decomposition of panel (c), (f) and (i), respectively.}
  \label{fig:ICO-sum}
 \end{center}
\end{figure}

\subsection{Body centered orthorhombic (ICO) lattice}
\begin{eqnarray}
{\bf a}_{1}=(a/2,b/2,c/2)
\\
{\bf a}_{2}=(-a/2,b/2,c/2)
\\
{\bf a}_{3}=(-a/2, -b/2, c/2)
.
\end{eqnarray}

(i) (0 0 1)
\begin{eqnarray}
&& d_{\rm th}=c
\\
&& {\bf k}_{\rm BIFP}(t)=\\
&& \hspace{5pt}
 \left\{
 \begin{array}{ll}
 (\frac{1}{4}+t, -\frac{1}{4}+t, -\frac{1}{4}+t) &  (C_{2})\\
  (\frac{1}{4}+t, \frac{1}{4}+t, -\frac{1}{4}+t) &  (C_{2})\\
  (\frac{1}{4}+\frac{s}{2}+t, -\frac{1}{4}+\frac{s}{2}+t, -\frac{1}{4}-\frac{s}{2}+t) &  (m_{x})\\
    (\frac{1}{4}+\frac{s}{2}+t, \frac{1}{4}-\frac{s}{2}+t, -\frac{1}{4}-\frac{s}{2}+t) &  (m_{y})
 \end{array}
\right.
\end{eqnarray}

(ii) (0 1 0)
\begin{eqnarray}
&& d_{\rm th}=b
\\
&& {\bf k}_{\rm BIFP}(t)=\\
&& \hspace{5pt}
 \left\{
 \begin{array}{ll}
 (\frac{1}{4}+t, -\frac{1}{4}+t, -\frac{1}{4}-t) &  (C_{2})\\
  (-\frac{1}{4}+t, -\frac{1}{4}+t, -\frac{1}{4}-t) &  (C_{2})\\
  (\frac{1}{4}-\frac{s}{2}+t, -\frac{1}{4}-\frac{s}{2}+t, -\frac{1}{4}-\frac{s}{2}-t) &  (m_{x})\\
    (-\frac{1}{4}+\frac{s}{2}+t, -\frac{1}{4}-\frac{s}{2}+t, -\frac{1}{4}-\frac{s}{2}-t) &  (m_{y})
 \end{array}
\right.
\end{eqnarray}

(iii) (1 0 0)
\begin{eqnarray}
&& d_{\rm th}=a
\\
&& {\bf k}_{\rm BIFP}(t)=\\
&& \hspace{5pt}
 \left\{
 \begin{array}{ll}
 (-\frac{1}{4}+t, -\frac{1}{4}-t, \frac{1}{4}-t) &  (C_{2})\\
  (-\frac{1}{4}+t, -\frac{1}{4}-t, -\frac{1}{4}-t) &  (C_{2})\\
  (-\frac{1}{4}-\frac{s}{2}+t, -\frac{1}{4}-\frac{s}{2}-t, \frac{1}{4}-\frac{s}{2}-t) &  (m_{x})\\
    (-\frac{1}{4}-\frac{s}{2}+t, -\frac{1}{4}-\frac{s}{2}-t, -\frac{1}{4}+\frac{s}{2}-t) &  (m_{y})
 \end{array}
\right.
\end{eqnarray}

\begin{figure}[h!]
 \begin{center}
  \includegraphics[scale=.36]{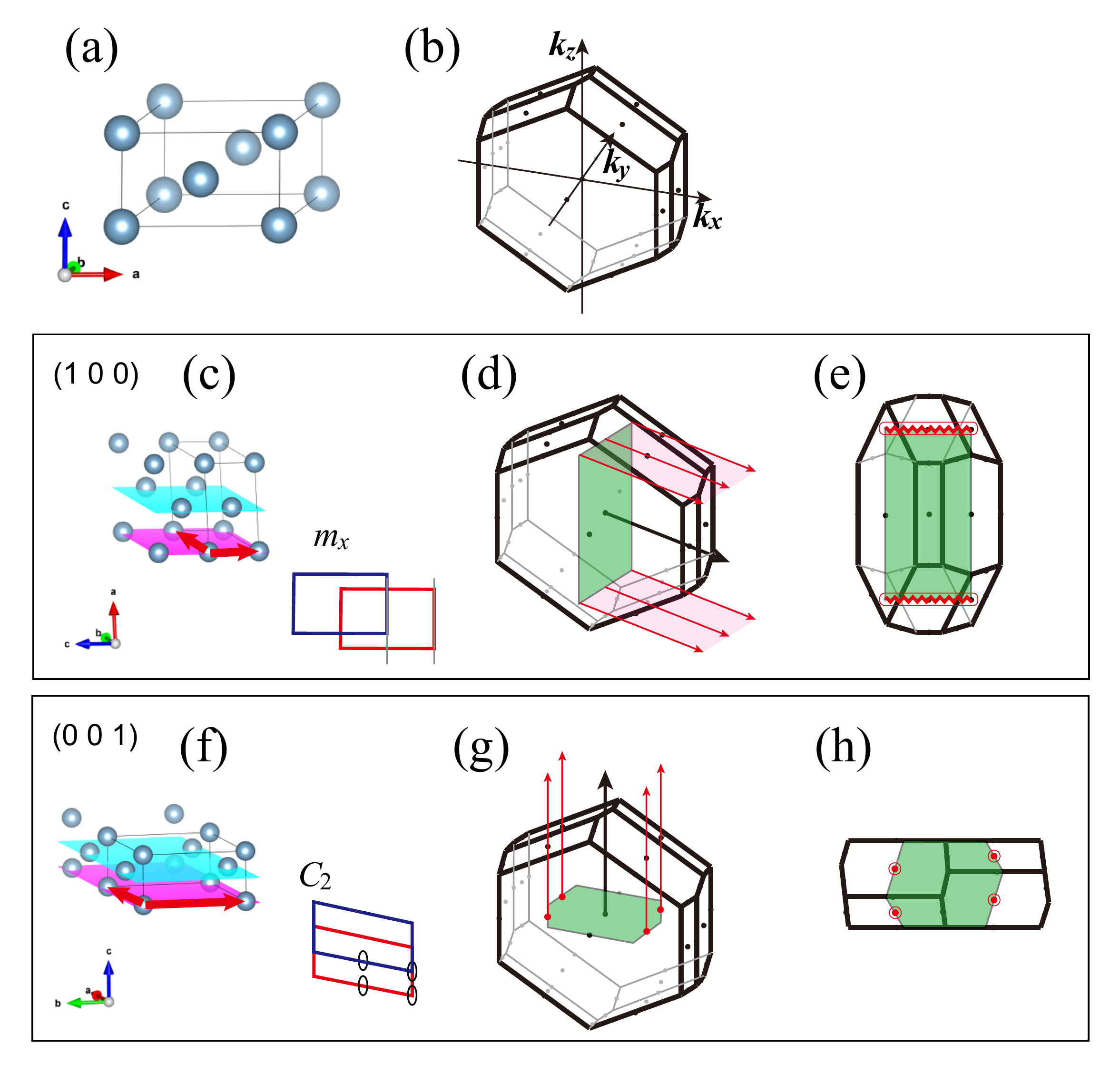}
  \caption{(a) Base-centered monoclinic lattice and (b) its corresponding BZ. (c) [(f)] Decomposition of the lattice into layers perpendicular to the (1 0 0) [(0 0 1)] direction. (d) (e) [(g) (h)] BIFP yielded by the decomposition of panel (c) [(f)].}
  \label{fig:BCM-sum}
 \end{center}
\end{figure}

\subsection{Base centered monoclinic (BCM) lattice}
\begin{eqnarray}
{\bf a}_{1}=(a/2, 0,-c/2)
\\
{\bf a}_{2}=(b\cos\gamma,b\sin\gamma,0)
\\
{\bf a}_{3}=(a/2,0,c/2)
.
\end{eqnarray}

(i) (0 0 1)
\begin{eqnarray}
&& d_{\rm th}=c 
\\
&& {\bf k}_{\rm BIFP}(t)=\\
&& \hspace{5pt}
 \left\{
 \begin{array}{ll}
 (\frac{1}{4}-t, 0, \frac{1}{4}+t) &  (C_{2})\\
 (\frac{1}{4}-t, \frac{1}{2}, \frac{1}{4}+t) &  (C_{2})
 \end{array}
\right.
\end{eqnarray}

(ii) (1 0 0)
\begin{eqnarray}
&& d_{\rm th}=a\sin\gamma 
\\
&& {\bf k}_{\rm BIFP}(t)=\\
&& \hspace{5pt}
 \left\{
 \begin{array}{ll}
 (\frac{1}{4}+\frac{a \cos\gamma}{2b}s+t, s, -\frac{1}{4}+\frac{a \cos\gamma}{2b}s+tt) &  (m_{x})
 \end{array}
\right.
\end{eqnarray}

\begin{figure}[h!]
 \begin{center}
  \includegraphics[scale=.36]{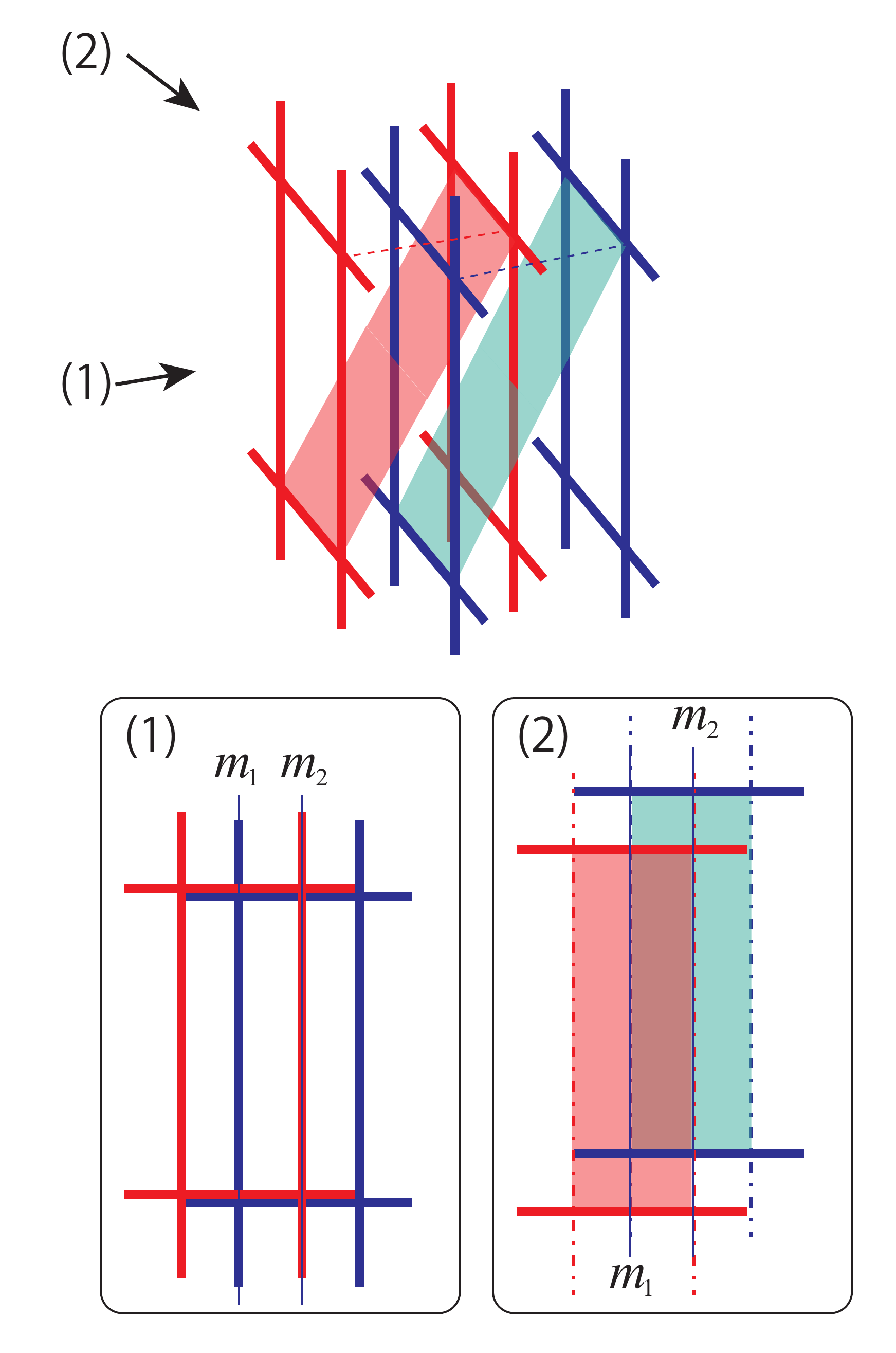}
  \caption{Schematic demonstration of the infinite number of decompositions yielded by the mirror symmetry. Red and Blue graphs represent two-dimensional layers having identical structure. When the lattice can be regarded from a certain direction (1) as the layers with uniform shift so that the different types of mirror planes ($m_{1}$ and $m_{2}$) of the neighboring layers are shared, there is always another direction [(2), for example] where the similar decomposition applies.}
  \label{fig:mirror}
 \end{center}
\end{figure}

\section{On the reflection-plane-shifted stacking}
\label{sec:infinite-decompose}
Among the in-plane symmetries, the mirror reflection $m$ is unique in that $\mathcal{K}_{\rm in}(m; {\bm \tau})$ can be non-empty for ${\bm \tau}={\bm \tau}(s)$ with $s$ being a continuous variable. In cases of three-dimensional crystals, this peculiar property yields the following. Suppose, seen from a certain direction, a certain three dimensional lattice can be decomposed into the layers stacked with the uniform shift ${\bm \tau}$ so that $\mathcal{K}_{\rm in}(m;{\bm \tau})$ is not empty; we can find an infinite number of possible decompositions so that $\mathcal{K}_{\rm in}(m;{\bm \tau})$ is not empty by changing the viewing direction. This point is illustrated in Fig.~\ref{fig:mirror}. Note that many of such alternative decompositions do not yield useful BIFP since their corresponding threshold distances $d_{\rm th}$s are smaller than the minimum interatomic distance. In Appendix~\ref{sec:list-BZ}, we have therefore shown only the representative BIFPs.

\end{appendix}


\begin{thebibliography}{999}
\bibitem{Tasaki-review-PTP1998} H. Tasaki, {\it From Nagaoka's Ferromagnetism to Flat-Band Ferromagnetism and Beyond}, Prog. Theor. Phys. \textbf{99}, 489 (1998).
\bibitem{Lieb-PRL1989} E. H. Lieb, {\it Two theorems on the Hubbard model}, Phys. Rev. Lett. \textbf{62}, 1201 (1989).
\bibitem{Mielke-JPhysA-1} A. Mielke, {\it Ferromagnetic ground states for the Hubbard model on line graphs}, J. Phys. A \textbf{24}, L73 (1991).
\bibitem{Mielke-JPhysA-2} A. Mielke, {\it Ferromagnetism in the Hubbard model on line graphs and further considerations}, J. Phys. A \textbf{24}, 3311 (1991).
\bibitem{Tasaki-PRL1992} H. Tasaki, {\it Ferromagnetism in the Hubbard models with degenerate single-electron ground states}, Phys. Rev. Lett. \textbf{69}, 1608 (1992).
\bibitem{Mielke-Tasaki-MathPhys1993} A. Mielke and H. Tasaki, {\it Ferromagnetism in the Hubbard model}, Commun. Math. Phys. \textbf{158}, 341 (1993).
\bibitem{Bergholtz-Liu-review2013} E. J. Bergholtz and Z. Liu, {\it Topological flat band models and fractional Chern insulators}, Int. J. Mod. Phys. B, \textbf{27}, 1331107 (2013).
\bibitem{Seitz-grouptheory1936} F. Seitz, {\it On the reduction of space groups}, Ann. Math. \textbf{37}, 17 (1936). 
\bibitem{Geim-Novoselov-graphene2004} K. S. Novoselov, A. K. Geim, S. V. Morozov, D. Jiang, Y. Zhang, S. V. Dobonos, I. V. Grigorieva, and A. A. Firsov, {\it Electric field effect in atomically thin carbon films}, Science \textbf{306}, 666 (2004).
\bibitem{Geim-Novoselov-thinlayer} K. S. Novoselov, D. Jiang, F. Schedin, T. J. Booth, V. V. Khotkevich, S. V. Morozov, and A. K. Geim, {\it Two-dimensional atomic crystals}, Proc. Natl. Acad. Sci. U. S. A. \textbf{102}, 10451 (2005).
\bibitem{Park-Lee-1Tprime-MoTe2} J. C. Park, S. J. Yun, H. Kim, J.-H. Park, S. H. Chae, S.-J. An, J.-G. Kim, S. M. Kim, K. K. Kim, and Y. H. Lee, {\it Phase-Engineered Synthesis of Centimeter-Scale 1T'- and 2H-Molybdenum Ditelluride Thin Films}, ACS NANO, \textbf{9}, 6548 (2015).
\bibitem{Zhou-Dresselhaus-1Tprime-MoTe2} L. Zhou, K. Xu, A. Zubair, A. D. Liao, W. Fang, F. Ouyang, Y.-H. Lee, K. Ueno, R. Saito, T. Palacios, J. Kong, M. S. Dresselhaus, {\it Large-Area Synthesis of High-Quality Uniform Few-Layer MoTe$_2$}, J. Am. Chem. Soc. \textbf{137}, 11892 (2015).
\bibitem{Naylor-Johnson-1Tprime-MoTe2-monolayer} C. H. Naylor, W. M. Parkin, J. Ping, Z. Gao, Y. R. Zhou, Y. Kim, F. Streller, R. W. Carpick, A. M. Rappe, M. Drndi\'{c}, J. M. Kikkawa, and A. T. C. Johnson, {\it Monolayer Single-Crystal 1T'-MoTe$_2$ Grown by Chemical Vapor
Deposition Exhibits Weak Antilocalization Effect}, Nano Lett. 16, 4297 (2016).
\bibitem{Li-Zhang-phosphorene-transistor-nnano}  L. Li, Y. Yu, G. J. Ye, Q. Ge, X. Ou, H. Wu, D. Feng, X. H. Chen, and Y. Zhang, {\it Black phosphorus field-effect transistors}, Nat. Nanotechnol. \textbf{9}, 372 (2014).
\bibitem{Liu-Tomanek-phosphorene-mobility-acsnano} H. Liu, A. T. Neal, Z. Zhu, Z. Luo, X. Xu, D. Tom\'{a}nek, and P. D. Ye, {\it Phosphorene: An unexplored 2D semiconductor with a high hole mobility}, ACS Nano \textbf{8}, 4033 (2014).
\bibitem{CastroNeto-graphene-review2009} A. H. Castro Neto, F. Guinea, N. M. R. Peres, K. S. Novoselov, and A. K. Geim, {\it The electronic properties of graphene}, Rev. Mod. Phys. \textbf{81}, 109 (2009).
\bibitem{Xu-Heinz-review-NPhys2014} X. Xu, W. Yao, D. Xiao, and T. F. Heinz, {\it Spin and pseudospins in layered transition metal dichalcogenides}, Nat. Phys. \textbf{10}, 343 (2014).
\bibitem{Qian-review-2DMater} X. Qian, Y. Wang, W. Li, J. Lu, and J. Li, {\it Modeling of stacked 2D materials and devices}, 2D Mater. \textbf{2}, 032003 (2015).
\bibitem{Tomanek-DMC-phosphorene} L. Schlenburger, A. D. Baczewski, Z. Zhu, J. Guan, and D. Tom\'{a}nek, {\it The Nature of the Interlayer Interaction in Bulk and Few-Layer Phosphorus}, Nano Lett., \textbf{15}, 8170 (2015).
\bibitem{McClure-model-graphite1957} J. W. McClure, {\it Band Structure of Graphite and de Haas-van Alphen Effect}, Phys. Rev. \textbf{108}, 612 (1957).
\bibitem{Slonczewski-Weiss-model-graphite1958} J. C. Slonczewski and P. R. Weiss, {\it Band structure of graphite}, Phys. Rev. \textbf{109}, 272 (1957).
\bibitem{McClure-rhombo-graphite1969} J. W. McClure, {\it Electron energy band structure and electronic properties of rhombohedral graphite}, Carbon \textbf{7}, 425 (1969).
\bibitem{Charlier-Gonze-multilayer-graphite1994} J. C. Charlier, X. Gonze, and J. P. Michenaud, {\it First-principles study of the stacking effect on the electronic properties of graphite(s)}, Carbon \textbf{32}, 289 (1994).
\bibitem{Kresse-graphite-BN-band1996} J. Furthm\"uller, J. Hafner, and G. Kresse, {\it Ab initio calculation of the structural and electronic properties of carbon and boron
nitride using ultrasoft pseudopotentials} Phys. Rev. B \textbf{50}, 15606 (1994). 
\bibitem{Coehoorn-Wold-TMDC-bandstruct-PES} R. Coehoorn, C. Haas, J. Dijkstra, c. J. F. Flipse, R. A. de Groot, and A. Wold, {\it Electronic structure of MoSez, MoS2, and WSez. I. Band-structure calculations and photoelectron spectroscopy}, Phys. Rev. B \textbf{35}, 6195 (1987).
\bibitem{Dawson-Bullett-TMTe2} W. G. Dawson and D. W. Bullett, {\it Electronic structure and crystallography of MoTe$_{2}$ and WTe$_{2}$}, J. Phys. C: Solid State Phys. \textbf{20}, 6159 (1987).
\bibitem{Partoens-multilayer2006} B. Partoens and F. M. Peeters, {\it From graphene to graphite: Electronic structure around the K point}, Phys. Rev. B \textbf{74}, 075404 (2006).
\bibitem{Liu-Shen-BN-stack} L. Liu, Y. P. Feng, and Z. X. Shen, {\it Structural and electronic properties of h-BN}, Phys. Rev. B \textbf{68}, 104102 (2003).
\bibitem{Aoki-Amawashi-SSComm} M. Aoki and H. Amawashi, {\it Dependence of band structures on stacking and field in layered graphene}, Solid State Commun. \textbf{142}, 123 (2007).
\bibitem{Chiu-Shyu-r-graphite} C.-W. Chiu, Y.-C. Huang, S.-C. Chen, M.-F. Lin, and
F.-L. Shyu, {\it Low-frequency electronic and optical properties of rhombohedral graphite} Phys. Chem. Chem. Phys. \textbf{13}, 6036 (2011).
\bibitem{MacDonald-graphene-chiraldecomp} H. Min and A. H. MacDonald, {\it Chiral decomposition in the electronic structure of graphene multilayers}, Phys. Rev. B \textbf{77}, 155416 (2008).
\bibitem{Suzuki2014} R. Suzuki, M. Sakano, Y. J. Zhang, R. Akashi, D. Morikawa, A. Harasawa, K. Yaji, K. Kuroda, K. Miyamoto, T. Okuda, K. Ishizaka, R. Arita, and Y. Iwasa, {\it Valley-dependent spin polarization in bulk MoS2
with broken inversion symmetry}, Nat. Nanotechnol. \textbf{9}, 611 (2014).
\bibitem{Akashi2015} R. Akashi, M. Ochi, R. Suzuki, S. Bord\'{a}cs, Y. Tokura, Y. Iwasa, and R. Arita. {\it Two-dimensional valley electrons and excitons in noncentrosymmetric 3R-MoS$_{2}$}, Phys. Rev. Applied \textbf{4}, 014002 (2015).
\bibitem{footnote-net-origin}Although the grid points of the net do not have to be placed at any atomic sites, we obey a convention to place them so that the rotational axes $C_{n}$ with maximum $n$ pass through the grid points.
\bibitem{footnote-1d} The one-dimensional variant of the net is called chain. Trivially, there is only one kind of chain, which is generated by a single primitive translation vector. The only compatible ``in-plane" operation is the mirror $m$ perpendicular to the chain or two-fold rotation $C_{2}$ around any axis normal to the chain, either of which yields $\mathcal{K}_{\rm in}(S; \tau$$=$$a/2)=\pi$ as demonstrated in Sec.~\ref{sec:one-dim}. The formal definition of $\mathcal{K}_{\rm in}$, characterizing where the interference affects the hybridization, appears later in Eq.~(\ref{eq:K-in-def}).
\bibitem{inui}  T. Inui, Y. Tanabe, and Y. Onodera, {\it Group Theory and Its Applications in Physics} (Springer-Verlag, Berlin, 1990)
\bibitem{comment-v12} Note that the same result also applies if we take $|i,{\bf k}\rangle_{v}$ as the basis functions instead of $|i,{\bf k}\rangle$, where the former are eigenstates of the layer-diagonal part of the Hamitonian $\mathcal{H}({\bm \tau})-h_{12}({\bm \tau})$. This is because $v_{1}$ and $v_{2}$ does not change the symmetry properties of the wave functions within the interference manifold.
\bibitem{Warner-Buchner-BNbilayer-topography2010} J. H. Warner, M. H. R\"ummeli, A. Bachmatiuk, and B. B\"uchner, {\it Atomic resolution imaging and topography of boron nitride sheets produced by chemical exfoliation}, ACS Nano \textbf{4}, 1299 (2010).
\bibitem{Ooi-Adams-variousBN2006} N. Ooi, A. Rairkar, L. Lindsley, and J. B. Adams, {\it Electronic structure and bonding in hexagonal boron nitride}, J. Phys.: Condens. Matter \textbf{18}, 97 (2006).
\bibitem{Constantinescu-Kuc-BNstacking-PRL2013} G. Constantinescu, A. Kuc, and T. Heine, {\it Stacking in bulk and bilayer hexagonal boron nitride}, Phys. Rev. Lett. \textbf{111}, 036104 (2013).
\bibitem{Liu-review2015} G.-B. Liu, D. Xiao, Y. Yao, X. Xu, and W. Yao, {\it Electronic structures and theoretical modelling of two-dimensional group-VIB transition metal dichalcogenides}, Chem. Soc. Rev. \textbf{44}, 2643 (2015).
\bibitem{Xiao-MoS2-PRL} D. Xiao, G. B. Liu, W. Feng, X. Xu, and Wang Yao,  {\it Coupled Spin and Valley Physics in Monolayers of MoS2 and Other Group-VI Dichalcogenides}, Phys. Rev. Lett. \textbf{108}, 196802 (2012).
 \bibitem{comment-valence} Note that we further need to examine the phase factor of Eq.~(\ref{eq:theorem-proof-multi}) to clarify whether the hybridization is allowed for the valence-top state (and one then finds that this is the case). As a corollary, we can also derive that the hybridizations are prohibited for the both states when the 2H bilayer adopt a shift of $-{\bm \tau}$ instead.
 \bibitem{Morita-review1986} A. Morita, {\it Semiconducting black phosphorus}, Appl. Phys. A: Solids Surf. \textbf{39}, 227 (1986).
\bibitem{Liu-Ye-review2015} H. Liu, Y. Du, Y. Deng and P. D. Ye, {\it Semiconducting black phosphorus: synthesis, transport properties and electronic applications}, Chem. Soc. Rev. \textbf{44}, 2732 (2015).
\bibitem{Ling-Dresselhaus-review2015} X. Ling, H. Wang, S. Huang, F. Xia, and M. S. Dresselhaus, {\it The renaissance of black phosphorus}, Proc. Natl. Acad. Sci. USA, \textbf{112}, 4523 (2015)
\bibitem{Carvalho-CastroNeto-review2016} {\it Phosphorene: from theory to applications} A. Carvalho, M. Wang, X. Zhu, A. S. Rodin, H. Su, and A. H. Castro Neto, Nat. Rev. Mater. \textbf{1}, 16061 (2016).
\bibitem{Herring-nonsymmolph} C. Herring, {\it Effect of Time-Reversal Symmetry on Energy Bands of Crystals}, Phys. Rev. \textbf{52}, 361 (1937).
\bibitem{Heine-textbook} V. Heine, {\it Group Theory in Quantum Mechanics: Introduction to the Present Usage} (Pergamon, New York, 1960) 2nd ed.
\bibitem{Keyes-P-measurement-PR1958}R. W. Keyes, {\it The Electrical Properties of Black Phosphorus}, Phys. Rev. \textbf{92}, 580 (1953).
\bibitem{Akahama-P-hall-measure-JPSJ1983} Y. Akahama, S. Endo, and S. Narita, {\it Electrical Properties of Black Phosphorus Single Crystals}, J. Phys. Soc. Jpn. \textbf{52}, 2148(1983).
\bibitem{PBEGGA} J. P. Perdew, K. Burke, and M. Ernzerhof, {\it Generalized Gradient Approximation Made Simple}, Phys. Rev. Lett. \textbf{77}, 3865 (1996).
\bibitem{Tran-Yang-P-calc-PRB2014} V. Tran, R. Soklaski, Y. Liang, and L. Yang, {\it Layer-controlled band gap and anisotropic excitons in few-layer black phosphorus}, Phys. Rev. B \textbf{89}, 235319 (2014).
\bibitem{Kim-graphene-pret-a-porte-PRB} K. Kim, Z. Lee, B. D. Malone, K. T. Chan, B. Alem\'an, W. Regan, W. Gannett, M. F. Crommie, M. L. Cohen, and A. Zettl, {\it Multiply folded graphene}, Phys. Rev. B {\textbf 83}, 245433 (2011).
\bibitem{Jiang-Wu-MoS2-fold-NNano}T. Jiang, H. Liu, D. Huang, S. Zhang, Y. Li, X. Gong, Y.-R. Shen, W.-T. Liu, and S. Wu, {\it Valley and band structure engineering of folded MoS$_{2}$ bilayers}, Nat. Nanotechnol. {\textbf 9}, 825 (2014).
\bibitem{Mizuguchi-Bi4O4S3-PRBR2012} Y. Mizuguchi, H. Fujihisa, Y. Gotoh, K. Suzuki, H. Usui, K. Kuroki, S. Demura, Y. Takano, H. Izawa, and O. Miura, {\it BiS$_{2}$-based layered superconductor Bi$_{4}$O$_{4}$S$_{3}$}, Phys. Rev. B \textbf{86}, 220510(R) (2012).
\bibitem{Mizuguchi-LaOFBiS2-JPSJ2012}  Y. Mizuguchi, S. Demura, K. Deguchi, Y. Takano, H. Fujihisa, Y. Gotoh, H. Izawa, and O. Miura, {\it Superconductivity in Novel BiS$_{2}$-Based Layered Superconductor LaO$_{1-x}$F$_{x}$BiS$_{2}$}, J. Phys. Soc. Jpn. \textbf{81}, 114725 (2012)
\bibitem{Wolowiec-Yazici-pressure-PRB2013} C. T. Wolowiec, D. Yazici, B. D. White, K. Huang, and M. B. Maple, {\it Pressure-induced enhancement of superconductivity and suppression of semiconducting behavior in LnO$_{0.5}$F$_{0.5}$BiS$_{2}$ ($Ln$=La,Ce) compounds}, Phys. Rev. B \textbf{88}, 064503 (2013).
\bibitem{Tomita-Mizuguchi-pressure-structure-JPSJ2014} T. Tomita, M. Ebata, H. Soeda, H. Takahashi, H. Fujihisa, Y. Gotoh, Y. Mizuguchi, H. Izawa, O. Miura, S. Demura, K. Deguchi, and Y. Takano, {\it Pressure-induced enhancement of superconductivity and structural transition in BiS$_{2}$-Layered LaO$_{1-x}$F$_{x}$BiS$_{2}$}, J. Phys. Soc. Jpn. \textbf{83}, 063704 (2014).
\bibitem{Guo-Yuan-EuBiS2F-pressure-structure-PRB2015} C. Y. Guo, Y. Chen, M. Smidman, S. A. Chen, W. B. Jiang, H. F. Zhai, Y. F. Wang, G. H. Cao, J. M. Chen, X. Lu, and H. Q. Yuan, {\it Evidence for two distinct superconducting phases in EuBiS$_{2}$F under pressure}, Phys. Rev. B \textbf{91}, 214512 (2015).
\bibitem{Ochi-BiS2-JPSJ} M. Ochi, R. Akashi, and K. Kuroki, {\it Strong bilayer coupling induced by the symmetry breaking in the monoclinic phase of BiS$_{2}$-based superconductors}, J. Phys. Soc. Jpn. {\textbf 85}, 094705 (2016).
\bibitem{Mak-Heinz-fewlayer-graphene-PRL2010} K. F. Mak, J. Shan, and T. F. Heinz, {\it Electronic structure of few-layer graphene: Experimental demonstration of strong dependence on stacking sequence}, Phys. Rev. Lett. \textbf{104}, 176404 (2010).
\bibitem{xTAPP} {\it xTAPP}, which is developed by Yoshihide Yoshimoto, Jun Yamauchi and Kanako Yoshizawa, is the extended version of {\it Tokyo Ab-initio Program Package (TAPP)}~\cite{TAPP-Oshiyama, TAPP-Yamauchi}; http://ma.cms-initiative.jp/en/application-list/xtapp?set\_language=en.
\bibitem{TAPP-Oshiyama} O. Sugino and A. Oshiyama, {\it Vacancy in Si: Successful description within the local-density approximation}, Phys. Rev. Lett. \textbf{68}, 1858 (1992).
\bibitem{TAPP-Yamauchi} J. Yamauchi, M. Tsukada, S. Watanabe, and O. Sugino, {\it First-principles study on energetics of c-BN(001) reconstructed surfaces}, Phys. Rev. B \textbf{54}, 5586 (1996).
\bibitem{Troullier-Martins} N. Troullier and J. L. Martins, {\it Efficient pseudopotentials for plane-wave calculations}, Phys. Rev. B \textbf{43}, 1993 (1991).
\bibitem{WIEN2k} P. Blaha, K. Schwarz, G. K. H. Madsen, D. Kvasnicka, and J. Luitz, WIEN2k: {\it An Augmented Plane Wave+Local Orbitals Program for Calculating Crystal Properties} (Karlheinz Schwarz, Techn. Universit\"{a}t Wien, Austria, 2001), ISBN 3-9501031-1-2; http://www.wien2k.at.
\bibitem{Ceperley-Alder} D. M. Ceperley and B. J. Alder, {\it Ground state of the electron gas by a stochastic method}, Phys. Rev. Lett. \textbf{45}, 566 (1980).
\bibitem{LDA-PZ} J. P. Perdew and A. Zunger, {\it Self-interaction correction to density-functional approximations for many-electron systems}, Phys. Rev. B \textbf{23}, 5048 (1981).
\bibitem{BN-Xray-exp} R. S. Pease, {\it An X-ray Study of Boron Nitride}, Acta Cryst. \textbf{5}, 356 (1952).
\bibitem{graphite-Xray-exp} Y. X. Zhao and I. L. Spain, {\it X-ray diffraction data for graphite to 20GPa}, Phys. Rev. B \textbf{40}, 993
(1989).
\bibitem{P-As-Xray-exp} I. Shirotani, S. Shiba, K. Takemura, O. Shimomura, and T. Yagi, {\it Pressure-induced phase transitions of phosphorus-arsenic alloys}, Physica B: Condens. Matter \textbf{190}, 169 (1993).
\bibitem{Zintl-Harder-LiH} E. Zintl and A. Harder, {\it Gitterdimensionen des Lithiumhydrids LiH und Lithiumdeuterids LiD}, Z. Phys. Chem.\textbf{27}--\textbf{28}, 478 (1935).
\bibitem{Nickels-Wallace-NaCl} J. E. Nickels, M. A. Fineman, and W. E. Wallace, {\it X-ray diffraction studies of sodium chloride--sodium bromide solid solutions}, J. Phys. Colloid Chem. \textbf{53}, 625 (1949).
\bibitem{BZ-list} W. Setyawan, S. Curtarolo, {\it High-throughput electronic band structure calculations: Challenges and tools}, Comp. Mat. Sci. \textbf{49}, 299 (2010).
\end{thebibliography}
\end{document}